\documentclass[%
 reprint,
superscriptaddress,
 amsmath,amssymb,
 aps,
]{revtex4-2}

\usepackage{graphicx}% Include figure files
\usepackage{ragged2e}
\usepackage{caption}

\usepackage{subcaption} % include subfigures
\usepackage{dcolumn}% Align table columns on decimal point
\usepackage{bm}% bold math
\setcitestyle{super,sort&compress}
\usepackage[most]{tcolorbox}
\usepackage{placeins}

\usepackage{graphicx}
\usepackage[percent]{overpic}

\usepackage[T1]{fontenc}
\usepackage[english]{babel} % Oppure 'american' o 'british'

\usepackage{xcolor}

\usepackage{hyperref}% add hypertext capabilities
\hypersetup{
    colorlinks=true,
    linkcolor=blue,
    filecolor=magenta,
    urlcolor=cyan,
}

\begin{document}

\preprint{APS/123-QED}

\title{Nonadiabatic polaron dynamics from machine-learned hopping trajectories}% Force 

\author{Daniela Mangano}
\email{daniela.mangano@univie.ac.at}
\affiliation{University of Vienna, Faculty of Physics and Center for Computational Materials Science, Vienna, Austria}
\affiliation{Department of Physics and Astronomy “Augusto Righi”, Alma Mater Studiorum - Università di Bologna, Bologna, 40127 Italy}

\author{Luca Leoni}
\affiliation{Department of Physics and Astronomy “Augusto Righi”, Alma Mater Studiorum - Università di Bologna, Bologna, 40127 Italy}

\author{Zhongfei Xu}
\affiliation{College of Environmental Science and
Engineering, North China Electric Power University, Beijing
102206, China}

\author{Li-Min Liu}
\affiliation{School of Physics, Beihang University,
Beijing 100191, China}

\author{Cesare Franchini}
\email{cesare.franchini@univie.ac.at}
\affiliation{University of Vienna, Faculty of Physics and Center for Computational Materials Science, Vienna, Austria}
\affiliation{Department of Physics and Astronomy “Augusto Righi”, Alma Mater Studiorum - Università di Bologna, Bologna, 40127 Italy}

\begin{abstract}
The formation and motion of small polarons are ubiquitous in energy materials and critically influence the photoinduced processes. Polaron transport is governed by thermally activated lattice dynamics and rare hopping events, requiring spatial and temporal scales beyond those accessible to conventional first-principle approaches but attainable with machine-learning molecular dynamics, which enables extensive sampling of atomic trajectories and polaron configurations over nanoseconds. Moreover, the Born-Oppenheimer approximation 
underlying standard ab initio molecular dynamics cannot describe the nonadiabatic coupling between electronic states mediated by nuclear motion, which controls energy exchange, charge localization, and excited-state relaxation. Here, we develop a simulation framework that combines machine-learning-assisted polaron dynamics with nonadiabatic molecular dynamics employing decoherence-induced surface hopping to bridge these complementary aspects of polaron physics, and determine how different polaron-hopping mechanisms affect excited-state dynamics. The enlarged trajectory space allows us to provide the first direct dynamical confirmation of the adiabatic/nonadiabatic classification of polaron hopping regimes originally proposed on the basis of static Marcus-theory arguments, and to connect individual hopping events directly to nonadiabatic electronic transitions. In bulk TiO$_2$, nonadiabatic hoppings occur preferentially when the energy of the polaron-induced gap state approaches the conduction-band manifold, promoting both electronic de-excitation and, via a Landau–Zener mechanism, thermally activated excitation from the ground state. In MgO, the predominantly adiabatic character of hole-polaron hopping suppresses these effects, leading to longer charge-carrier lifetimes and an absence of thermally activated excitations. At the reduced rutile TiO$_2$(110) surface, polaron hopping events are much more rare than in bulk because the excess electron is strongly stabilized in the subsurface region, leading to correspondingly longer de-excitation times. Crucially, hopping from subsurface to reactive surface is predominantly nonadiabatic, identifying nonadiabatic electron-phonon coupling as an essential ingredient in polaron migration associated photocatalytic activity.

\end{abstract}

\maketitle

\section{INTRODUCTION}

Small polarons are quasiparticles formed when an excess charge (electron 
or hole) localizes in a polarizable material and couples to a surrounding 
lattice distortion~\cite{landau1933electron,shluger1993small,
franchini2021polarons}. Their formation and dynamics critically influence 
charge transport, optical and excited-state properties, surface chemistry, 
catalysis, and superconductivity in a broad range of materials, including 
transition-metal oxides and halide 
perovskites~\cite{morita2023models,pastor2022electronic,
meggiolaro2020polarons}.

In energy-conversion and energy-storage materials, their role can be 
either beneficial or detrimental depending on the physical process 
involved. In photovoltaic and light-harvesting systems, carrier 
self-trapping reduces mobility and diffusion lengths and enhances 
nonradiative recombination, thereby limiting charge extraction and device 
performance~\cite{ambrosio2025charge}. At the same time, small-polaron 
hopping can provide an effective electronic conduction mechanism in 
otherwise poorly conducting oxides~\cite{brunin2019high}. The interplay 
between localization and transport is particularly important in 
photocatalysis: polarons interact strongly with 
adsorbates~\cite{Reticcioli2019} and can substantially modify 
charge-carrier dynamics and surface 
reactivity~\cite{zhang2019state,tanner2022tio2,sombut2022role}. 
Localized electrons and holes can directly participate in surface redox 
reactions, including proton reduction and water 
oxidation~\cite{tanner2022tio2}, but excessive localization may trap 
photogenerated carriers in the bulk and hinder their transport toward 
reactive surface sites, reducing the quantum efficiency of the 
photocatalytic process. The overall role of polarons therefore results 
from a delicate balance between charge localization, charge separation, 
and carrier migration toward chemically active 
regions~\cite{rousseau2020theoretical,chen2020small,wu2026potential}. 
Understanding not only where polarons form, but also how they move and 
exchange energy with their environment, is thus essential for establishing 
microscopic design principles for functional energy materials.

\begin{figure*}[htbp]
    \centering
     \begin{subfigure}{.49\textwidth}
         \centering
         \begin{overpic}[width=\linewidth]{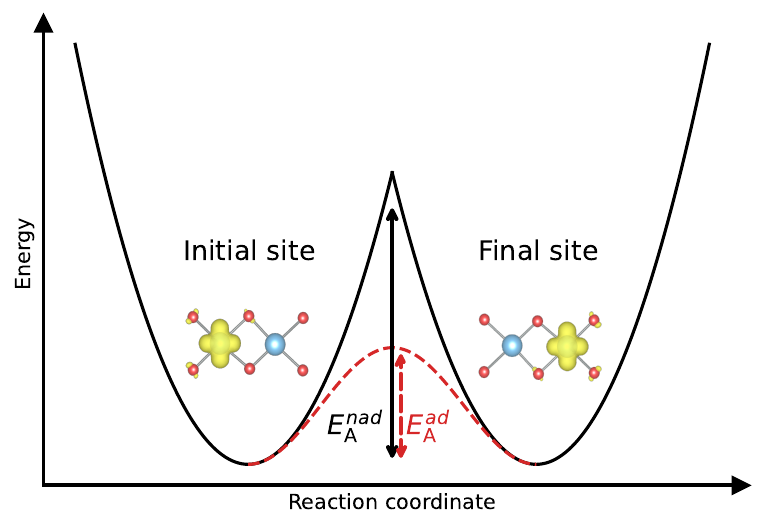}
            \put(15,62){\large (a)}
        \end{overpic}
         \phantomsubcaption
         \label{fig:Fig1a_Marcus}
    \end{subfigure}
    \begin{subfigure}{.49\textwidth}
         \centering
         \begin{overpic}[width=\linewidth]{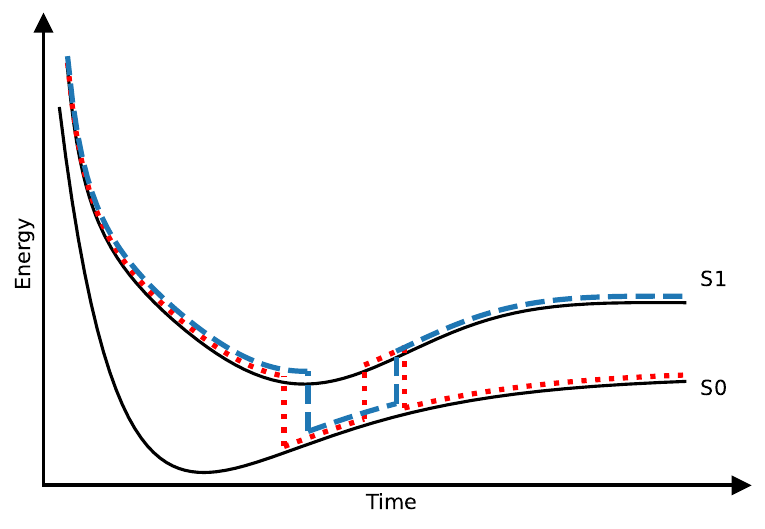}
            \put(15,62){\large (b)}
        \end{overpic}
         \phantomsubcaption
         \label{fig:Fig1b_sh}
    \end{subfigure}
        \caption{\justifying (a) Schematic representation of adiabatic 
        and nonadiabatic polaron hopping according to the 
        Deskins--Dupuis classification. The transition-state energy 
        landscape is inspired by Marcus 
        theory~\cite{marcus1956theory,marcus2020electron}. Stronger 
        electronic coupling between the initial and final localized 
        states in the adiabatic regime lowers the activation energy 
        barrier ($E_{A}^{ad}$) compared with the nonadiabatic hopping 
        process ($E_{A}^{nad}$). (b) Schematic representation of the 
        decoherence-induced surface hopping approach employed in this 
        work. The system evolves on a single adiabatic potential-energy 
        surface at a time (solid lines S0 and S1) and can switch 
        stochastically to a different surface according to the 
        nonadiabatic coupling strength. Since transitions are 
        stochastic, the integration is repeated multiple times along 
        the same nuclear trajectory, as represented by the dashed and 
        dotted lines.}
        \label{fig:Marcus_SH}
\end{figure*}

Density functional theory (DFT) has played a central role in 
establishing the microscopic picture of small-polaron formation and 
stability~\cite{reticcioli2020small,dai2025polarons}. Accurate charge 
localization and electronic gaps can be obtained using hybrid functionals 
or DFT+$U$, where an on-site Hubbard-like Coulomb interaction is 
introduced. These approaches provide access to static polaron properties, 
including the relative stability of different localized configurations 
and hopping activation barriers, which can for instance be evaluated 
using the nudged elastic band (NEB) 
method~\cite{jonsson1998nudged,henkelman2000climbing,morita2023models}. 
Such calculations, however, typically describe individual transitions 
along predefined reaction pathways. Polaron transport at finite 
temperature is intrinsically dynamical: it results from thermal lattice 
fluctuations and from a statistical sequence of hopping events involving 
different local environments, sites, and microscopic pathways.

Small-polaron hopping can proceed in qualitatively different electronic 
regimes. According to the seminal work of Deskins and 
Dupuis~\cite{deskins2007electron}, thermally activated hopping can be 
adiabatic or nonadiabatic depending on the electronic coupling between 
the initial and final localized states, as schematically illustrated in 
Fig.~\ref{fig:Fig1a_Marcus}. In the adiabatic regime, 
sufficiently strong electronic coupling allows the charge to continuously 
follow the evolution of the nuclear configuration. In the nonadiabatic 
regime, instead, the electronic transition occurs between more weakly 
coupled localized states and cannot be described solely in terms of 
motion on a single electronic potential-energy surface. Distinguishing 
these regimes from static quantities alone is difficult, particularly 
in realistic finite-temperature environments where both the electronic 
levels and their couplings continuously fluctuate with the lattice. The 
theoretical framework connecting Marcus-theory activation energetics to 
the adiabatic/nonadiabatic crossover has been extensively developed for 
condensed-phase electron transfer in biological and electrochemical 
systems~\cite{blumberger2015,oberhofer2010}, and provides the conceptual 
foundation for the classification of polaron hopping regimes adopted here.

Molecular dynamics (MD) therefore provides a natural framework for 
investigating polaron transport at finite temperature, since it allows 
thermally activated hopping events to emerge directly from the dynamical 
evolution of the 
lattice~\cite{kowalski2010charge,setvin2014direct}. In particular, 
\textit{ab initio} molecular dynamics has provided important microscopic 
insight into polaron formation and 
motion~\cite{zhang2021dynamics,reticcioli2018formation}. Its 
computational cost, however, severely restricts both the accessible 
simulation time and system size. This represents a major limitation for 
small-polaron transport because hopping events can be rare and 
statistically meaningful transport properties require the sampling of 
many transitions over extended spatial and temporal scales. A realistic 
description of polaron mobility and of the distribution of competing 
hopping pathways therefore requires going substantially beyond the scales 
routinely accessible to direct \textit{ab initio} molecular dynamics.

Machine-learning molecular dynamics (MLMD) offers a route to overcome 
this limitation by replacing repeated first-principles 
electronic-structure calculations with machine-learned models trained on 
DFT data, thereby extending accessible simulation times while retaining 
close-to-first-principles accuracy~\cite{Friederich2021}. Applying this 
strategy to charged and localized electronic states remains challenging, 
however, because conventional machine-learned interatomic potentials do 
not explicitly encode the position and evolution of the localized charge. 
Several approaches to incorporate charged states and long-range electronic 
effects into machine-learning potentials are currently under active 
development~\cite{Fan2026,Ko2021}.

Among them, the recently developed LEOPOLD framework is specifically 
designed to describe small-polaron 
dynamics~\cite{birschitzky2025machine,Leoni2026}. LEOPOLD combines 
machine-learned interatomic potentials with equivariant graph neural 
networks~\cite{batatia2022mace,batzner20223} to predict the time 
evolution of site-selective polaron occupation matrices, allowing the 
instantaneous position of the localized charge to be identified 
dynamically. By learning molecular-dynamics polaron trajectories, 
LEOPOLD makes it possible to follow polaron motion on nanosecond time 
scales and to obtain statistically meaningful sampling of hopping 
pathways and finite-temperature mobilities. The framework has been 
applied to electron polarons in bulk rutile TiO$_2$ and hole polarons 
in MgO over a broad temperature 
range~\cite{birschitzky2025machine}, and has more recently been 
extended to electron-polaron transport at the reduced rutile 
TiO$_2$(110) surface~\cite{Leoni2026}.

These developments overcome an important sampling bottleneck, but leave 
a second fundamental limitation. LEOPOLD, like conventional ground-state 
DFT-based molecular dynamics, evolves the system within the 
Born--Oppenheimer (BO) approximation. Within this framework, the 
electronic subsystem is assumed to adjust instantaneously to the nuclear 
coordinates and the system evolves on a single adiabatic 
potential-energy surface. Explicit transitions between different 
electronic states are therefore not included. Indeed, previous LEOPOLD 
simulations showed that, according to static activation-energy 
considerations, most hopping processes in rutile TiO$_2$ can be 
classified as adiabatic, while a fraction of the observed events displays 
signatures associated with nonadiabatic 
hopping~\cite{birschitzky2025machine,Leoni2026}. The latter processes, 
however, cannot be fully described within a ground-state 
Born--Oppenheimer framework.

This limitation becomes particularly important when polaron motion is 
coupled to electronic excitation and relaxation. Lattice fluctuations 
accompanying a hopping event can modify both the energy and spatial 
character of the localized polaron state and may bring it into resonance 
with extended electronic states. Under such conditions, electronic and 
nuclear degrees of freedom can become strongly coupled and transitions 
between different electronic states become possible. A proper description 
of these processes requires going beyond the Born--Oppenheimer 
approximation through nonadiabatic molecular dynamics 
(NAMD)~\cite{prezhdo2021modeling,akimov2013theoretical,
doltsinis2006molecular,barbatti2011nonadiabatic,mai2018nonadiabatic}.

Among the available NAMD methods, trajectory surface hopping (SH), 
originally introduced by Tully~\cite{tully1990molecular}, is one of 
the most widely employed approaches. In surface hopping, nuclei evolve 
classically on a single potential-energy surface, while transitions 
between electronic states are allowed according to algorithm-specific 
hopping criteria that generally depend on the nonadiabatic couplings. 
A schematic representation of the method is shown in 
Fig.~\ref{fig:Fig1b_sh}. Such an explicit treatment of electronic 
transitions becomes necessary whenever the Born--Oppenheimer separation 
between electronic and nuclear motion breaks down.

Nonadiabatic effects are crucial for a wide variety of dynamical 
phenomena, including carrier recombination at surfaces and in condensed 
phases~\cite{xu2022nonadiabatic}, photochemistry and radiationless 
transitions, laser- and electron-induced chemistry, dynamics at metal 
surfaces~\cite{dou2020nonadiabatic}, and electron transfer in molecular 
systems~\cite{akimov2014nonadiabatic}, biological 
matter~\cite{groenhof2020observe,polli2010conical}, and interfacial and 
electrochemical environments. In recent years, NAMD has provided 
important microscopic insight into ultrafast excited-state processes in 
systems where electronic and nuclear motion are strongly coupled. 
Photoexcited processes in solid-state materials have been extensively 
investigated in 
perovskites~\cite{li2018time,li2021ab,li2019anharmonicity} and 
transition-metal oxides~\cite{xu2022nonadiabatic,cheng2023time}. NAMD 
techniques have also been widely employed to determine charge-carrier 
relaxation and recombination times, with particular attention devoted 
to polaron formation and 
recombination~\cite{li2024formation,cheng2022photoinduced}, and to the 
influence of adsorbates on the excited-state dynamics of 
polaron-containing materials~\cite{xu2023polaron}.

Much less is known, however, about the direct interplay between 
\emph{polaron hopping} and nonadiabatic excited-state dynamics. 
Existing NAMD studies of polarons in TiO$_2$ and related oxides have 
relied on short \textit{ab initio} molecular dynamics trajectories, 
which provide limited statistical sampling of hopping events and do not 
allow a systematic distinction between hopping pathways with 
qualitatively different electronic 
character~\cite{zhang2021dynamics,xu2022nonadiabatic,xu2023polaron}. 
In particular, it remains unclear how the microscopic character of 
individual hopping events modifies the nonadiabatic coupling, how 
hopping affects carrier excitation and relaxation, and whether distinct 
hopping pathways exhibit qualitatively different adiabatic or 
nonadiabatic behavior. Addressing these questions requires 
simultaneously accessing a sufficiently broad ensemble of thermally 
activated polaron trajectories and treating the electronic transitions 
induced by nuclear motion.

We focus on two prototypical oxide systems that host distinct types of 
polaronic charge carriers and are both of direct relevance to energy 
applications. Rutile TiO$_2$ hosts electron polarons and is widely 
employed as a photocatalyst and photoanode for solar-driven water 
splitting~\cite{Oi2016,Jiang2017} and in 
photovoltaics~\cite{marcelis2024titanium}, making the microscopic 
understanding of its polaron dynamics directly relevant to its 
photocatalytic performance. MgO is an ionic crystal in which hole 
polarons associated with Mg vacancies act as the dominant charge 
carriers~\cite{smith2017intrinsic}. As a low-cost material with high 
thermal stability and chemical resistance~\cite{Mahiba2026}, MgO is 
utilized in thermochemical energy storage and has been investigated as 
an electrode material for 
supercapacitors~\cite{Mastronardo2017,Goyal2025}. Together, these two 
systems offer complementary polaron physics (electron versus hole, 
covalent oxide versus ionic crystal) and provide a broad basis for 
establishing general principles of nonadiabatic polaron dynamics.

Here, we combine the extensive configurational and temporal sampling 
enabled by machine-learning polaron dynamics with explicit nonadiabatic 
molecular dynamics to investigate, for the first time, how individual 
thermally activated polaron hopping events drive nonadiabatic electronic 
transitions in bulk rutile TiO$_2$, bulk MgO, and the reduced 
TiO$_2$(110) surface.

Long trajectories generated with LEOPOLD provide a statistically rich 
ensemble of thermally activated polaron hopping events. These 
trajectories are subsequently used to evaluate the nonadiabatic 
couplings and to perform decoherence-induced surface-hopping 
simulations, thereby allowing transitions between different electronic 
states. This combined approach enables us to correlate individual 
polaron hopping events with the time evolution of the polaronic states, 
the strength of the nonadiabatic coupling, electronic excitation and 
de-excitation processes, charge-carrier lifetimes, and decoherence 
times. Temperature effects are investigated at 400, 500, and 600~K for 
bulk TiO$_2$ and MgO and at 600 and 700~K for the reduced rutile 
TiO$_2$(110) surface.

Our results reveal a direct and quantitative connection between the 
microscopic character of individual polaron hopping events and 
nonadiabatic excited-state dynamics. The enlarged trajectory space 
allows us to provide the first dynamical confirmation of the 
adiabatic/nonadiabatic classification of polaron hopping, and to show 
that nonadiabatic transitions are favored when the polaronic gap 
state approaches the conduction-band manifold, resulting in enhanced 
nonadiabatic coupling and de-excitation times that decrease by nearly 
an order of magnitude between 400 and 600\,K in bulk rutile. 
Nonadiabatic hopping events also promote thermally activated excitations 
from the ground state via a Landau--Zener mechanism, an effect absent 
in MgO where hole-polaron hopping is predominantly adiabatic. At the 
reduced TiO$_2$(110) surface, polaron hopping is orders of magnitude 
slower than in the bulk because the electron polaron is preferentially 
stabilized in the subsurface region, resulting in de-excitation times 
two orders of magnitude longer than in the bulk. Most importantly, 
subsurface-to-surface migration, the pathway of greatest relevance for 
photocatalysis, is found to be predominantly nonadiabatic, identifying 
nonadiabatic electron--lattice coupling as a key ingredient for enabling 
polaron migration toward the reactive surface layer.

\section{Computational workflow}

In this section, we describe the computational protocol employed to analyze the nonadiabatic character of polaron hopping events extracted from  MLMD trajectories. The workflow combines three computational tools: the Vienna \textit{Ab initio} Simulation Package (VASP)~\cite{kresse1996efficient, kresse1996efficiency, kresse1999ultrasoft}, LEOPOLD~\cite{birschitzky2025machine,Leoni2026}, and PYXAID~\cite{akimov2013pyxaid, akimov2014advanced}, together with in-house post-processing tools. The long-time polaron trajectories were generated with LEOPOLD using machine-learning models trained on first-principles data obtained with VASP. These trajectories were computed and characterized in our previous studies of bulk MgO and rutile TiO$_2$~\cite{birschitzky2025machine} and of the reduced rutile TiO$_2$(110) surface~\cite{Leoni2026}, and are used here as the starting point for the nonadiabatic analysis. 

Selected portions of the MLMD trajectories containing polaron hopping events are subsequently re-evaluated with VASP: for each geometry along the selected trajectory, a self-consistent calculation is performed to obtain the electronic energies and wave functions. Wave-function overlaps between electronic states at consecutive time steps are then evaluated using in-house post-processing tools and corrected for the arbitrary phases of the electronic wave functions. The corresponding nonadiabatic couplings (NACs) are obtained from the time-dependent overlaps using a finite-difference scheme. Together with the electronic energies, these couplings define the time-dependent electronic Hamiltonian employed in the subsequent nonadiabatic dynamics. The resulting Hamiltonians are provided as input to PYXAID~\cite{akimov2013pyxaid,akimov2014advanced} to propagate the electronic dynamics along the prescribed nuclear trajectories.

\begin{figure}[h!]
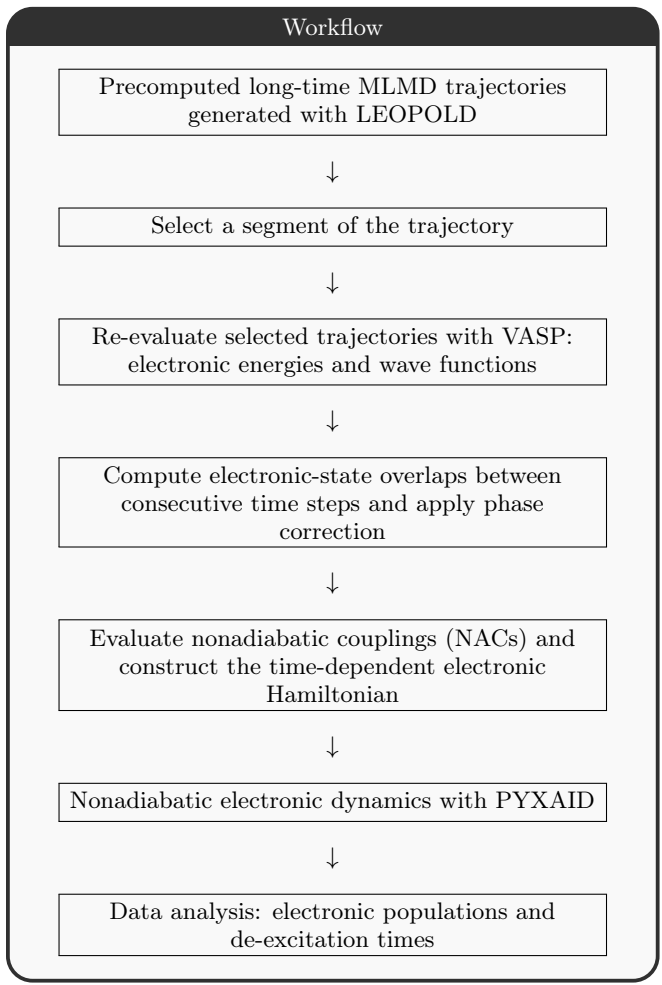

    \centering
\begin{tcolorbox}[arc=3mm, colback=gray!5!white, colframe=darkgray!75!black, title=Workflow]

\fbox{%
    \parbox{0.93\textwidth}{%
    \centering
        Precomputed long-time MLMD trajectories generated with LEOPOLD 
    }%
}
\[ \downarrow \] 
\fbox{% 
    \parbox{0.93\textwidth}{% 
    \centering 
        Select a segment of the trajectory  
    }% 
}
%\medskip
\[ \downarrow  \]
%\medskip
\fbox{%
    \parbox{0.93\textwidth}{%
    \centering
        Re-evaluate selected trajectories with VASP: electronic energies and wave functions
    }%
}
\[ \downarrow  \]
%\centering
\fbox{%
    \parbox{0.93\textwidth}{%
    \centering
    Compute electronic-state overlaps between consecutive time steps and apply phase correction
    }%
}
%\medskip
\[ \downarrow \] 
%\medskip 
\fbox{% 
\parbox{0.93\textwidth}
{% 
\centering 
    Evaluate nonadiabatic couplings (NACs) and construct the time-dependent electronic Hamiltonian 
}% 
}
%\medskip
\[ \downarrow  \]
%\medskip
%\centering
\fbox{%
    \parbox{0.93\textwidth}{%
    \centering
       Nonadiabatic electronic dynamics with PYXAID
    }%
}
\[ \downarrow  \]
\fbox{%
    \parbox{0.93\textwidth}{%
    \centering
       Data analysis: electronic populations and de-excitation times
    }%
}
%\end{center}

\end{tcolorbox}

\caption{\justifying Schematic representation of the computational workflow adopted in this study. Polaron hopping events are first selected from precomputed long-time MLMD trajectories generated with LEOPOLD. The selected nuclear trajectories are subsequently re-evaluated with VASP to obtain the electronic energies and wave functions. Wave-function overlaps between consecutive time steps and the corresponding nonadiabatic couplings  are then constructed. The resulting energies and NACs are provided to PYXAID to simulate the nonadiabatic electronic dynamics and extract quantities such as the time-dependent electronic populations and characteristic de-excitation times. } 

    \label{fig:workflow}
\end{figure}

This procedure allows us to characterize the electronic population dynamics associated with individual polaron hopping events and, in particular, to distinguish between predominantly adiabatic and nonadiabatic hopping processes. A schematic representation of the complete computational workflow is shown in Fig.~\ref{fig:workflow}. 

\medskip

\begin{figure*}[htbp]
    \centering
     \begin{subfigure}[t]{.38\textwidth}
         \centering
         \begin{overpic}[width=\linewidth]{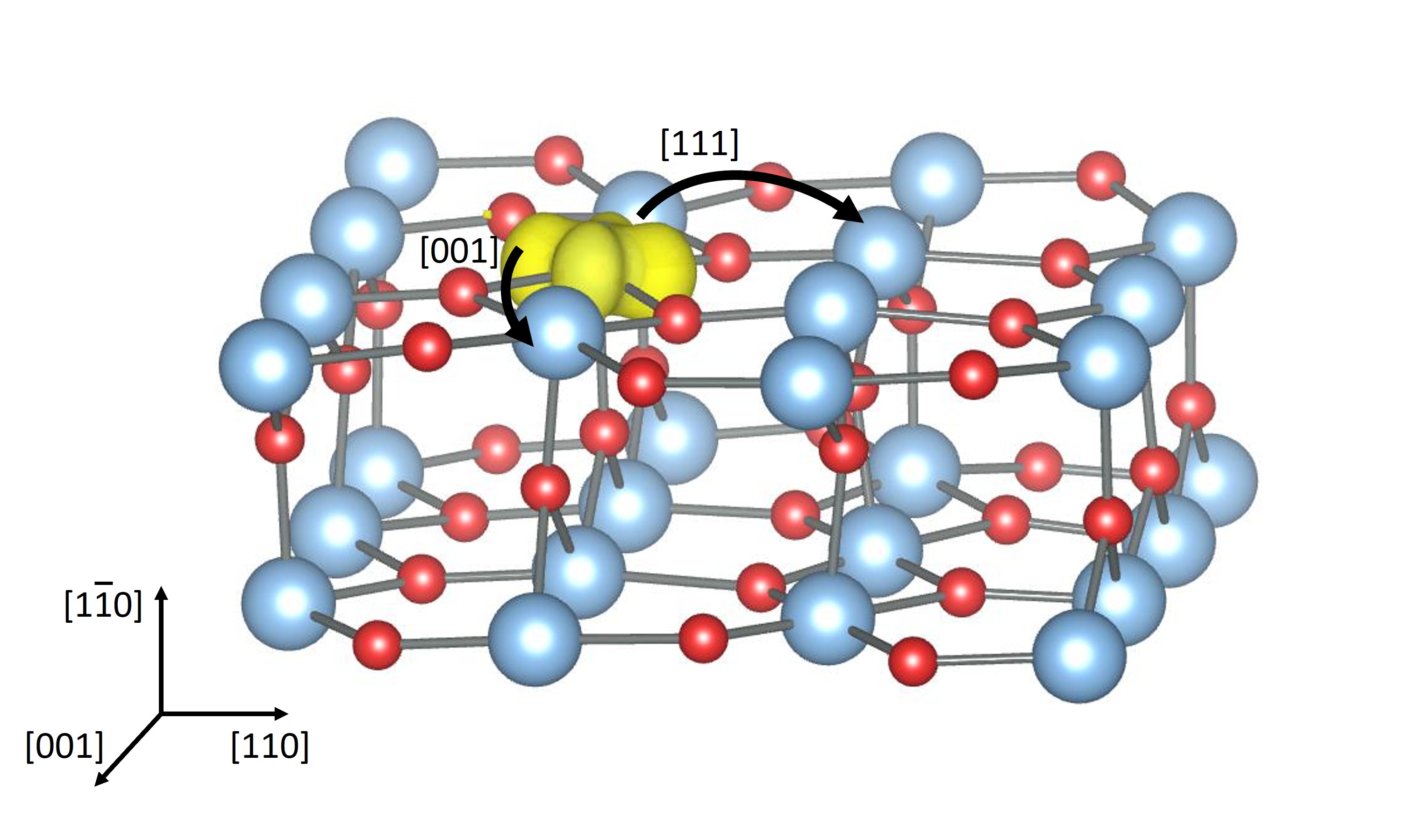}
            \put(6,55){\large (a)}
        \end{overpic}
         \phantomsubcaption
         \label{fig:Fig3a_TiO2bulk}
    \end{subfigure}
     \hspace{-0.8cm}
\begin{subfigure}[t]{.35\textwidth}
         \centering
         \begin{overpic}[width=\linewidth]{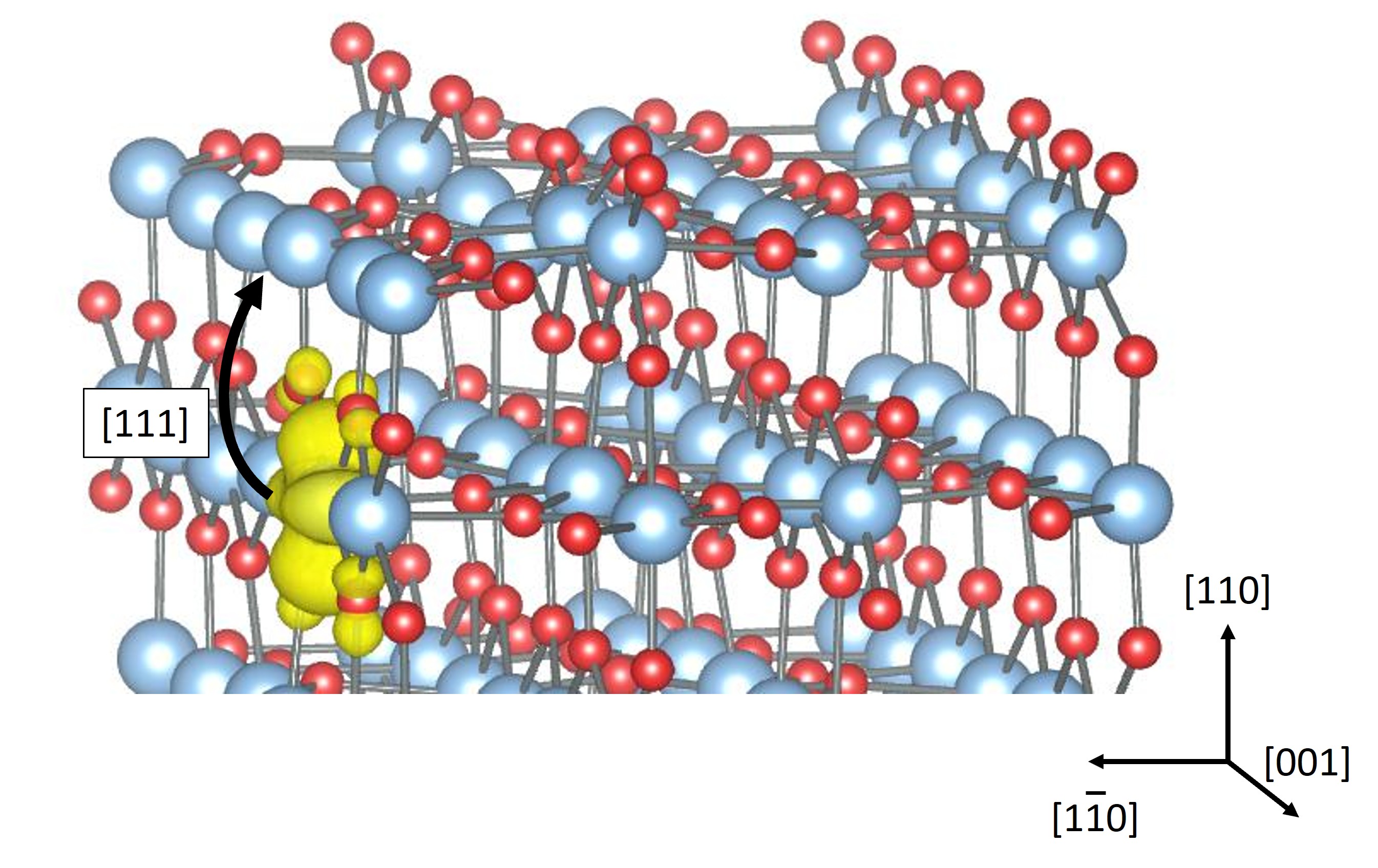}
            \put(6,60){\large (b)}
        \end{overpic}
         \phantomsubcaption
         \label{fig:Fig3b_TiO2surf}
    \end{subfigure}
    \begin{subfigure}[t]{.28\textwidth}
         \centering
         \begin{overpic}[width=\linewidth]{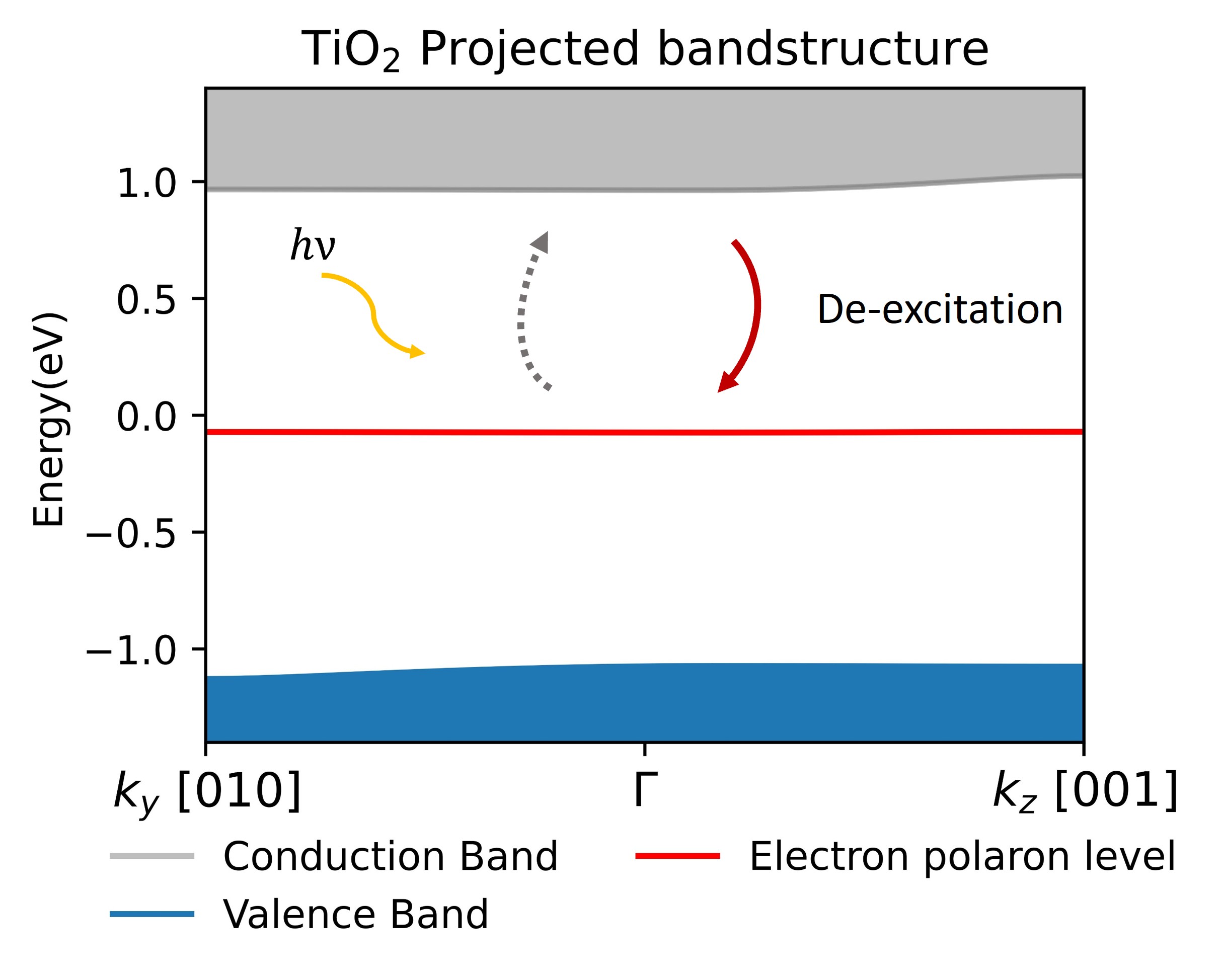}
            \put(-6,75){\large (c)}
        \end{overpic}
         \phantomsubcaption
         \label{fig:Fig3c_TiO2bands}
    \end{subfigure}

\begin{subfigure}{.38\textwidth}
         \centering
         \begin{overpic}[width=\linewidth]{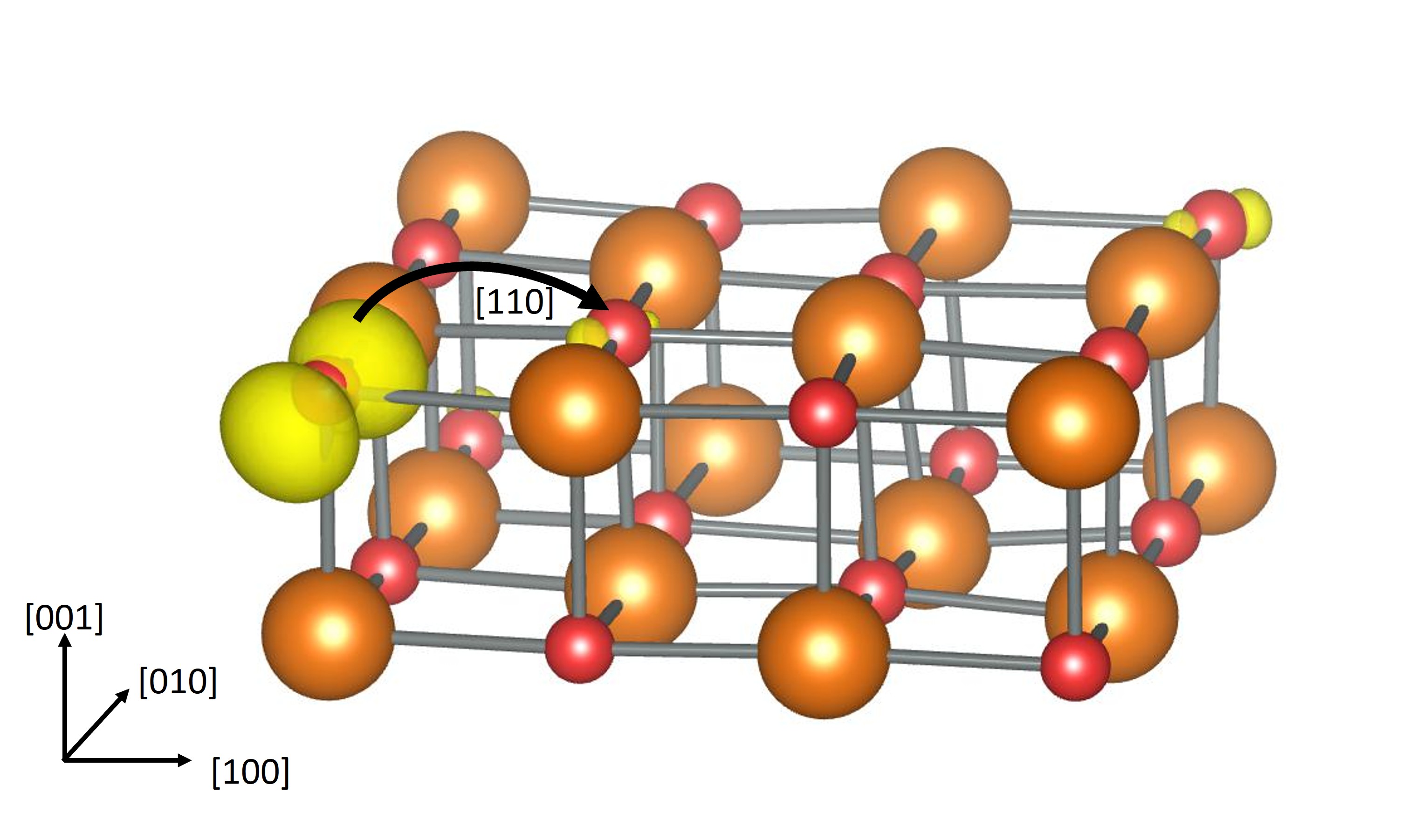}
            \put(6,50){\large (d)}
        \end{overpic}
         \phantomsubcaption
         \label{fig:Fig3d_MgObulk}
    \end{subfigure}
    \begin{subfigure}{.26\textwidth}
         \centering
         \begin{overpic}[width=\linewidth]{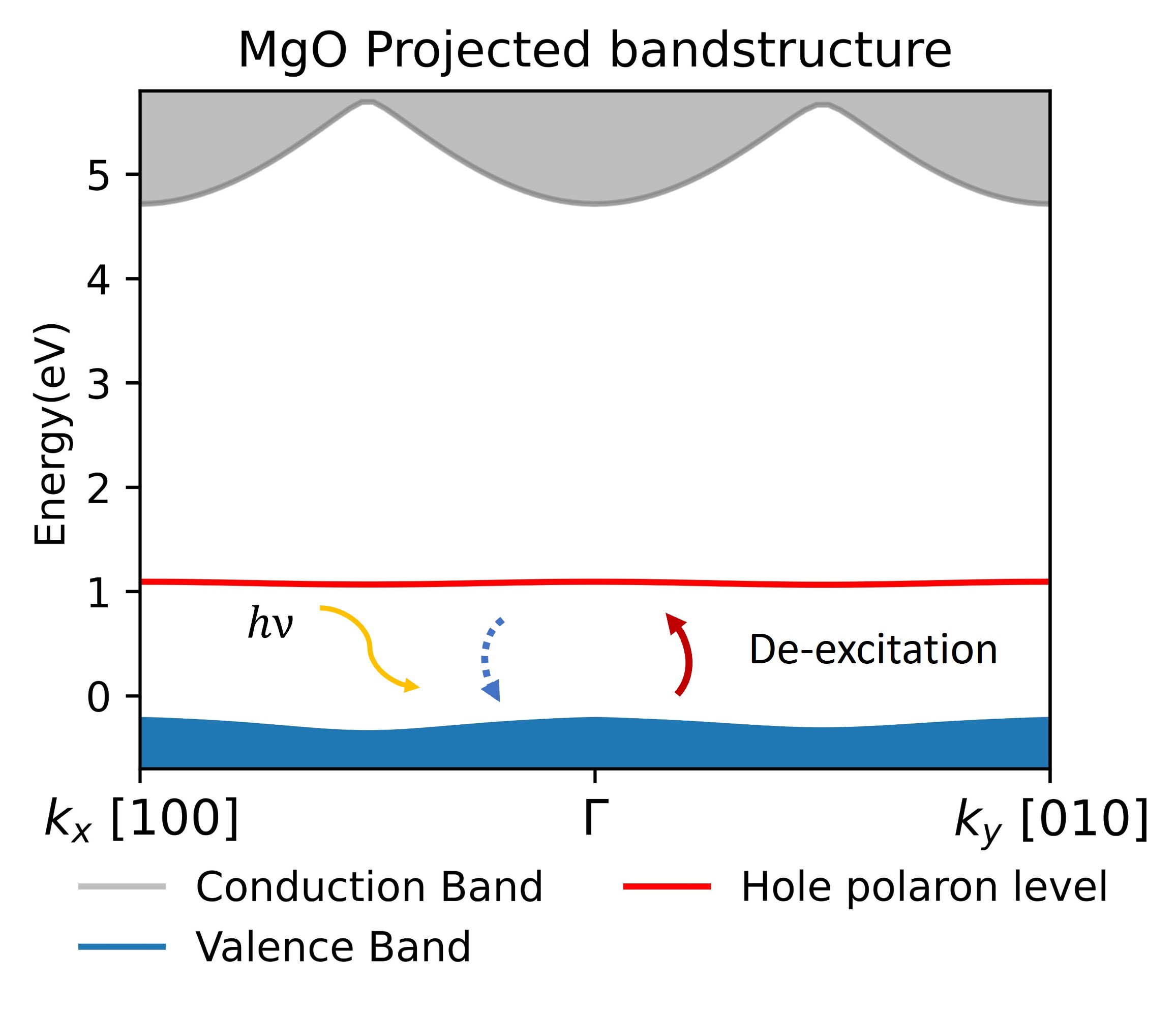}
            \put(-6,75){\large (e)}
        \end{overpic}
         \phantomsubcaption
         \label{fig:Fig3e_MgObands}
    \end{subfigure}
    
        \caption{\justifying Structure and charge density of (a) bulk and (b) surface rutile $\text{TiO}_{2}$ with an electron polaron in Ti d-orbitals; (d) sketch of bulk MgO, with a hole polaron in O p-orbitals. For every system the representative hopping paths are shown. In rutile bulk the most frequent polaron hopping event is along [001] direction; another possible event, along [111], is represented, as well. It has been classified as nonadiabatic by Deskins and Dupuis \cite{deskins2007electron}. Regarding TiO$_2$ surface it is shown the subsurface to surface polaron hopping, along [111] path. In MgO the most probable polaron hopping event is along [110] direction. Figures (c) and (e) show the projected band structures for bulk rutile and MgO, respectively, with the Fermi level set to zero. Also, the arrows provide a schematic representation of the excitation/de-excitation processes for the two materials. In $\text{TiO}_{2}$ an excitation promotes the electron from the gap state to the conduction band. In MgO the excited hole is promoted in the valence band.
}
        \label{fig:systems}
\end{figure*}

As anticipated in the Introduction, we consider hole polarons in bulk MgO and electron polarons in bulk rutile TiO$_2$, as well as electron polarons at the rutile TiO$_2$(110) surface. Bulk hopping events are analyzed at 400, 500, and 600~K, whereas surface hopping events are considered at 600 and 700~K. In the following, we first describe the structural models and common electronic-structure setup, followed by the generation and selection of the MLMD hopping trajectories, the evaluation of the electronic-state overlaps and NACs, and finally the nonadiabatic dynamics performed with PYXAID.

\subsection{Structural models and electronic-structure calculations}

We consider three representative oxide systems: bulk MgO, bulk rutile TiO$_2$, and the reduced rutile TiO$_2$(110) surface. The bulk systems host different types of small polarons: an excess-electron polaron in rutile TiO$_2$ and a hole polaron in MgO. At the reduced TiO$_2$(110) surface, we consider a single excess-electron polaron generated in the presence of an oxygen vacancy. A schematic representation of the corresponding structural models and polaronic states is shown in Figs. ~\ref{fig:Fig3a_TiO2bulk}, ~\ref{fig:Fig3b_TiO2surf} and ~\ref{fig:Fig3d_MgObulk}. %~\ref{fig:systems}.
Bulk rutile TiO$_2$ is described using a $2\times2\times6$ supercell containing 288 atoms. One excess electron is introduced into the supercell to form an electron small polaron. The excess charge localizes predominantly on a Ti $3d$ orbital, resulting in a Ti$^{3+}$-like center; the mid-gap state, about 1\,eV below the conduction band, can be seen in the  projected band structure in Fig. \ref{fig:Fig3c_TiO2bands}. Bulk MgO is modeled using a $2\times2\times2$ supercell containing 64 atoms. In this case, one electron is removed from the system, giving rise to a hole polaron that localizes predominantly on an O $2p$ orbital and forms an O$^{-}$-like center~\cite{nabi2025impact}. In this case the polaron level is 1\,eV above the valence band, see the projected band structure in Fig. \ref{fig:Fig3e_MgObands}. The reduced rutile TiO$_2$(110) surface is modeled by an asymmetric five-layer slab using a $6\times3$ surface supercell and a vacuum region of 15~\AA\ along the surface-normal direction. A bridging-oxygen vacancy is introduced to generate excess charge. Since a neutral oxygen vacancy releases two electrons, one electron is compensated by a homogeneous positive background charge, leaving a single excess electron available to form a small polaron. The resulting slab contains 539 atoms.

As in our previous study of surface-polaron dynamics~\cite{Leoni2026}, the upper three layers are allowed to evolve, whereas the two bottom layers are kept fixed at their bulk positions. The electronic-structure calculations are performed using the VASP~\cite{kresse1996efficient,kresse1996efficiency,kresse1999ultrasoft}. Exchange and correlation are treated within the Perdew--Burke--Ernzerhof (PBE) generalized-gradient approximation, and all calculations are spin-polarized. The localized polaronic states are described within the rotationally invariant DFT+$U$ approach. For rutile TiO$_2$, an effective Hubbard parameter $U_{\mathrm{eff}}=3.9$~eV is applied to the Ti $3d$ orbitals, consistently with previous work and with values obtained from constrained random-phase-approximation calculations~\cite{setvin2014direct}. For MgO, $U_{\mathrm{eff}}=10.0$~eV is applied to the O $2p$ states, within the range previously shown to provide an appropriate description of localized hole polarons in MgO~\cite{falletta2022hubbard}. These values yield well-localized polaronic states with energy levels inside the corresponding band gaps. A plane-wave kinetic-energy cutoff of 400~eV is employed for the TiO$_2$ systems and 500~eV for MgO. Owing to the large supercells, the Brillouin zone is sampled at the $\Gamma$ point for all systems. Further details concerning the structural models and the underlying first-principles calculations used to construct the LEOPOLD training sets can be found in Refs.~\cite{birschitzky2025machine,Leoni2026}.

\subsection{Machine-learned molecular-dynamics trajectories}

All MD trajectories analyzed in the present work were generated in our previous studies using the LEOPOLD framework. They are therefore not recomputed here, but constitute the starting point for the subsequent electronic-structure and nonadiabatic analysis. In the following, we summarize the aspects of the MLMD simulations that are relevant for the present study; a detailed description of the machine-learning architecture, training procedure, database construction, and validation is given in Refs.~\cite{birschitzky2025machine,Leoni2026} and their corresponding Supplementary Materials. 

LEOPOLD is a polaron-aware machine-learning framework based on an equivariant message-passing neural-network architecture derived from the NequIP architecture~\cite{Batzner2022}. In addition to the atomic species and positions, the model explicitly encodes the polaronic charge state and is trained to reproduce both the potential-energy surface and the local magnetic moments obtained from reference DFT+$U$ calculations. The latter provide a local descriptor of the polaronic state and allow its position to be followed dynamically during the simulation. The original formulation of LEOPOLD for bulk systems was subsequently extended to surface geometries~\cite{Leoni2026}, including the treatment of slab models and selective atomic dynamics. In its present formulation, LEOPOLD therefore provides a unified framework for simulating long-time polaron dynamics in both bulk and surface systems.

All MLMD trajectories considered here were propagated in the canonical ($NVT$) ensemble using a Nos\'e--Hoover thermostat and a time step of 1~fs. At each MD step, the atomic positions are propagated according to the machine-learned forces, while the predicted local magnetic moments are used to track the instantaneous localization of the polaron. This simultaneous description of the atomic motion and of the localized charge degree of freedom allows thermally activated polaron hopping events to be followed over nanosecond time scales. 

For bulk MgO and rutile TiO$_2$, the original simulations covered temperatures from 100 to 600~K in steps of 100~K and extended to the nanosecond time scale~\cite{birschitzky2025machine}. In the present work, we analyze selected hopping events extracted from the trajectories at 400, 500, and 600~K. For MgO, trajectories of up to 10~ns were available over the investigated temperature range, whereas for pristine rutile TiO$_2$ the accessible trajectory length decreased at the highest temperatures, reaching approximately 4~ns at 500~K and 1.3~ns at 600~K. 

For the reduced rutile TiO$_2$(110) surface, four independent production trajectories were generated at each temperature between 300 and 700~K~\cite{Leoni2026}. The different simulations were initialized with the polaron localized at distinct sites, thereby improving the sampling of the accessible surface and subsurface configurations. Individual trajectories extended over several nanoseconds, ranging from approximately 1--3~ns at 700~K to up to 9~ns at the lowest temperatures. In the present study, we consider hopping events extracted from the trajectories at 600 and 700~K. 

The nanosecond time scales accessible with LEOPOLD provide a broad sampling of the thermally activated polaron dynamics and, importantly for the present work, a large ensemble of distinct hopping events. Individual events are identified from the time evolution of the polaron localization and selected for the electronic-structure and nonadiabatic analysis described in the following.

\subsection{Selection and processing of polaron hopping trajectories}
The hopping trajectories analyzed in the present work were selected from the long-time MLMD simulations described above. The selection was designed to sample the distinct hopping mechanisms identified in our previous studies, including both the dominant diffusion pathways and less frequent transitions with a qualitatively different electronic character. To identify the hopping events we track the evolution of polaron position from the local magnetic moments.

In our previous studies~\cite{birschitzky2025machine}, for bulk rutile TiO$_2$ some representative hopping events were considered, namely the ones along the [001], [110], and [111] directions.

Within the Marcus-theory analysis (see Fig.~\ref{fig:Marcus_SH}), the dominant [001] pathway was associated with the adiabatic regime, whereas the [110] and [111] processes were assigned to the nonadiabatic regime. These hopping events are therefore included in this work in order to directly assess their dynamical character within the explicit nonadiabatic treatment~\cite{birschitzky2025machine}. 

For bulk MgO, symmetry reduces nearest-neighbor polaron transfer to a single inequivalent hopping pathway along the [110] direction. Only in few cases the
hopping occur suddenly and among non nearest neighbors
Oxygens. 

At the reduced rutile TiO$_2$(110) surface, the electron polaron is preferentially localized in the subsurface region~\cite{cheng2021co,reticcioli2020small, reticcioli2018formation, kruger2008defect, kruger2012intrinsic, setvin2014direct}. The most frequent hopping events occur between neighboring Ti sites along the subsurface Ti rows in the [001] direction, although with a substantially lower hopping frequency than in bulk rutile TiO$_2$~\cite{Leoni2026}. At 700~K, the MLMD trajectories also sample interlayer hopping events in which the polaron migrates between subsurface and surface Ti sites, predominantly along a [111]-type pathway. Both classes of surface hopping events are considered in the present analysis.

For every system at every investigated temperature we selected a 4.5~ps trajectory interval; being the time step 1~fs, we got a time-resolved sequence of 4500 nuclear configurations. % containing the hopping events. 
All 4500 configurations are retained and used for the subsequent electronic-structure calculations.
The selected interval contains the hopping event together with the surrounding nuclear dynamics, thereby allowing the evolution of the electronic states before, during, and after the charge transfer to be resolved. Each geometry along the selected trajectory segment is subsequently re-evaluated self-consistently with VASP to obtain the electronic energies and wave functions required for the construction of the time-dependent electronic Hamiltonian.

\subsection{Electronic structure along the hopping trajectories and NACs}

For each nuclear configuration along the selected trajectories, sampled at a time interval of $\Delta t=1$~fs, a self-consistent electronic-structure calculation is performed with VASP to obtain the Kohn--Sham eigenvalues and wave functions. The resulting wave functions are subsequently processed using in-house routines to evaluate the time-overlap matrix between electronic states $i$ and $j$ at consecutive time steps, 
\begin{equation} S_{ij}(t,t+\Delta t) = \left\langle \psi_i(t) \middle| \psi_j(t+\Delta t) \right\rangle . \label{eq:overlap} \end{equation} 
The time-overlap matrices provide the information required to evaluate the nonadiabatic couplings between the electronic states along the prescribed nuclear trajectory. Because the phases of the Kohn--Sham wave functions obtained independently at consecutive time steps are arbitrary, a phase-correction procedure is applied to ensure a consistent phase convention along the trajectory. The phase correction is performed using an in-house implementation following the procedure adopted in the Libra code~\cite{shakiba2022libra,akimov2018simple}. For each adiabatic state $i$ the phase correction is factor given by

\begin{equation}
   f_{i}=\frac{\left\langle\psi_i(t)\middle|\psi_i(t+\Delta t)\right\rangle}{|\left\langle\psi_i(t)\middle|\psi_i(t+\Delta t)\right\rangle|}. 
\end{equation}

 After phase correction, the NACs are evaluated using the finite-difference expression~\cite{hammes1994proton} 
\begin{equation} 
d_{ij}(t) = \frac{ \left\langle\psi_i(t)\middle|\psi_j(t+\Delta t)\right\rangle - \left\langle\psi_i(t+\Delta t)\middle|\psi_j(t)\right\rangle } {2\Delta t}. \label{eq:NAC_Tully} 
\end{equation} 
The quantities $d_{ij}$ represent the time-derivative nonadiabatic couplings between the adiabatic electronic states and provide the off-diagonal terms of the time-dependent electronic Hamiltonian. In the adiabatic basis, the Hamiltonian is constructed as~\cite{akimov2013pyxaid, jaeger2012decoherence, Jain2022} 
\begin{equation} 
H_{ij}(t) = \epsilon_i(t)\delta_{ij} - i\hbar d_{ij}(t), \label{eq:Hamiltonian} 
\end{equation} 
where $\epsilon_i(t)$ is the instantaneous Kohn--Sham eigenvalue of electronic state $i$ and $\delta_{ij}$ is the Kronecker delta. The NACs quantify the mixing between adiabatic electronic states induced by the nuclear motion. Large values of $d_{ij}$ indicate regions of the nuclear trajectory where the electronic states are strongly coupled and where nonadiabatic transitions are therefore more likely to occur. The time-dependent Hamiltonians constructed from the electronic energies and NACs are subsequently used as input for the nonadiabatic dynamics performed with PYXAID, as described in the following section.

The NACs in Eq.\ref{eq:NAC_Tully} quantify the coupling between electronic states induced by nuclear motion, measuring the extent to which the electronic and nuclear degrees of freedom are coupled. This coupling can drive nonadiabatic transitions that play a key role in non-radiative relaxation processes.

\subsection{Nonadiabatic dynamics with PYXAID}
The time-dependent electronic Hamiltonians constructed from the Kohn--Sham energies and nonadiabatic couplings are used as input to PYXAID ~\cite{akimov2013pyxaid,akimov2014advanced} to perform the nonadiabatic molecular dynamics simulations. The calculations are carried out within the classical path approximation (CPA) ~\cite{augustin1978classical,c1998mixed}, in which the nuclear dynamics is assumed to be unaffected by the evolution of the electronic degrees of freedom. Within this approximation, the electronic Hamiltonian depends parametrically on the classical nuclear coordinates, which evolve along a predetermined MD trajectory. The nuclear trajectory can therefore be generated independently of the subsequent electronic dynamics ~\cite{tully1971trajectory,chng2026ab}. This framework is particularly well suited to the present approach, in which the long-time nuclear trajectories are obtained beforehand from the LEOPOLD MLMD simulations. 

Electronic transitions are treated using the decoherence-induced surface hopping method~\cite{jaeger2012decoherence}. In addition to stochastic transitions between adiabatic electronic states, DISH accounts for the loss of electronic coherence induced by the interaction with the fluctuating nuclear environment, an important ingredient in the description of charge-carrier dynamics in condensed-phase systems. 

\begin{figure}[htbp]
    \centering
    \includegraphics[width=1.0\linewidth]{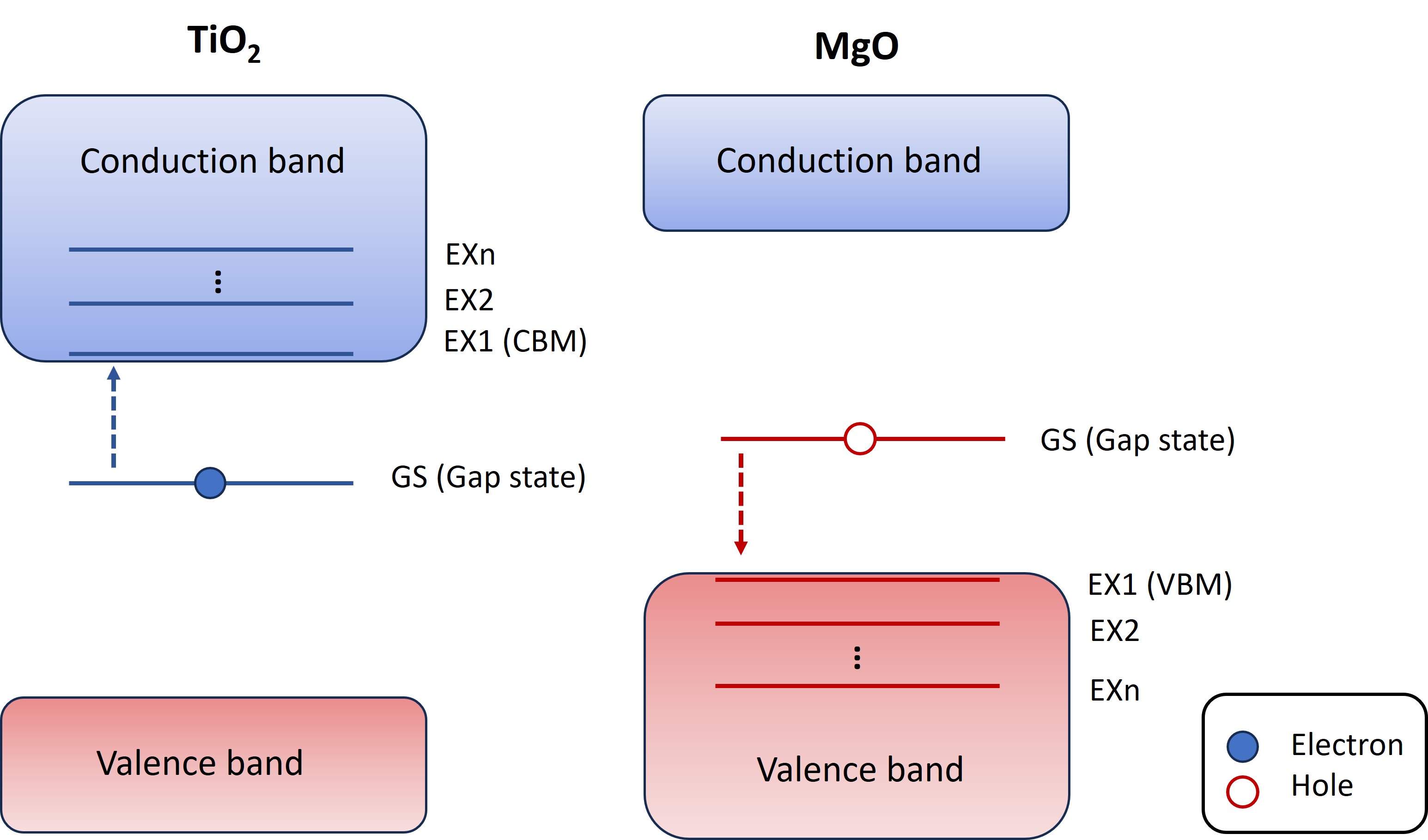}
    \caption{\justifying In both materials the ground state is represented by the polaron localized at the gap state. The excited states of rutile $\text{TiO}_{2}$ and MgO are obtained by promoting the electron/hole in the conduction/valence band, respectively. We consider 5 states in total, namely the GS and the first four excitations, EX1-EX4.}
    \label{fig:Fig4_Excited_states}
\end{figure}

For all systems, the electronic active space comprises five states: the ground state (GS) and the first four excited states (EX1--EX4), as illustrated schematically in Fig.~\ref{fig:Fig4_Excited_states}. For bulk and surface rutile TiO$_2$, the GS corresponds to the localized electron-polaron state, with the excess electron occupying the in-gap level. The first excited state, EX1, is obtained by promoting the electron from the polaronic gap state to the conduction-band minimum (CBM), while EX2--EX4 correspond to successively higher conduction-band states. 

For MgO, the GS corresponds to a localized hole-polaron state. Since the dynamics is formulated in terms of the hole degree of freedom, the excited states are constructed by transferring the hole from the localized polaronic state to extended valence-band states. EX1 corresponds to a hole at the valence-band maximum (VBM), whereas EX2--EX4 correspond to states progressively deeper in the valence band. 

To run the nonadiabatic dynamics simulations, in each system 150 distinct nuclear configurations are used as initial conditions. These are sampled every fourth frame from the beginning of the 4.5~ps trajectory segment, corresponding to starting points at times ($t_{in}$=0,4,$\ldots$,596)~fs. 
Different initial conditions are therefore selected independently of the timing of the polaron hopping events to initiate the NAMD simulations. For each initial condition, 1000 stochastic surface-hopping realizations are propagated.

Each NAMD trajectory is thus propagated for 3.9~ps along the corresponding precomputed time-dependent Hamiltonian.
The 1000 realizations are averaged to account for the probabilistic nature of the surface-hopping algorithm. The resulting population dynamics are subsequently averaged over the 150 starting configurations, reducing the dependence on the particular nuclear geometry chosen as the initial condition. In total, this procedure yields an ensemble of ($1.5\times10^{5}$) surface-hopping trajectories for the statistical evaluation of the time-dependent electronic populations and relaxation dynamics.

\begin{figure*}
    \centering
    \includegraphics[width=1.0\linewidth]{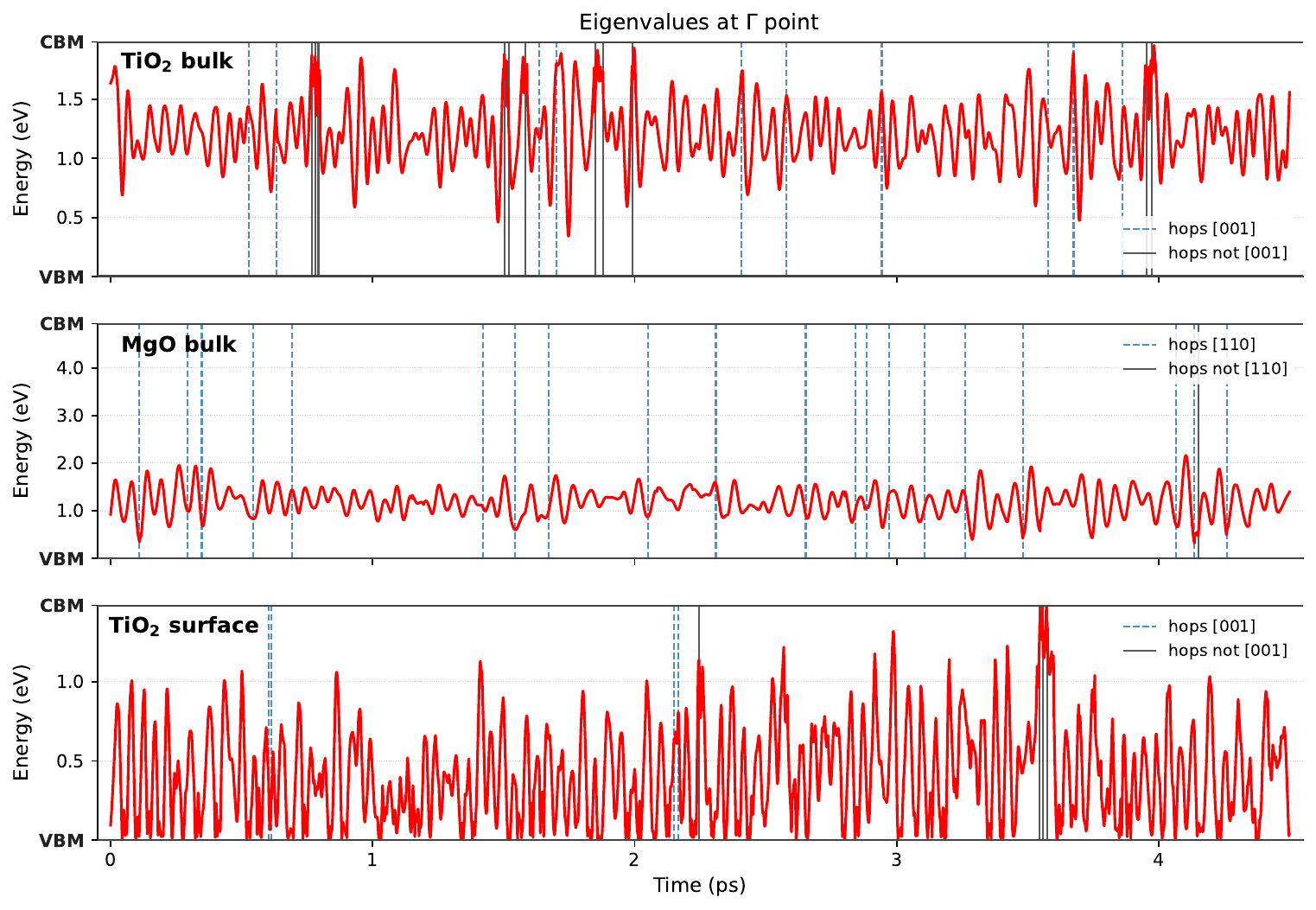}
    \caption{\justifying 
    Band energy of polaron level for the considered systems. The average values of the CBM and VBM energies are indicated by horizontal black lines. The polaron is in the spin up/down channel for the $\text{TiO}_{2}$ and MgO materials, respectively. The vertical lines represent the polaron hoppings. (Top) Simulation of bulk $\text{TiO}_{2}$ at T=500\,K. Hoppings occur when the gap level and the CBM approach. (Center) Band energies during the MgO simulation at T=500\,K. Also in this case polaron hoppings generally occur when the gap level and VBM approach. (Bottom) $\text{TiO}_{2}$ surface, at 700\,K; hoppings are more rare than in bulk system. In this simulation we observe the polaron moving from subsurface to surface and back for two times.
    }
    \label{fig:Fig5_Eigenvals}
\end{figure*}

\section{RESULTS}
To understand the influence of polaron motion on the charge carrier recombination in photoexcited processes we study at first the evolution of the band energies and the NACs as a function of time. Next, the results regar\-ding the electron/hole dynamics are presented.

\subsection{Analysis of the eigenvalues}
\label{sec:eigenvalues}

Figure~\ref{fig:Fig5_Eigenvals} shows the time evolution  of the polaronic gap level along representative trajectories at 500\,K for bulk rutile TiO$_2$ and MgO, and at 700\,K for the reduced TiO$_2$(110) surface. Results at other temperatures are provided in Fig.~S1.

In bulk rutile TiO$_2$, the gap state undergoes strong oscillations driven by thermal lattice fluctuations, and polaron hopping events are systematically accompanied by a pronounced narrowing of the gap between the polaronic level and the CBM (Fig.~\ref{fig:Fig5_Eigenvals}, top panel). This behavior reflects the lattice reorganization that accompanies the charge transfer: as the polaron moves from one Ti site to another, the local distortion (elongated Ti--O bonds at the initial site and a compressed octahedron at the final site) evolves through a transition state in which the polaronic orbital delocalizes transiently and the associated self-trapping energy is partially released. Within the Marcus--Holstein 
picture of small-polaron hopping~\cite{marcus1956theory,marcus2020electron}, this  corresponds to the crossing of the two parabolic potential-energy surfaces associated with the initial and final polaron sites (see Fig.~\ref{fig:Fig1a_Marcus}): at the crossing point, the reorganization energy is minimized and the gap between the two electronic levels is smallest, making the in-gap level approach the CBM. The magnitude of this approach and the degree of level mixing at the transition state determine whether the subsequent charge transfer proceeds adiabatically or 
nonadiabatically~\cite{deskins2007electron, blumberger2015, oberhofer2010}.

\begin{figure*}[htbp]
    \centering
    \includegraphics[width=1.9\columnwidth]{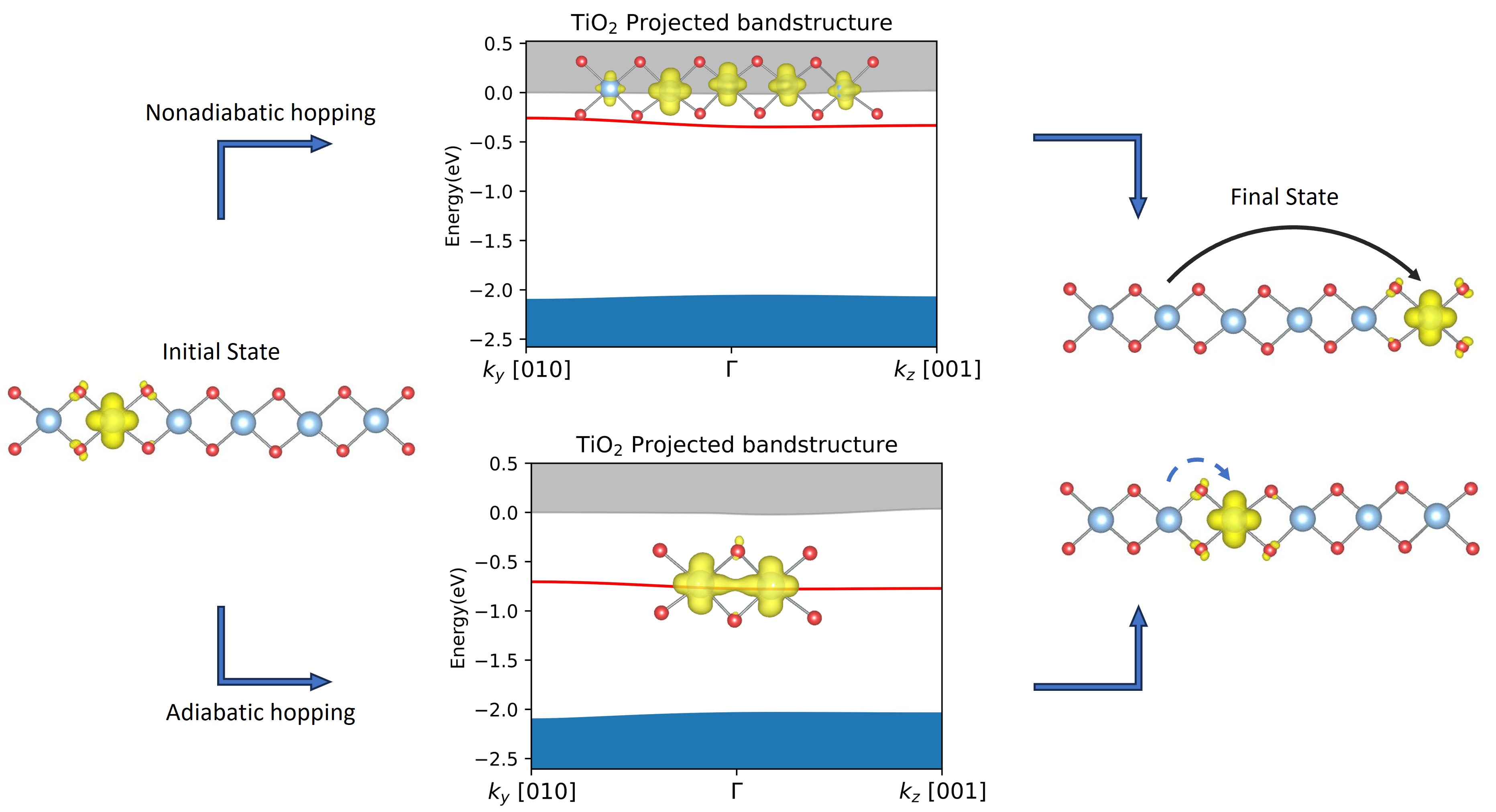}
    \caption{\justifying Schematic illustration of the two distinct polaron hopping regimes in 
    rutile TiO$_2$. (Top) During nonadiabatic hopping events (non-[001] directions), the polaron charge delocalizes across multiple Ti sites at the transition state and the polaronic defect level approaches the CBM, leading to strong nonadiabatic coupling. (Bottom) During adiabatic hopping ([001] direction), the polaron migrates smoothly between neighboring Ti sites while remaining localized, and the in-gap level stays well separated from the conduction band throughout the transfer. The zero of the energy has been set on the CBM.}
    \label{fig:Fig6_Two_hops_scheme}
\end{figure*}

The eigenvalue trajectories reveal a clear distinction between the two hopping classes identified in our previous LEOPOLD 
studies~\cite{birschitzky2025machine,Leoni2026}. During [001] hopping events --- the dominant pathway in bulk rutile, connecting nearest-neighbor Ti sites along the crystallographic $c$-axis --- the gap level approaches the CBM only moderately and the polaron charge migrates smoothly between the initial and final Ti sites, always relocalizing at the nearest available lattice site. The energy gap between the polaronic level and the CBM remains finite throughout the transfer, consistent with the adiabatic regime in which the electronic subsystem follows the nuclear motion continuously. By contrast, during non-[001] hopping events the energy gap narrows substantially and in some cases the polaron charge delocalizes 
transiently across the conduction band at the transition state, before relocalizing at a site that is not necessarily the nearest neighbor to the initial one. These contrasting behaviors, illustrated schematically in 
Fig.~\ref{fig:Fig6_Two_hops_scheme}, constitute a first dynamical 
signature distinguishing the two hopping classes as adiabatic and nonadiabatic, respectively, and are consistent with the static classification proposed by Deskins and Dupuis~\cite{deskins2007electron} on the basis of activation-energy arguments. The present analysis extends that classification to a dynamical, finite-temperature setting and --- by virtue of the extended MLMD trajectories --- captures the full statistical distribution of hopping events rather than isolated 
transition-state configurations.

The analogous picture holds for MgO, where the hole-polaron level approaches the VBM during hopping events (Fig.~\ref{fig:Fig5_Eigenvals}, center panel). However, in MgO the approach to the VBM is generally less pronounced than the gap-level--CBM approach in rutile, reflecting the predominantly adiabatic character of hole-polaron hopping in MgO, which occurs almost exclusively between nearest-neighbor oxygen sites along the [110] direction: the alignment of the O~$2p$ orbitals at the initial and final sites facilitates a smooth, continuous charge transfer. During the rare non-[110] hopping events, the excess charge is transiently shared among three oxygen sites, indicating a more delocalized transition state consistent with a partial breakdown of the adiabatic picture.

At the reduced TiO$_2$(110) surface (Fig.~\ref{fig:Fig5_Eigenvals}, bottom panel), hopping events are far less frequent than in the bulk, reflecting the strong stabilization of the electron polaron in the subsurface Ti layer~\cite{reticcioli2018formation,setvin2014direct,kruger2008defect,
kruger2012intrinsic}. 
At 600\,K, the single observed hopping is a [001]-type event within the subsurface layer with only a modest narrowing of the polaronic level--CBM gap (Fig.~S1c); the event is between neighboring Ti sites, analogous to the adiabatic events in the bulk. At 700\,K, interlayer hopping events along the [111] pathway --- in which the polaron migrates from the subsurface to the surface Ti layer --- are accompanied by strong defect-level oscillations and pronounced approaches to the CBM, mirroring the behavior of non-[001] nonadiabatic events in the bulk. This already suggests that subsurface-to-surface hopping at the rutile TiO$_2$(110) surface is 
predominantly nonadiabatic in character, a conclusion that will be confirmed quantitatively through the NAC analysis and the excited-state population dynamics presented in the following sections.

\begin{figure*}[hbtp]
    \centering
     \begin{subfigure}{.49\textwidth}
         \centering
         \begin{overpic}[width=\linewidth]{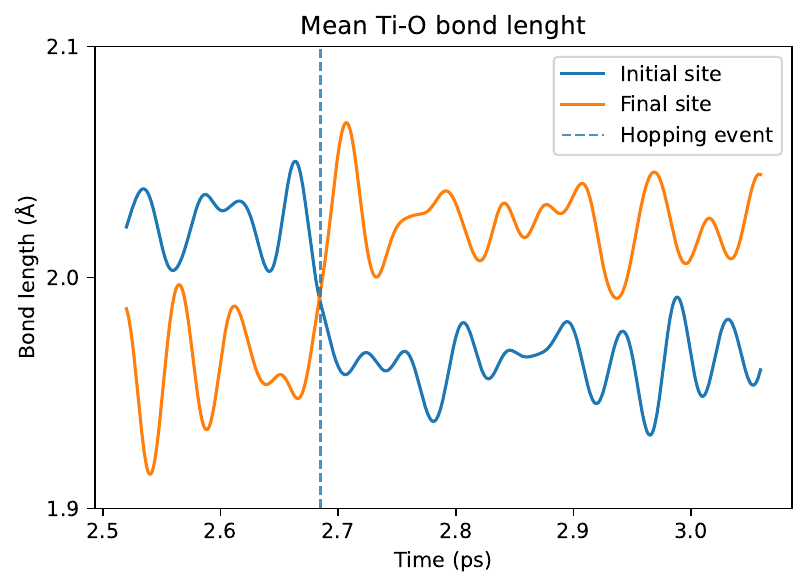}
            \put(15,62){\large (a)}
        \end{overpic}
         \phantomsubcaption
         \label{fig:Fig7a_Ti-Obonds}
    \end{subfigure}
    \begin{subfigure}{.49\textwidth}
         \centering
         \begin{overpic}[width=\linewidth]{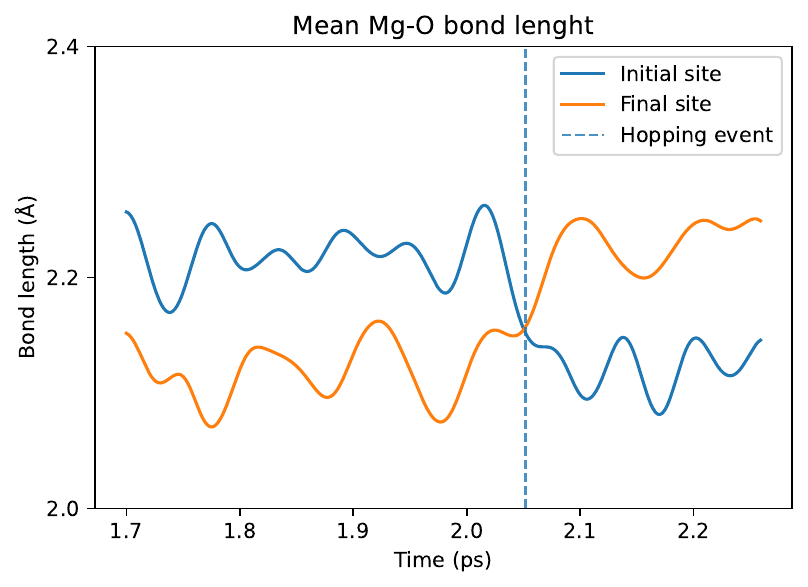}
            \put(15,62){\large (b)}
        \end{overpic}
         \phantomsubcaption
         \label{fig:Fig7b_Mg-Obonds}
    \end{subfigure}
        \caption{\justifying (a) Average Ti--O bond length during a polaron hopping event along the [001] direction in the simulation of bulk rutile at 400\,K. As the polaron moves from the initial to the final site, the bonds around the initial site contract while those around the final site expand, reflecting the transfer of the local lattice distortion. The vertical dashed line marks the hopping event. (b) Average Mg--O bond length during a polaron hopping event in the simulation of bulk MgO at 500\,K, showing analogous behavior. }
        \label{fig:Bond_length}
\end{figure*}

To investigate the lattice distortions associated with polaron hopping, we analyze the average Ti–O and Mg–O bond lengths for the sites involved in each hopping event (Fig.~\ref{fig:Bond_length}). The average bond length, \textit{l}, is defined as

\[l=\frac{1}{6}\sum_{i=1}^{6} d_{i}\]

where $d_{i}$ is the distance between the polaron-hosting atom (Ti in $\text{TiO}_{2}$ and O in MgO) and each of its six nearest-neighbor atoms. For both systems, the hopping event is accompanied by a transfer of the local lattice distortion. As the polaron moves from the initial to the final site, the average bond length around the initial site decreases, whereas that around the final site increases. The data shown in Fig.~\ref{fig:Bond_length} are obtained from the MLMD simulations. Although LEOPOLD identifies polaron hopping by tracking the evolution of the spin matrix, these results indicate that the associated local structural distortions, despite being relatively small, also provide a reliable descriptor of the hopping process.

\begin{figure*}[htbp]
    \centering    
    \begin{subfigure}{\textwidth}
         \centering
         
         \begin{overpic}[width=\linewidth]{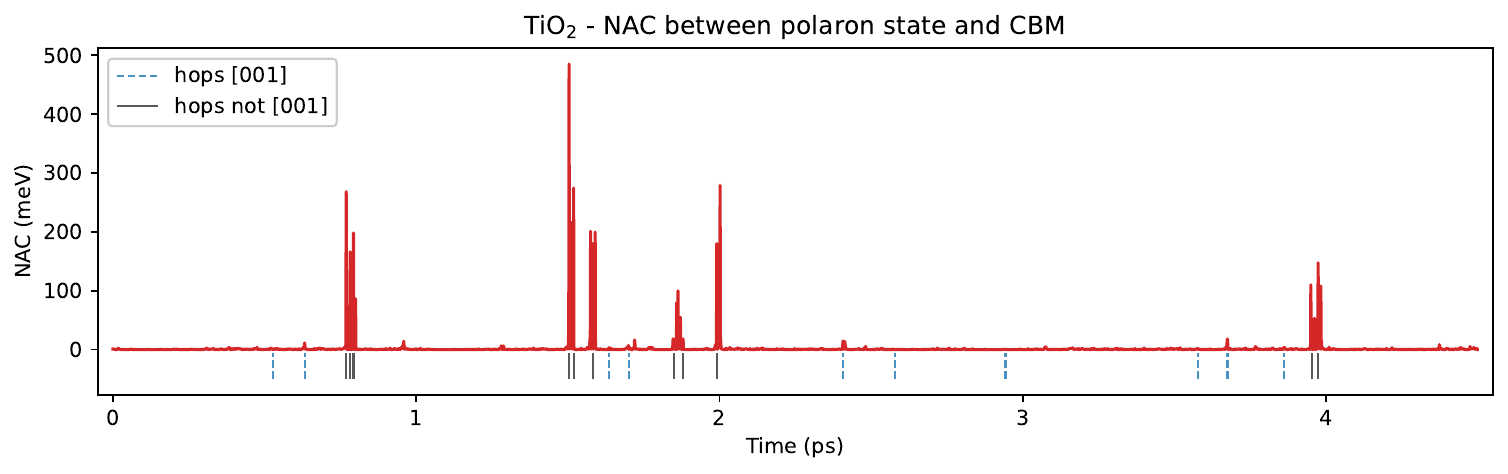}

            \put(8,29){\large (a)}
        \end{overpic}
        \phantomsubcaption
        \label{fig:Fig8a_NAC_TiO2}
    \end{subfigure}
  %\vspace{0.5cm}
    \begin{subfigure}{\textwidth}
         \centering
         
         \begin{overpic}[width=\linewidth]{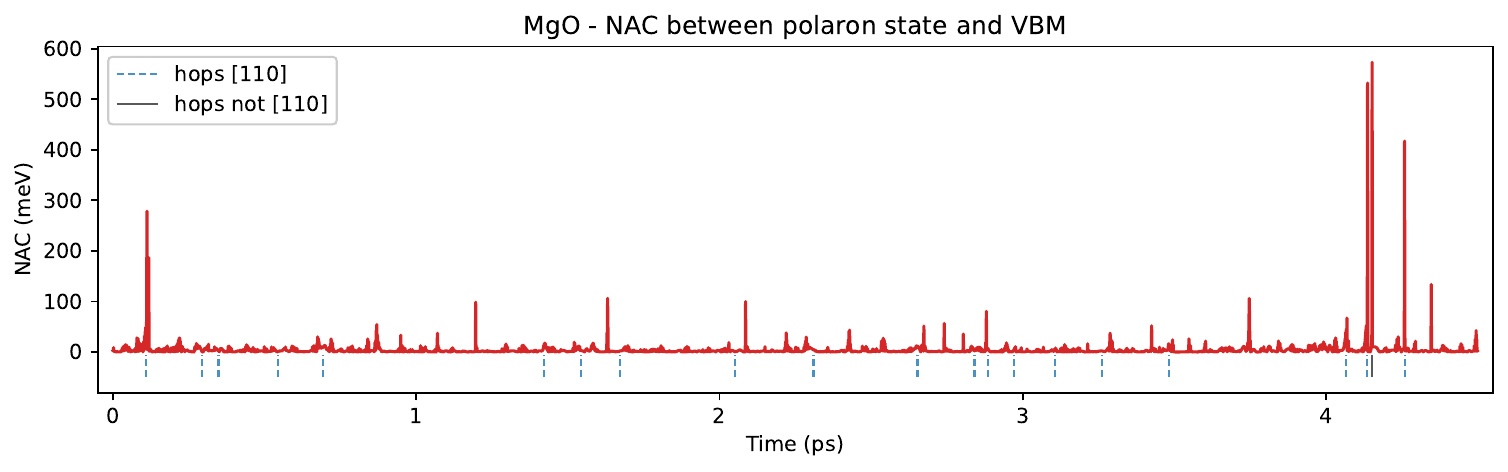}
            \put(8,29){\large (b)}
        \end{overpic}
        \phantomsubcaption
         \label{fig:Fig8b_NAC_MgO}
    \end{subfigure}
    %\vspace{0.5cm}
    \begin{subfigure}{\textwidth}
         \centering

         \begin{overpic}[width=\linewidth]{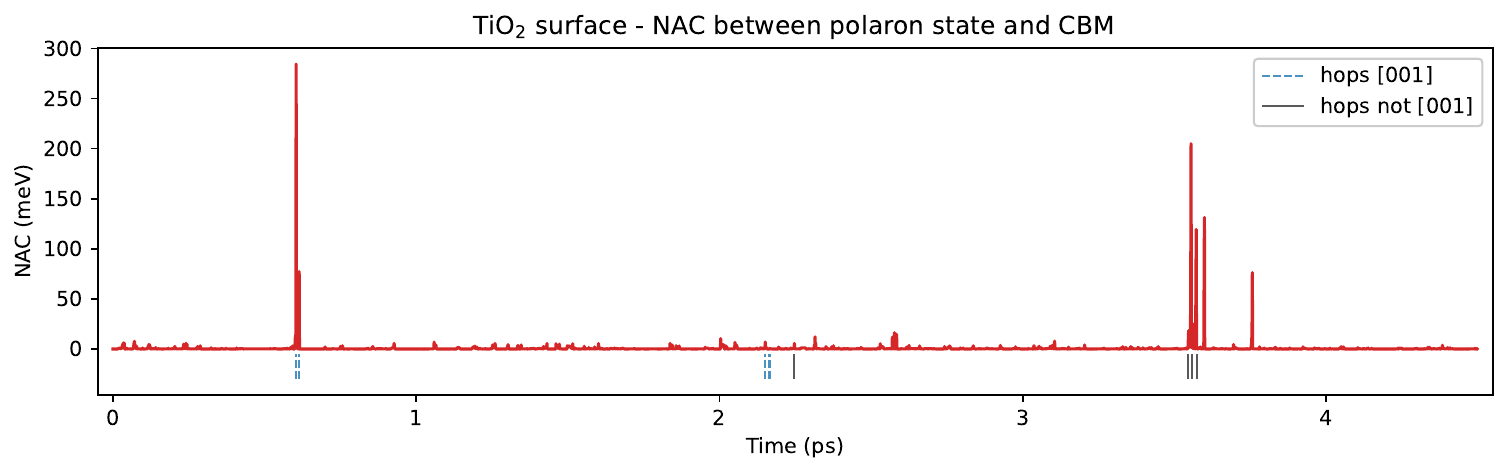}
            \put(8,29){\large (c)}
        \end{overpic}
        \phantomsubcaption
         \label{fig:Fig8c_NAC_TiO2-surf}
    \end{subfigure}
        \caption{\justifying Absolute value of the NAC between (a) the polaron state and the conduction-band minimum for bulk TiO$_2$ and (b) the polaron state and the valence-band maximum for bulk MgO, as a function of time at 500\,K. For both systems the NAC fluctuates around a few meV; sharp peaks reaching hundreds of meV are associated with polaron hopping events, particularly non-[001] hops in rutile. (c) NAC between the polaron state and the CBM for the TiO$_2$(110) surface at 700\,K; the strong peaks in the second half of the simulation are associated with [111] interlayer hopping events between the subsurface and surface Ti layers.}
        \label{fig:NACs}
\end{figure*}

%----------------------------------------------------------

\subsection{Nonadiabatic couplings analysis}
\label{sec:NACs}
The NAC can be evaluated numerically using a finite-difference approach, as introduced in Eq.~\ref{eq:NAC_Tully}, which approximates the time derivative of the electronic wavefunctions from their overlap at consecutive time steps. An alternative expression for the NAC between electronic states $i$ and $j$ is give by

\begin{equation}
d_{ij}= \langle \psi_{i} \left|\frac{\partial}{\partial t}\right| \psi_{j} \rangle 
= \langle \psi_{i} |{\nabla_{R}}| \psi_{j} \rangle \dot{R} 
= \frac{\langle \psi_{i}|\nabla_{R}H |\psi_{j} \rangle}{\epsilon_{j}-\epsilon_{i}}\dot{R}
\label{eq:NAC}
\end{equation}

which quantifies the coupling between the electronic and nuclear degrees of freedom induced by the nuclear motion. As evident from Eq.~\ref{eq:NAC}, the NAC magnitude is governed by three factors: the nonadiabatic electron–nuclear coupling between two electronic states
 $\langle \psi_{i}|\nabla_{R}H |\psi_{j} \rangle$, the nuclear velocity $\dot{R}$, and inversely, by the energy gap $\epsilon_{j} - \epsilon_{i}$ between the two 
states. Consequently, large NACs are expected when the energy gap between the polaronic level and the band edge narrows, precisely the condition identified in Sec.~\ref{sec:eigenvalues} as characteristic of nonadiabatic hopping events --- and when the nuclear motion is rapid, as during the lattice reorganization 
accompanying a hopping event. The NACs therefore provide a direct, quantitative link between the structural dynamics of polaron hopping and the probability of electronic transitions between the polaronic state and the extended band states.

Fig.~\ref{fig:NACs} shows the absolute value of the NAC between the polaron state and the CBM for bulk and surface rutile TiO$_2$, and between the polaron state and the VBM for MgO, as a function of time. Vertical lines mark the times at which polaron hopping events occur. Results at all other investigated temperatures are provided in Fig.~S2.

In bulk TiO$_2$ (Fig.~\ref{fig:Fig8a_NAC_TiO2}), the NAC fluctuates around a baseline of a few meV throughout the simulation, punctuated by sharp peaks reaching hundreds of meV. These peaks are correlated with polaron hopping events, reflecting the increase in electron--phonon coupling and the reduction of the defect-level--CBM gap that accompany the charge transfer. Crucially, not all hopping events produce equally strong NAC peaks: usually [001] hopping events generate weak NAC oscillations, whereas non-[001] events are responsible for the largest peaks. This is a direct consequence of Eq.~\ref{eq:NAC}: during [001] hopping the gap state remains well separated from the CBM (large $\epsilon_j - \epsilon_i$), suppressing the NAC even when the nuclear velocity is large, whereas during non-[001] hopping the narrowing of the gap amplifies the coupling substantially. This provides further dynamical evidence that the 
two hopping classes are physically distinct: the markedly different NAC response to [001] and non-[001] hopping events is fully consistent with their classification as adiabatic and nonadiabatic, respectively, and corroborates the picture emerging from the eigenvalue analysis of Sec.~\ref{sec:eigenvalues}. 
Taken together, the eigenvalue and NAC analyses establish a self-consistent dynamical picture in which the adiabatic/nonadiabatic character of each hopping event is directly encoded in the time evolution of the electronic structure~\cite{zhang2021dynamics,xu2022nonadiabatic}. 
To further investigate the behavior of the NACs for a localized polaron and during a hopping event, Fig.~S3 shows the NAC matrix elements between the polaron state and the higher-energy conduction-band states in bulk TiO$_2$ at 500 K. Comparing the localized and hopping configurations reveals a pronounced increase in the NAC strength during the hopping process.

In MgO (Fig.~\ref{fig:Fig8b_NAC_MgO}), the relationship between hopping events and NAC fluctuations is less pronounced. The NAC remains limited to a few meV for the majority of the simulation, with only occasional peaks reaching hundreds of meV. 
Inspection of these anomalous peaks reveals that they are moslty associated with rare non-nearest-neighbor hopping events in which the excess charge is transiently shared among multiple oxygen sites, as noted in Sec.~\ref{sec:eigenvalues}. The generally weak NAC in MgO reflects the predominantly adiabatic character of hole-polaron hopping: since the polaron level remains well separated from the VBM throughout the dynamics and the O~$2p$ orbital alignment along the [110] direction facilitates smooth charge transfer, the electron--phonon coupling 
matrix element and the associated NAC remain small. 

\begin{figure*}[htbp]
    \centering
     \begin{subfigure}{.465\textwidth}
         \centering
         \begin{overpic}[width=\linewidth]{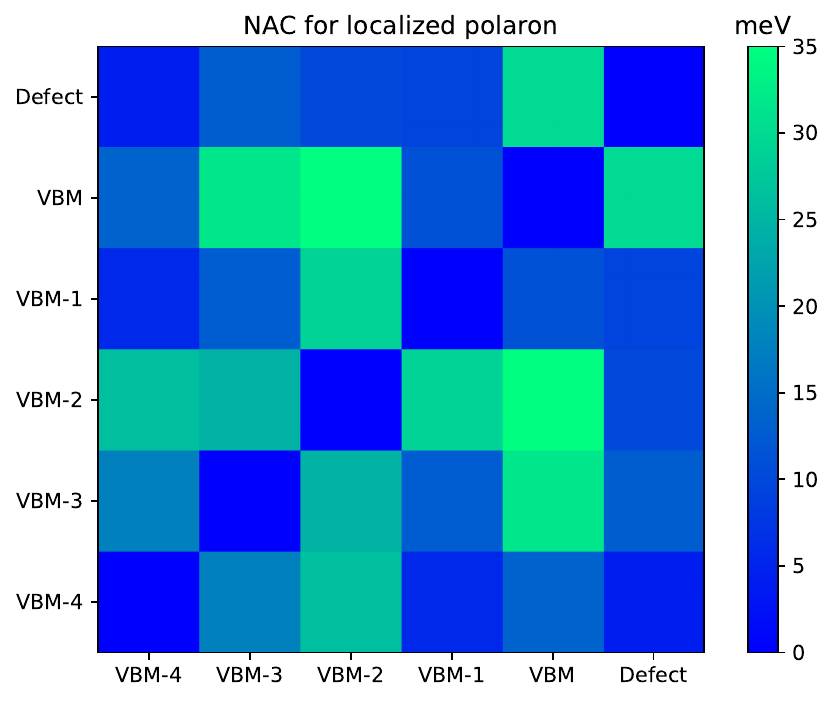}
            \put(12,73.5){\large \textcolor{white}{(a)}}
        \end{overpic}
        \phantomsubcaption
         \label{fig:Fig9a_Colormap_localized}
    \end{subfigure}
    \begin{subfigure}{.47\textwidth}
         \centering
         \begin{overpic}[width=\linewidth]{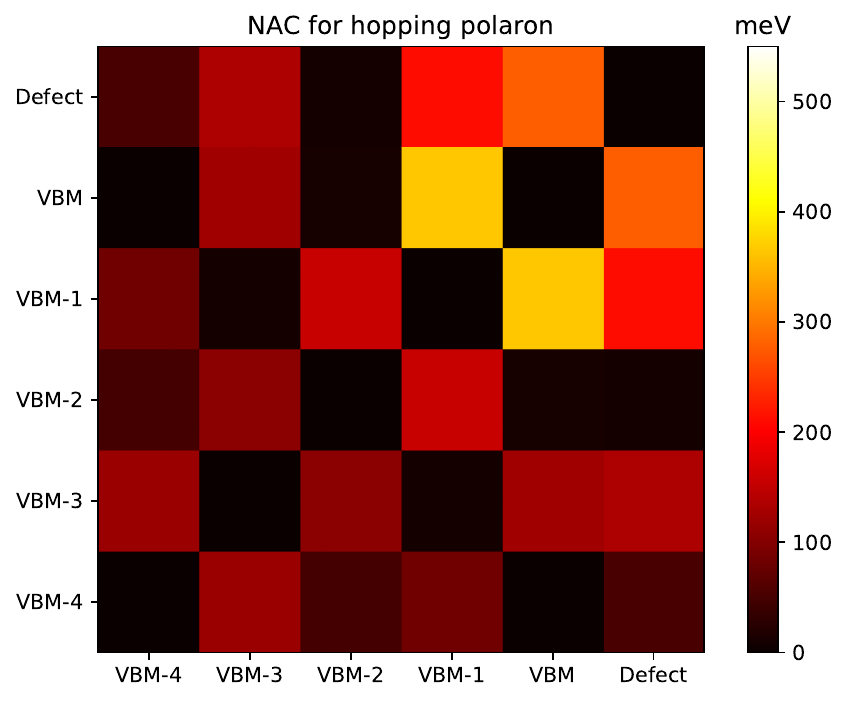}
            \put(12,72){\large \textcolor{white}{(b)}}
        \end{overpic}
        \phantomsubcaption
         \label{fig:Fig9b_Colormap_hopping}
    \end{subfigure}
        \caption{\justifying NAC colormap of MgO when the polaron is (a) localized at a site or (b) undergoing a hopping event. The color scale represents the absolute value of the NAC strength. Different color scales are used for the localized and hopping configurations, as the corresponding NAC strengths differ by approximately one order of magnitude.
        }
        \label{fig:NAC_colormap}
\end{figure*}

Fig.~\ref{fig:NAC_colormap} shows the full NAC matrix between the polaron state and the nearby valence-band states for MgO at 500\,K, comparing a configuration in which the polaron is localized at a single site with one in which it is undergoing a hopping event. In particular, we chose to investigate one of the [110] hopping events with a strong coupling between the in-gap polaron state and the VBM. In this case the NAC strength increases by approximately one order of magnitude during the hop, confirming that even in MgO, polaron motion is the primary driver of nonadiabatic coupling, albeit 
at a much weaker level than in rutile.

The rutile TiO$_2$(110) surface presents a striking contrast between the two temperatures considered. At 700\,K (Fig.~\ref{fig:Fig8c_NAC_TiO2-surf}), strong NAC peaks, reaching up to $\sim$300\,meV, are observed and are directly associated with the [111] interlayer hopping events in which the polaron migrates between the subsurface and surface Ti layers. These peaks closely mirror the behavior of non-[001] nonadiabatic events in the bulk, confirming the nonadiabatic character 
of subsurface-to-surface polaron migration. At 600\,K, by contrast, the NAC between the polaron state and the CBM remains below 10\,meV throughout the entire simulation (Fig.~S2c), reflecting the absence of interlayer hopping events at this temperature and the correspondingly weak electron--phonon coupling. This order-of-magnitude difference in NAC strength between 600 and 700\,K at the 
surface, far more dramatic than the temperature dependence observed in the bulk systems, points directly to the dominant role of subsurface-to-surface hopping as the gateway to strong nonadiabatic coupling at the TiO$_2$(110) surface.

\subsection{Excited states dynamics}
\label{sec:excited_states}

We now examine the electronic population dynamics obtained from the DISH simulations, focusing on how polaron hopping events drive de-excitation and, in the reverse process, how they can promote electronic excitations from the ground state. Since the excited-state po\-pu\-la\-tion dynamics are not uniformly well described by a single exponential decay across all systems and temperatures 
we adopt a consistent definition of the de-excitation time across all systems: $\tau$ is defined as the time at which the EX1 population decreases to $1/e$ of its initial value.

\begin{figure}[htbp]
    \centering
    \includegraphics[width=0.97\linewidth]{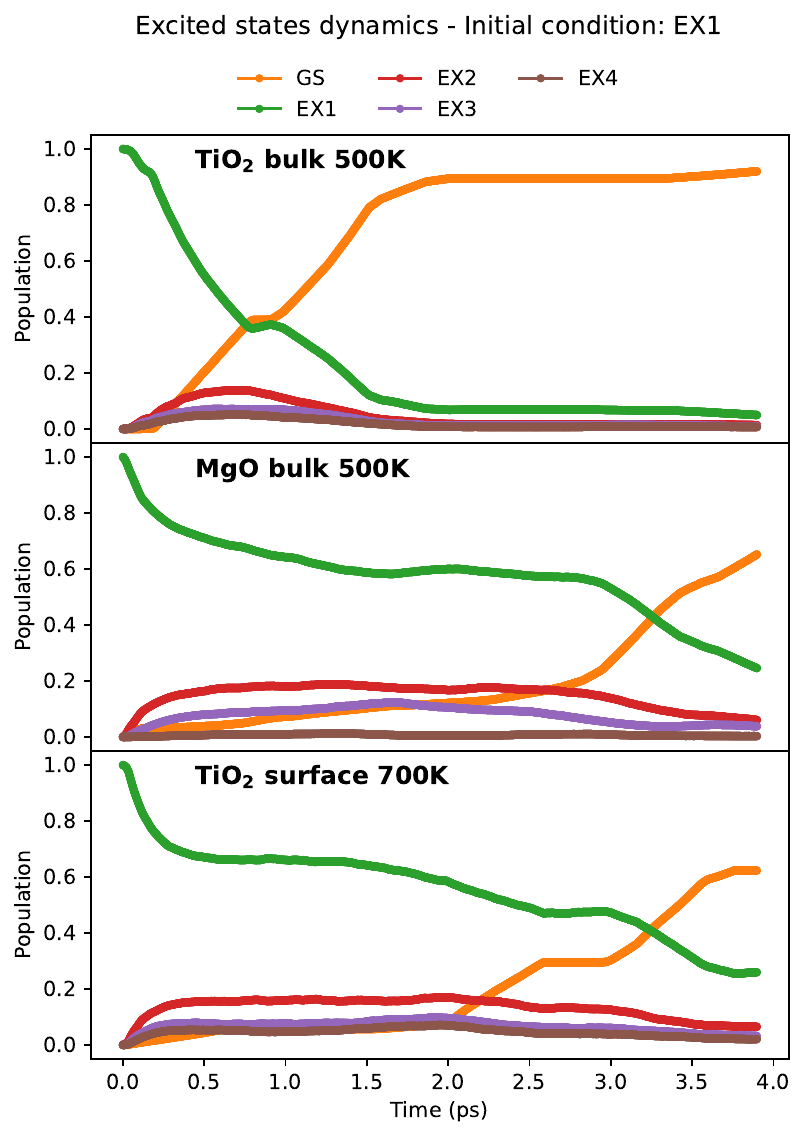}
    \caption{\justifying Evolution of the ensemble-averaged electronic populations for (top) bulk TiO$_2$, (center) bulk MgO, and (bottom) the reduced TiO$_2$(110) surface, at 500\,K (bulk systems) and 700\,K (surface), with the system initialized in the first excited state EX1. In all cases the dominant transition is toward the ground state; contributions from intermediate excited states are also visible, reflecting the multi-state character of the relaxation process.}
    \label{fig:Fig10_NAMD}
\end{figure}

\textit{Bulk rutile TiO$_2$.}
The ensemble-averaged de-excitation dynamics of bulk rutile TiO$_2$ at 500\,K are shown in the top panel of Fig.~\ref{fig:Fig10_NAMD}. 
The EX1 population decays pre\-do\-mi\-nan\-tly toward the GS, with minor 
contributions from intermediate excited states, reflecting the multi-state character of the relaxation process. The de-excitation times are 3.01, 0.87, and 0.46\,ps at 400, 500, and 600\,K, respectively, showing a strong temperature dependence that directly mirrors the increase in hopping frequency with temperature. More frequent polaron hoppings, in particular the nonadiabatic events discussed in the previus sections, generate more frequent NAC peaks: this accelerates the EX1$\rightarrow$GS relaxation. This connection between polaron hopping rate and charge-carrier lifetime is consistent with prior observations that defect motion accelerates excited-state 
dynamics~\cite{li2017sulfur, zhang2021dynamics}.

\textit{Bulk MgO.}
The de-excitation dynamics of MgO at 500\,K are shown in the center panel of Fig.~\ref{fig:Fig10_NAMD}. The EX1 population decays more slowly and less abruptly than in rutile, with de-excitation times of 4.29, 3.40, and 2.64\,ps at 400, 500, and 600\,K, respectively. These values are approximately four to six times longer than the corresponding rutile values at the same temperatures, reflecting the predominantly adiabatic character of hole-polaron hopping in MgO and the correspondingly weaker NAC. The smoother population decay, without the sudden discontinuities seen in 
rutile, confirms that individual hopping events have a much weaker impact on the excited-state population in MgO than in rutile. For the system at 400\,K, the simulation time of 3.9\,ps was insufficient to reach the 1/e decay threshold; the dynamics were therefore extended by iterating the nonadiabatic Hamiltonians~\cite{zhang2021dynamics, li2020anti}.

\textit{Reduced rutile TiO$_2$(110) surface.}
The de-excitation dynamics at the rutile TiO$_2$(110) surface reveal a striking contrast with the bulk behavior. At 700\,K, the EX1 population reaches the 1/e threshold after 3.35\,ps (bottom panel of 
Fig.~\ref{fig:Fig10_NAMD}), comparable to the bulk MgO 
timescale. At 600\,K, however, the EX1 decay is so slow that the 1/e 
threshold cannot be reached within the simulation time, and the 
de-excitation lifetime is estimated to be at least one to two orders of magnitude longer than in the bulk --- on the scale of tens to hundreds of picoseconds. This dramatic slowdown reflects the combination of a much lower hopping frequency at the surface (one event at 600\,K versus 48 in the bulk, see Fig.~S1) and the correspondingly weak NAC: without frequent nonadiabatic hopping events to drive electronic transitions, the polaron-mediated relaxation channel is effectively suppressed. The pronounced temperature dependence of the surface de-excitation time --- from hundreds of picoseconds at 600\,K to 3.35\,ps at 700\,K --- is therefore a direct consequence of the onset of interlayer [111] hopping events at 700\,K, which activate the nonadiabatic relaxation channel.

The de-excitation dynamics at all other investigated temperatures are 
reported in the Supplementary Materials (Figs.~S4). The qualitative picture is consistent across temperatures: bulk rutile TiO$_2$ exhibits clear polaron-hopping-induced population decay at all temperatures considered, while MgO displays a smoother de-excitation process with a weaker dependence on individual hopping events.

\subsection{Analysis of individual surface hopping trajectories}

To complement the ensemble-averaged analysis, we also examined the populations along a few representative individual trajectories, each starting from a different initial geometry, namely starting from initial time t$_{\text{in}}$=0, 200,
and 500\,fs. This allows us to identify how specific polaron hopping events influence the excited-state dynamics, information that is largely masked by ensemble averaging.

\begin{figure*}[htbp]
    \centering
     \begin{subfigure}{.49\textwidth}
         \centering
        
         \begin{overpic}[width=\linewidth]{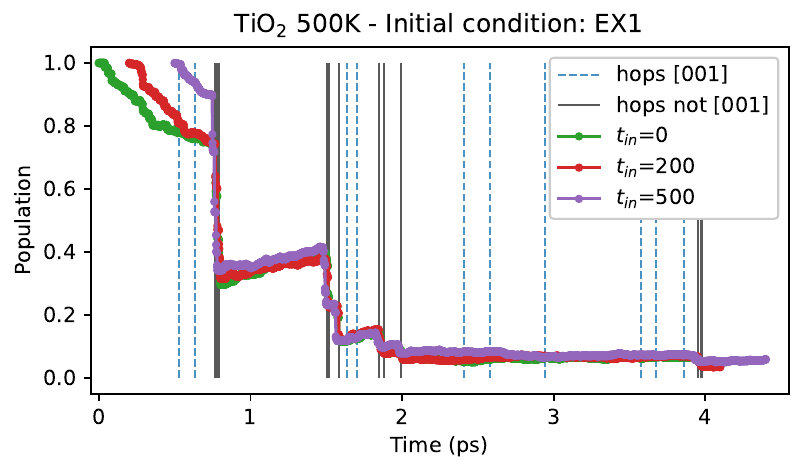}
            \put(0,53){\large (a)}
        \end{overpic}
         \phantomsubcaption
\label{fig:Fig11a_3iconds_TiO2-500K_EX1}
    \end{subfigure}
     %\hspace{0.cm}
\begin{subfigure}{.49\textwidth}
         \centering
       
         \begin{overpic}[width=\linewidth]{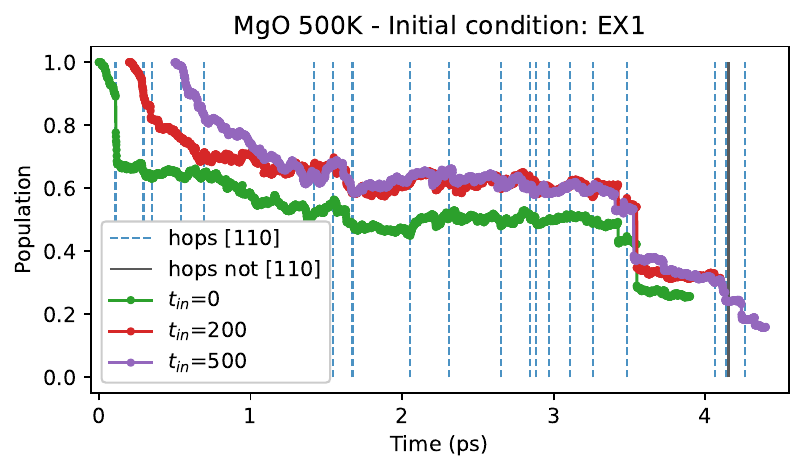}
            \put(0,53){\large (b)}
        \end{overpic}
        \phantomsubcaption
         \label{fig:Fig11b_3iconds_MgO-500K_EX1}
    \end{subfigure}
%\vspace{0.5cm}

\begin{subfigure}{.49\textwidth}
         \centering
         
                  \begin{overpic}[width=\linewidth]{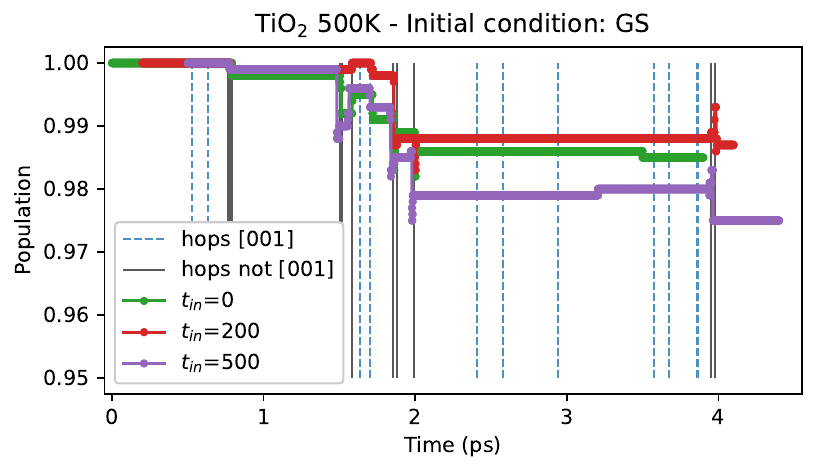}
            \put(0,53){\large (c)}
        \end{overpic}
        \phantomsubcaption
        \label{fig:Fig11c_3iconds_TiO2-500K_GS}
    \end{subfigure}
%\hspace{0.5cm}     
    \begin{subfigure}{.49\textwidth}
         \centering
         
         \begin{overpic}[width=\linewidth]{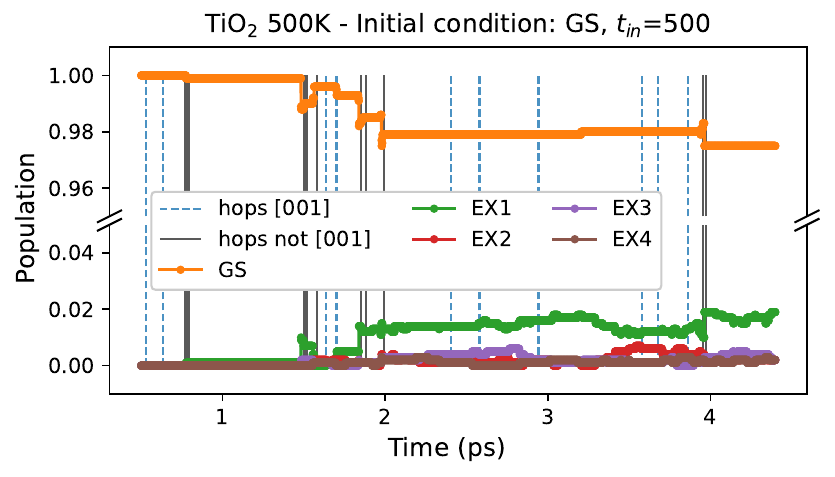}
            \put(0,53){\large (d)}
        \end{overpic}
        \phantomsubcaption
        \label{fig:Fig11d_1iconds_TiO2-500K_GS}
    \end{subfigure}

\begin{subfigure}{.49\textwidth}
         \centering
         
                  \begin{overpic}[width=\linewidth]{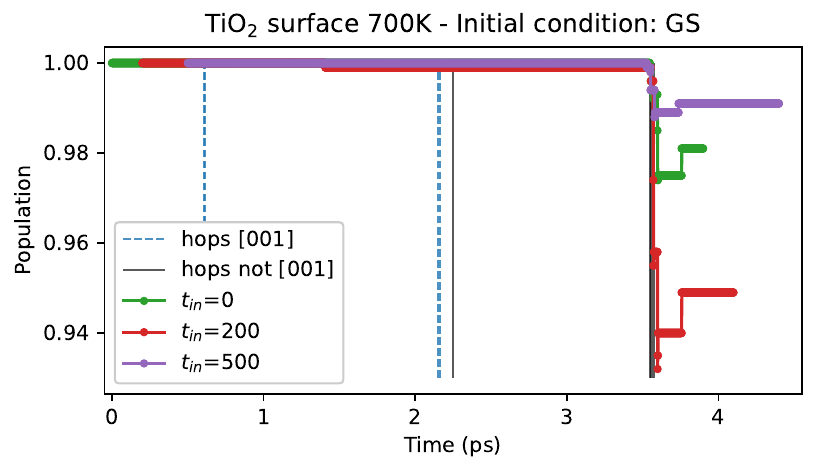}
            \put(0,53){\large (e)}
        \end{overpic}
        \phantomsubcaption
        \label{fig:Fig11e_3iconds_TiO2surf_GS}
    \end{subfigure}
%\hspace{0.5cm}     
    \begin{subfigure}{.49\textwidth}
         \centering
         
         \begin{overpic}[width=\linewidth]{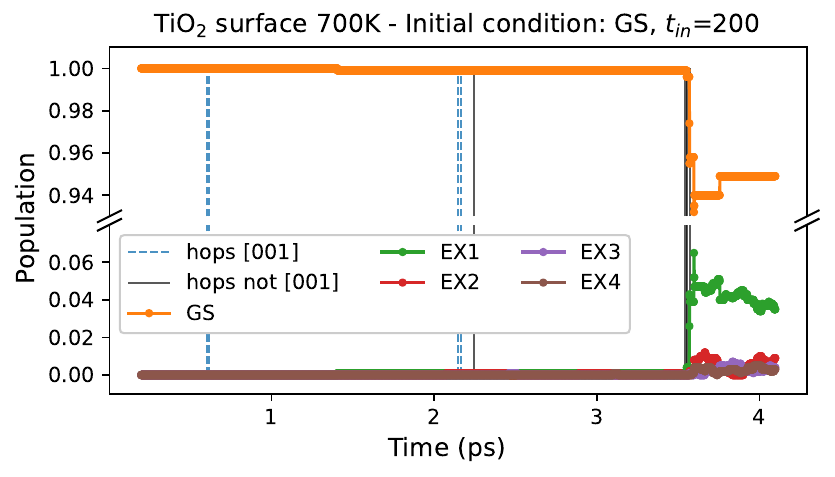}
            \put(0,53){\large (f)}
        \end{overpic}
        \phantomsubcaption
        \label{fig:Fig11f_1iconds_TiO2surf_GS}
    \end{subfigure}  
        \caption{\justifying Individual surface-hopping trajectories for representative initial geometries ($t_{\rm in} = 0$, 200, and 500\,fs). (a) Bulk TiO$_2$ at 500\,K, EX1 initial condition: sudden population drops are associated with non-[001] polaron hopping events. (b) Bulk MgO at 500\,K, EX1 initial condition: population drops are present but smoother than in rutile. (c) Bulk TiO$_2$ at 500\,K, GS initial condition: small population decreases indicate thermally activated GS$\rightarrow$EX transitions driven by nonadiabatic hopping. (d) Single trajectory ($t_{\rm in} = 500$\,fs) showing the GS$\rightarrow$EX excitation in detail. (e) Reduced TiO$_2$(110) surface at 700\,K, GS initial condition: GS$\rightarrow$EX transitions are associated with [111] interlayer hopping events. (f) Single trajectory ($t_{\rm in} = 200$\,fs) showing the surface GS$\rightarrow$EX excitation in detail.}
        \label{fig:single_iconds}
\end{figure*}

\begin{figure}
    \centering
    \includegraphics[width=0.99\linewidth]{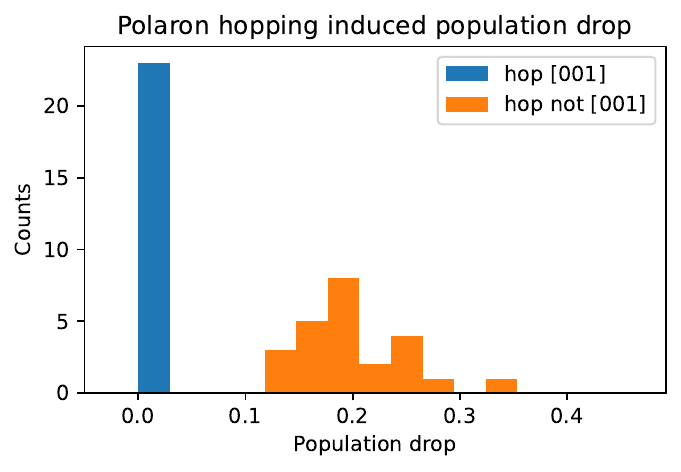}
    \caption{\justifying Statistical distribution of EX1 population drops induced by polaron hopping events in bulk rutile TiO$_2$ at 500\,K, computed across all surface-hopping trajectories. Hoppings along [001] (blue) never produce a population discontinuity, while hoppings along non-[001] directions (orange) produce a broad distribution of non-zero population drops, confirming their classification as adiabatic and nonadiabatic, respectively.
    }
    \label{fig:Fig12_Histograms}
\end{figure}

\textit{De-excitation driven by polaron hopping in TiO$_2$.}
When the bulk TiO$_2$ system is initialized in EX1, the population for representative individual trajectories (Fig.~\ref{fig:Fig11a_3iconds_TiO2-500K_EX1}) exhibits distinct sudden drops, directly correlated with non-[001] polaron hopping events. By contrast, [001] hopping events produce no discernible discontinuity in the EX1 population, consistent with the weak NAC associated with these adiabatic events. To quantify this behavior statistically, for each hopping event across all NAMD simulations with different initial conditions we evaluated the EX1 population drop, defined as the population difference before and after each polaron hopping event,   and collected the results in a histogram (Fig.~\ref{fig:Fig12_Histograms}). The [001] events produce a population drop indistinguishable from zero in all 
trajectories, whereas non-[001] events produce a broad distribution of non-zero drops, confirming their predominantly nonadiabatic character.

These findings provide the first direct dynamical confirmation of the 
adiabatic/nonadiabatic classification originally proposed by Deskins and Dupuis~\cite{deskins2007electron} on the basis of static activation-energy arguments. While their classification relied on the magnitude of the electronic coupling at the transition state, the criterion adopted here is inherently dynamical: it is based on the measurable impact of individual hopping events on the excited-state population during finite-temperature simulations. Moreover, whereas the original classification considered only the [001] and [111] hopping pathways, the present approach is fully general and can be extended to hopping events along any crystallographic direction 
without additional static calculations.

In MgO (Fig.~\ref{fig:Fig11b_3iconds_MgO-500K_EX1}), individual hopping events produce at most weak discontinuities in the EX1 population, and the overall EX1$\rightarrow$GS process proceeds gradually, consistent with the predominantly adiabatic character of hole-polaron hopping along the [110] direction. The alignment of O~$2p$ orbitals at the initial and final sites allows for smooth charge transfer with little perturbation to the electronic excited-state manifold.

The de-excitation dynamics for representative trajectories at other temperatures for bulk TiO$_2$ and MgO and for TiO$_2$(110) surface are provided in Fig.~S5.

\textit{Polaron-hopping-induced excitations from the ground state.}
A particularly striking and novel finding is that nonadiabatic polaron hopping in bulk rutile TiO$_2$ can drive electronic transitions not only downward (de-excitation) but also upward: when the system is initialized in the GS, non-[001] hopping events produce a 1--3\% decrease in the GS population (Fig.~\ref{fig:Fig11c_3iconds_TiO2-500K_GS}), corresponding to thermally activated excitations toward EX1 and EX2. 
We emphasize that these population changes, though small in absolute 
terms, are statistically well converged, because for every representative trajectory we perform 1000 realizations of the surface hopping algorithm. %they represent averages over an ensemble of $1.5 \times 10^{5}$ surface-hopping trajectories, 
In addition, their direct correlation with specific non-[001] hopping events 
(Fig.~\ref{fig:Fig11c_3iconds_TiO2-500K_GS}, ~\ref{fig:Fig11d_1iconds_TiO2-500K_GS}, ~\ref{fig:Fig11e_3iconds_TiO2surf_GS},
~\ref{fig:Fig11f_1iconds_TiO2surf_GS}) confirms that they are genuine 
physical signals rather than numerical noise of the DISH algorithm.
For the representative trajectory with $t_{\rm in} = 500$\,fs (Fig.~\ref{fig:Fig11d_1iconds_TiO2-500K_GS}), the GS population decreases by 2\%, with 1.5\% transferred to EX1 and 0.5\% to EX2. These excitations are absent for [001] hopping events, for which the NAC remains too weak to drive upward transitions.

This behavior can be rationalized in terms of a Landau--Zener 
transition~\cite{Landau:1932vnv, Zener1932, Glasbrenner2023}. The nonadiabatic polaron hopping induces strong phonon oscillations, increasing the NAC strength, and driving the polaronic in-gap state toward the conduction-band minimum; as a result, the energy gap between the two states narrows and the system approaches an avoided crossing. In this regime, the adiabatic approximation breaks down and the finite probability of a nonadiabatic transition becomes non-negligible, allowing the phonon vibrational energy deposited during the hopping event to be transferred to the electronic subsystem and promote the carrier to an excited state. The probability of such a transition increases as the minimum gap decreases, which explains why nonadiabatic hoppings --- characterized by a stronger approach of the 
polaronic level to the CBM --- are the events most directly responsible for thermally activated electronic excitations in bulk rutile TiO$_2$.

In MgO, no GS$\rightarrow$EX transitions are observed at any temperature: the VBM and the polaron level do not approach sufficiently during hopping, and the NAC between the involved states remains too weak to drive upward transitions (Fig.~S6c).

At the reduced rutile TiO$_2$(110) surface, analogous GS$\rightarrow$EX excitations are observed at 700\,K, where a 1--5\% change in the GS population (Fig.~\ref{fig:Fig11e_3iconds_TiO2surf_GS}) is associated with [111] interlayer hopping events between the subsurface and surface Ti layers. 
For a representative trajectory with $t_{\rm in} = 200$\,fs 
(Fig.~\ref{fig:Fig11f_1iconds_TiO2surf_GS}), the GS population decreases by 4\%, 
the largest excitation amplitude observed across all systems. This 
underscores the particularly strong nonadiabatic character of 
subsurface-to-surface polaron migration, which combines a large defect-level approach to the CBM with the rapid nuclear motion associated with interlayer charge transfer.

Polaron-hopping-promoted excitations from the GS are confirmed at 600\,K in bulk TiO$_2$ (Figs.~S6a,~S6b), where 1--3\% of the 
ground-state population is transferred to excited states, with approximately 1.3\% promoted to EX1 and 0.7\% to EX2 for a representative trajectory. No such thermally activated excitations are observed in MgO at any temperature or at the reduced rutile TiO$_2$(110) surface at 600\,K (Figs.~S6c,~S6d), where the NAC strength remains insufficient to drive GS$\rightarrow$EX transitions.

\subsection{Decoherence}
\label{sec:decoherence}

In systems hosting polarons, decoherence plays a central role in the 
excited-state dynamics because the strong coupling between the excess 
charge carrier and the lattice vibrations leads to rapid loss of electronic phase coherence. The pure dephasing time between two electronic states is estimated from the Gaussian fit of the decoherence function~\cite{akimov2013pyxaid}:

\begin{equation}
f(t) = e^{-\frac{1}{2}(t/\tau)^2}
\end{equation}

For bulk TiO$_2$, the dephasing time between the polaron state and the CBM is 2.83, 2.72, and 2.49\,fs at 400, 500, and 600\,K, respectively. At the rutile TiO$_2$(110) surface, the corresponding values are 2.91 and 2.80\,fs at 600 and 700\,K. For MgO, the dephasing time between the GS and EX1 is 3.06, 3.08, and 2.51\,fs at 400, 500, and 600\,K, respectively. In all systems, the dephasing time decreases slightly with increasing temperature, consistent with enhanced thermal fluctuations of the defect-level energy gap at higher temperatures.

\begin{figure}[htbp]
    \centering
\includegraphics[width=0.9\linewidth]{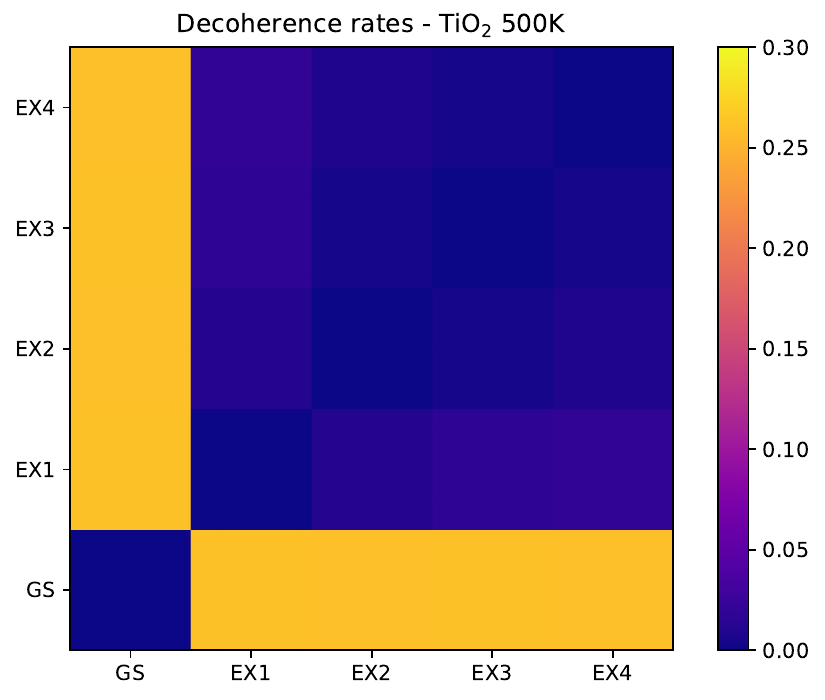}
    \caption{\justifying Decoherence rate matrix for bulk TiO$_2$ at 500\,K, shown as a color map. Decoherence rates are proportional to the inverse of the dephasing time. Values are averaged over all initial geometries. Pairs of states involving the polaronic defect level show the highest decoherence rates, reflecting the strong fluctuations of the defect-level energy gap induced by polaron motion. Pairs of extended band states remain coherent for longer, as their energy differences oscillate less strongly in time.}
    \label{fig:Fig13_Decoherence_rates}
\end{figure}

A key structural feature of the decoherence is the separation of timescales: the polaronic defect state decoheres from all other states on a timescale of a few femtoseconds, whereas pairs of extended band states remain coherent for tens of femtoseconds. This is illustrated in Fig.~\ref{fig:Fig13_Decoherence_rates}, which shows the 
decoherence rate matrix for TiO$_2$ bulk at 500\,K. The large decoherence rates involving the defect level reflect the strong fluctuations of the defect-level--CBM gap induced by polaron motion and lattice vibrations, whereas the slower decoherence between extended band states reflects the more stable energy gaps between those levels. Analogous considerations hold also for MgO (Fig. S7a), in which we record a fast decoherence between the polaronic state and any other valence  state, and for TiO$_2$(110) surface (Fig. S7b).

This separation of timescales has a direct consequence for the excited-state dynamics through the quantum Zeno 
effect~\cite{itano1990quantum,kilina2013quantum}: very rapid decoherence between the polaron state and the first excited state would largely suppress surface-hopping transitions between these two states if they were the only states in the active space. The inclusion of additional excited states (EX2--EX4) provides alternative relaxation pathways through which the system can evolve from EX1 to GS via intermediate 
states~\cite{akimov2014advanced,akimov2014nonadiabatic}, explaining why the de-excitation process in both TiO$_2$ and MgO involves multiple electronic states rather than a direct EX1$\rightarrow$GS transition. This also justifies the choice of a five-state active space adopted in this work.

\begin{table*}[htbp]
\begin{ruledtabular}
\begin{tabular}{cccc|ccc|cc}
 
 & &{$\text{TiO}_{2}$ bulk}& & & MgO bulk  & &\multicolumn{2}{c}{$\text{TiO}_{2}$ surface} \\

& 400\,K &  500\,K & 600\,K &  400\,K &   500\,K &  600\,K  & 600\,K& 700\,K\\ \hline
 
 N. hoppings in 4.5\,ps& 16& 21& 48 & 12 & 21 & 21& 1& 6\\
 De-excitation time (ps) & 3.01& 0.87& 0.46 & 4.29 & 3.40& 2.64& $10^{2}$-$10^{3}$ &3.35\\
Dephasing time (fs) & 2.83& 2.72 & 2.49 & 3.06&3.08 &2.51&2.91 &2.80 \\
\end{tabular}
\caption{\label{tab:summary} \justifying Comparison of the main parameters across the investigated systems. The de-excitation time $\tau$ is defined as the time at which the EX1 population reaches $1/e$ of its initial value; the value for the TiO$_2$(110) surface at 600\,K is a lower-bound estimate based on the slow population decay observed within the simulation time. The dephasing times correspond to the GS--EX1 pair.}
\end{ruledtabular}
\end{table*}

\section{Discussion}
\label{sec:discussion}

The results presented in this work establish a direct, quantitative connection between the microscopic character of individual polaron hopping events and nonadiabatic excited-state dynamics, providing a unified picture that integrates thermally activated polaron transport with charge-carrier relaxation and excitation. We discuss here the key physical findings and their implications, comparing with prior work and highlighting the aspects that are specific to the multiscale framework adopted.

\textit{Polaron hopping as a driver of nonadiabatic dynamics.}
A central finding of this work is that polaron hopping is the primary 
driver of nonadiabatic electronic transitions in both bulk rutile TiO$_2$ 
and at the reduced TiO$_2$(110) surface: every observed population 
discontinuity in the excited-state dynamics can be traced to a specific 
hopping event, and the magnitude of the resulting transition probability 
is directly governed by the NAC strength at the moment of the hop. This 
polaron-mediated relaxation mechanism is consistent with the general 
observation that defect motion accelerates excited-state 
dynamics~\cite{li2017sulfur}, and with prior NAMD studies of photoexcited 
polarons in TiO$_2$ that identified large NAC peaks during 
hopping~\cite{zhang2021dynamics,xu2022nonadiabatic,xu2023polaron}. The 
present work extends those studies in two important respects. First, by 
combining MLMD trajectories with NAMD, we access a much larger ensemble 
of thermally activated hopping events than is possible with AIMD-based 
approaches~\cite{zhang2021dynamics}, enabling a statistically meaningful 
separation of the two hopping classes and a direct mapping between hopping pathway and nonadiabatic transition probability. Second, we demonstrate that the polaron-mediated nonadiabatic channel operates not only during de-excitation but also in the reverse direction: nonadiabatic hopping events can transfer phonon vibrational energy to the electronic subsystem, promoting thermally activated GS$\rightarrow$EX transitions via a Landau--Zener mechanism~\cite{Landau:1932vnv,Zener1932,Glasbrenner2023} when the polaron in-gap state approaches the CBM. 
This bidirectional coupling between lattice dynamics and electronic 
excitation is a qualitatively new finding with no counterpart in prior 
NAMD studies of polaron-containing oxides.
The computed de-excitation timescales for bulk rutile TiO$_2$ 
(0.46--3.01\,ps over the 400--600\,K range) are consistent with 
sub-picosecond to few-picosecond carrier lifetimes observed 
experimentally in transient absorption and time-resolved photoemission 
studies of rutile TiO$_2$~\cite{zhang2019state,tanner2022tio2}, 
supporting the physical validity of the present multiscale framework.

\textit{Dynamical confirmation and generalization of the adiabatic/nonadiabatic classification.}
The static classification of polaron hopping regimes proposed by Deskins 
and Dupuis~\cite{deskins2007electron} on the basis of Marcus-theory 
activation energy has been widely used to interpret polaron transport 
in rutile TiO$_2$. The theoretical framework connecting the 
adiabatic/nonadiabatic crossover to the electronic coupling at the 
transition state was further developed in the context of condensed-phase 
electron transfer~\cite{blumberger2015,oberhofer2010}, but its dynamical 
validation at finite temperature has remained elusive. The present work 
provides the first direct dynamical confirmation of this classification: 
[001] hopping events produce negligible NAC peaks and zero excited-state 
population drops, while non-[001] events produce large NAC peaks and 
significant population discontinuities, in quantitative agreement with 
the adiabatic and nonadiabatic labels assigned by static criteria. 
Importantly, the dynamical approach adopted here is more general than the 
static one: it does not require the identification of specific transition 
states or the calculation of electronic coupling matrix elements, and can 
be straightforwardly applied to hopping events along any crystallographic 
direction, including those not considered in the original 
classification~\cite{deskins2007electron}. This makes it a powerful 
complement to static Marcus-theory analyses, particularly in systems where the relevant hopping pathways are not known a priori.

\textit{Material comparison: rutile TiO$_2$ versus MgO.}
The contrasting behavior of rutile TiO$_2$ and MgO reflects the 
fundamentally different nature of electron and hole polarons in these 
two materials. In rutile, the electron polaron on the Ti $3d$ manifold 
can access multiple hopping pathways with very different electronic 
character, and the proximity of the defect level to the CBM during 
non-[001] hops makes nonadiabatic transitions highly probable. In MgO, 
the hole polaron on the O $2p$ manifold is constrained to the [110] 
nearest-neighbor pathway by the orbital geometry, and the large gap 
between the polaron level and the VBM suppresses the NAC throughout. 
The resulting de-excitation times differ by a factor of four to six at 
the same temperature; also thermal excitations, mostly GS$\rightarrow$EX1 and GS$\rightarrow$EX2 less frequently, which are observed 
in TiO$_2$, are entirely absent in MgO. This contrast is also consistent 
with the general observation that hole polarons are less mobile than 
electron polarons in transition-metal and ionic 
oxides~\cite{Redondo2024,Tao2023,Jiang2025}: lower hopping frequency 
translates directly into weaker nonadiabatic coupling and longer 
charge-carrier lifetimes. This difference reflects the more localized character of the O~$2p$ 
hole state in MgO compared to the Ti~$3d$ electron state in rutile, 
which leads to a larger polaron binding energy, a higher activation 
barrier for hopping, and a smaller electronic coupling between 
neighboring sites.

\textit{Surface polarons and photocatalytic implications.}
The results for the reduced rutile TiO$_2$(110) surface carry the most 
direct implications for photocatalysis. The strong stabilization of the 
electron polaron in the subsurface Ti layer leads to a hopping rate orders of magnitude lower than in the bulk~\cite{reticcioli2018formation,
setvin2014direct}, and correspondingly to de-excitation times of tens to 
hundreds of picoseconds at 600\,K, at least two orders of magnitude 
longer than in the bulk. This means that photogenerated carriers trapped 
as subsurface polarons have a much longer excited-state lifetime, which 
could in principle be beneficial for charge separation in photocatalytic 
applications. However, the bottleneck for photocatalytic activity is the 
migration of the polaron from the subsurface to the reactive surface layer. Our results demonstrate that this migration --- the [111] interlayer hopping process --- is predominantly nonadiabatic and is accompanied by strong NAC peaks and GS$\rightarrow$EX transitions. % observed across all systems studied. 
This means that subsurface-to-surface polaron migration is not only a thermally activated process but also an intrinsically quantum mechanical one, in which nonadiabatic electron--lattice coupling is an essential ingredient. The implication for photocatalysis is that strategies to enhance surface reactivity should focus not only on lowering the thermodynamic barrier for subsurface-to-surface migration but also on maximizing the nonadiabatic electron--phonon coupling along the [111] pathway, for instance through surface engineering, dopant introduction, or adsorbate-mediated polaron stabilization~\cite{Reticcioli2019, cheng2021co, xu2023polaron}.
We note that the temperatures considered for the surface (600 and 700\,K) 
are higher than typical photocatalytic operating conditions; they are 
required to observe interlayer hopping events within the accessible 
simulation time. The qualitative mechanism identified here --- namely, 
that subsurface-to-surface polaron migration is predominantly nonadiabatic 
and is the key gateway to strong nonadiabatic coupling at the surface --- 
is expected to persist at lower temperatures, albeit with a substantially 
longer characteristic timescale that would require even more extended 
trajectories to sample directly.

\textit{Temperature dependence.}
Across all systems, increasing temperature accelerates the de-excitation 
process by increasing the polaron hopping rate, which in turn generates 
more frequent NAC peaks and shorter charge-carrier lifetimes (Table 
\ref{tab:summary}). This temperature dependence is most pronounced in 
bulk rutile TiO$_2$, where the de-excitation time decreases by almost 
an order of magnitude between 400 and 600\,K (from 3.01 to 0.46\,ps), 
reflecting the strong increase in hopping frequency (from 16 to 48 events 
per 4.5\,ps). In MgO the temperature dependence is weaker, consistent with the adiabatic character of hopping and the correspondingly weaker coupling between hopping rate and NAC strength. At the surface, the temperature dependence is discontinuous rather than gradual: the de-excitation time drops by two orders of magnitude between 600 and 700\,K because the qualitative character of the dominant hopping pathway changes, from in-plane [001] subsurface hops (adiabatic, weak NAC) to interlayer [111] hops (nonadiabatic, strong NAC). This non-monotonic behavior highlights that at surfaces, the relevant control parameter for nonadiabatic dynamics is not simply temperature but the specific hopping pathway that becomes thermally accessible.

\section{Conclusions}

We have presented a multiscale computational framework that combines 
machine-learning molecular dynamics with decoherence-induced surface 
hopping to investigate nonadiabatic polaron dynamics in bulk rutile 
TiO$_2$, bulk MgO, and the reduced rutile TiO$_2$(110) surface. By 
building nonadiabatic simulations on top of nanosecond MLMD trajectories 
generated with the LEOPOLD framework, we overcome two complementary 
limitations of existing approaches: the statistical sampling bottleneck 
of AIMD-based nonadiabatic methods and the inability of ground-state 
MLMD to describe electronic transitions between adiabatic states. The 
resulting framework provides simultaneous access to thermally activated 
polaron hopping statistics and to the nonadiabatic electronic dynamics 
driven by individual hopping events.

The main physical findings can be summarized as follows:

\textit{Dynamical classification of polaron hopping regimes.}
We provide the first direct dynamical confirmation that polaron hopping 
in rutile TiO$_2$ occurs in two distinct regimes corresponding to the 
adiabatic ([001] direction) and nonadiabatic (non-[001] directions) 
classes originally identified by Deskins and Dupuis~\cite{deskins2007electron} from static calculations. The dynamical 
criterion, based on the NAC strength and the excited-state population 
response during individual hopping events, is more general than the 
static one and can be applied to any hopping pathway without prior 
knowledge of the transition state geometry.

\textit{Bidirectional polaron--driven electronic transitions.}
Nonadiabatic hopping events in bulk rutile TiO$_2$ drive both 
de-excitation (EX1$\rightarrow$GS) and thermally activated excitation 
(GS$\rightarrow$EX1), the latter via a Landau--Zener mechanism in which 
phonon vibrational energy is transferred to the electronic subsystem 
as the defect level approaches the CBM. This bidirectional coupling is 
absent in MgO, where the predominantly adiabatic character of hole-polaron hopping and the larger defect-level--VBM gap suppress both the NAC and the associated electronic transitions.

\textit{Surface polarons and photocatalytic relevance.}
At the reduced rutile TiO$_2$(110) surface, the strong subsurface 
stabilization of the electron polaron leads to de-excitation times 
two orders of magnitude longer than in the bulk at 600\,K. Most 
importantly, the subsurface-to-surface migration pathway, the process 
most directly relevant to photocatalytic surface reactivity --- is 
found to be predominantly nonadiabatic, with the strongest NAC peaks 
and GS$\rightarrow$EX1 excitation amplitudes observed across all systems. 
This identifies nonadiabatic electron--lattice coupling as an essential 
ingredient for enabling polaron migration toward the reactive surface 
layer, with direct implications for the design of photocatalytic 
materials based on reduced TiO$_2$.

\textit{Outlook.}
The primary computational bottleneck of the present workflow is the 
evaluation of the time-overlap matrices $\langle \psi_i(t)|\psi_j(t+\Delta t)\rangle$, which requires a self-consistent DFT calculation at every time step of the selected trajectory segments. Replacing this step with a machine-learned model, for instance using the E(3)-equivariant NAMD framework recently 
proposed by Zhang et al.~\cite{zhang2025advancing} or the Hamiltonian 
interpolation approach of Wu et al.~\cite{wu2026efficient}, would 
remove this bottleneck and enable nonadiabatic simulations over the full 
nanosecond MLMD trajectories, with the potential to resolve rare 
nonadiabatic events that are currently inaccessible. Such an extension 
would also make the framework applicable to larger supercells and more 
complex surface geometries, opening the way to a truly predictive 
multiscale description of nonadiabatic polaron dynamics in 
technologically relevant energy materials.

\FloatBarrier

\begin{acknowledgments}
This research was partially funded by the Austrian Science Fund (FWF) 10.55776/F8100 project TACO. 
The computational results presented have been obtained with the Austrian Scientific Cluster (ASC). We acknowledge the CINECA award under the ISCRA initiative, for the availability of high-performance computing resources and support.

\end{acknowledgments}

\section*{data availability}
A representative example illustrating how to reproduce the results presented in this work will be made available 
in a public repository (Zenodo) upon publication.

\nocite{*}

\bibliography{bibliography}% Produces the bibliography via BibTeX.

@article{blumberger2015,
    author = {Blumberger, Jochen},
    title = {Recent Advances in the Theory and Molecular Simulation
of Biological Electron Transfer Reactions},
    journal = {Chemical Reviews},
    volume = {115},
    number = {20},
    pages = {11191-11238},
    year = {2015},
    month = {10},
    issn = {0009-2665},
    doi = {10.1021/acs.chemrev.5b00298},
    url = {https://doi.org/10.1021/acs.chemrev.5b00298}
}

@article{oberhofer2010,
    author = {Oberhofer, Harald and Blumberger, Jochen},
    title = {Electronic coupling matrix elements from charge constrained density functional theory calculations using a plane wave basis set},
    journal = {The Journal of Chemical Physics},
    volume = {133},
    number = {24},
    pages = {244105},
    year = {2010},
    month = {12},
    issn = {0021-9606},
    doi = {10.1063/1.3507878},
    url = {https://doi.org/10.1063/1.3507878}
}

@Article{Batzner2022,
author={Batzner, Simon
and Musaelian, Albert
and Sun, Lixin
and Geiger, Mario
and Mailoa, Jonathan P.
and Kornbluth, Mordechai
and Molinari, Nicola
and Smidt, Tess E.
and Kozinsky, Boris},
title={E(3)-equivariant graph neural networks for data-efficient and accurate interatomic potentials},
journal={Nature Communications},
year={2022},
month={May},
day={04},
volume={13},
number={1},
pages={2453},
issn={2041-1723},
doi={10.1038/s41467-022-29939-5},
url={https://doi.org/10.1038/s41467-022-29939-5}
}

@article{Reticcioli2019,
  title = {Interplay between Adsorbates and Polarons: CO on Rutile ${\mathrm{TiO}}_{2}(110)$},
  author = {Reticcioli, Michele and Sokolovi\ifmmode \acute{c}\else \'{c}\fi{}, Igor and Schmid, Michael and Diebold, Ulrike and Setvin, Martin and Franchini, Cesare},
  journal = {Phys. Rev. Lett.},
  volume = {122},
  issue = {1},
  pages = {016805},
  numpages = {6},
  year = {2019},
  month = {Jan},
  publisher = {American Physical Society},
  doi = {10.1103/PhysRevLett.122.016805},
  url = {https://link.aps.org/doi/10.1103/PhysRevLett.122.016805}
}

@article{akimov2013pyxaid,
  title = {The PYXAID Program for Non-Adiabatic Molecular Dynamics in Condensed Matter Systems},
  volume = {9},
  ISSN = {1549-9626},
  url = {http://dx.doi.org/10.1021/ct400641n},
  DOI = {10.1021/ct400641n},
  number = {11},
  journal = {Journal of Chemical Theory and Computation},
  publisher = {American Chemical Society (ACS)},
  author = {Akimov,  Alexey V. and Prezhdo,  Oleg V.},
  year = {2013},
  month = Oct,
  pages = {4959–4972}
}

@article{akimov2014advanced,
  title = {Advanced Capabilities of the PYXAID Program: Integration Schemes,  Decoherence Effects,  Multiexcitonic States,  and Field-Matter Interaction},
  volume = {10},
  ISSN = {1549-9626},
  url = {http://dx.doi.org/10.1021/ct400934c},
  DOI = {10.1021/ct400934c},
  number = {2},
  journal = {Journal of Chemical Theory and Computation},
  publisher = {American Chemical Society (ACS)},
  author = {Akimov,  Alexey V. and Prezhdo,  Oleg V.},
  year = {2014},
  month = Jan,
  pages = {789–804}
}

@article{xu2022nonadiabatic,
  title = {Nonadiabatic Dynamics of Polaron Hopping and Coupling with Water on Reduced TiO2},
  volume = {13},
  ISSN = {1948-7185},
  url = {http://dx.doi.org/10.1021/acs.jpclett.1c04231},
  DOI = {10.1021/acs.jpclett.1c04231},
  number = {3},
  journal = {The Journal of Physical Chemistry Letters},
  publisher = {American Chemical Society (ACS)},
  author = {Xu,  Zhong-Fei and Tong,  Chuan-Jia and Si,  Ru-tong and Teobaldi,  Gilberto and Liu,  Li-Min},
  year = {2022},
  month = Jan,
  pages = {857–863}
}

@ARTICLE{birschitzky2025machine,
  title={Machine learning small polaron dynamics},
  author={Birschitzky, Viktor C and Leoni, Luca and Reticcioli, Michele and Franchini, Cesare},
  ISSN = {1079-7114},
  url = {http://dx.doi.org/10.1103/PhysRevLett.134.216301},
  DOI = {10.1103/physrevlett.134.216301},
  journal={Physical Review Letters},
  volume={134},
  number={21},
  pages={216301},
  year={2025},
  publisher={APS}
}

@article{hammes1994proton,
  title = {Proton transfer in solution: Molecular dynamics with quantum transitions},
  volume = {101},
  ISSN = {1089-7690},
  url = {http://dx.doi.org/10.1063/1.467455},
  DOI = {10.1063/1.467455},
  number = {6},
  journal = {The Journal of Chemical Physics},
  publisher = {AIP Publishing},
  author = {Hammes-Schiffer,  Sharon and Tully,  John C.},
  year = {1994},
  month = Sep,
  pages = {4657–4667}
}

@article{jaeger2012decoherence,
  title = {Decoherence-induced surface hopping},
  volume = {137},
  ISSN = {1089-7690},
  url = {http://dx.doi.org/10.1063/1.4757100},
  DOI = {10.1063/1.4757100},
  number = {22},
  journal = {The Journal of Chemical Physics},
  publisher = {AIP Publishing},
  author = {Jaeger,  Heather M. and Fischer,  Sean and Prezhdo,  Oleg V.},
  year = {2012},
  month = Oct 
}

@ARTICLE{deskins2007electron,
  title={Electron transport via polaron hopping in bulk Ti O 2: A density functional theory characterization},
  ISSN = {1550-235X},
  url = {http://dx.doi.org/10.1103/PhysRevB.75.195212},
  DOI = {10.1103/physrevb.75.195212},
  author={Deskins, N Aaron and Dupuis, Michel},
  journal={Physical Review B},
  volume={75},
  number={19},
  pages={195212},
  year={2007},
  publisher={APS}
}

@article{kresse1996efficient,
  title = {Efficient iterative schemes for ab initio total-energy calculations using a plane-wave basis set},
  volume = {54},
  ISSN = {1095-3795},
  url = {http://dx.doi.org/10.1103/PhysRevB.54.11169},
  DOI = {10.1103/physrevb.54.11169},
  number = {16},
  journal = {Physical Review B},
  publisher = {American Physical Society (APS)},
  author = {Kresse,  G. and Furthm\"{u}ller,  J.},
  year = {1996},
  month = Oct,
  pages = {11169–11186}
}

@article{kresse1996efficiency,
  title = {Efficiency of ab-initio total energy calculations for metals and semiconductors using a plane-wave basis set},
  volume = {6},
  ISSN = {0927-0256},
  url = {http://dx.doi.org/10.1016/0927-0256(96)00008-0},
  DOI = {10.1016/0927-0256(96)00008-0},
  number = {1},
  journal = {Computational Materials Science},
  publisher = {Elsevier BV},
  author = {Kresse,  G. and Furthm\"{u}ller,  J.},
  year = {1996},
  month = Jul,
  pages = {15–50}
}

@article{kresse1999ultrasoft,
  title = {From ultrasoft pseudopotentials to the projector augmented-wave method},
  volume = {59},
  ISSN = {1095-3795},
  url = {http://dx.doi.org/10.1103/PhysRevB.59.1758},
  DOI = {10.1103/physrevb.59.1758},
  number = {3},
  journal = {Physical Review B},
  publisher = {American Physical Society (APS)},
  author = {Kresse,  G. and Joubert,  D.},
  year = {1999},
  month = Jan,
  pages = {1758–1775}
}

@article{franchini2021polarons,
  title = {Polarons in materials},
  volume = {6},
  ISSN = {2058-8437},
  url = {http://dx.doi.org/10.1038/s41578-021-00289-w},
  DOI = {10.1038/s41578-021-00289-w},
  number = {7},
  journal = {Nature Reviews Materials},
  publisher = {Springer Science and Business Media LLC},
  author = {Franchini,  Cesare and Reticcioli,  Michele and Setvin,  Martin and Diebold,  Ulrike},
  year = {2021},
  month = Mar,
  pages = {560–586}
}

@incollection{reticcioli2020small,
  title={Small Polarons in Transition Metal Oxides},
  author={Reticcioli, Michele and Diebold, Ulrike and Kresse, Georg and Franchini, Cesare},
  booktitle={Handbook of Materials Modeling: Applications: Current and Emerging Materials},
  pages={1035--1073},
  year={2020},
  publisher={Springer},
  doi={10.1007/978-3-319-50257-1_52-1}
}

@article{li2018time,
  title = {Time-Domain ab Initio Analysis Rationalizes the Unusual Temperature Dependence of Charge Carrier Relaxation in Lead Halide Perovskite},
  volume = {3},
  ISSN = {2380-8195},
  url = {http://dx.doi.org/10.1021/acsenergylett.8b01608},
  DOI = {10.1021/acsenergylett.8b01608},
  number = {11},
  journal = {ACS Energy Letters},
  publisher = {American Chemical Society (ACS)},
  author = {Li,  Wei and Tang,  Jianfeng and Casanova,  David and Prezhdo,  Oleg V.},
  year = {2018},
  month = Oct,
  pages = {2713–2720}
}

@article{li2021ab,
  title = {Ab initio nonadiabatic molecular dynamics of charge carriers in metal halide perovskites},
  volume = {13},
  ISSN = {2040-3372},
  url = {http://dx.doi.org/10.1039/d1nr01990b},
  DOI = {10.1039/d1nr01990b},
  number = {23},
  journal = {Nanoscale},
  publisher = {Royal Society of Chemistry (RSC)},
  author = {Li,  Wei and She,  Yalan and Vasenko,  Andrey S. and Prezhdo,  Oleg V.},
  year = {2021},
  pages = {10239–10265}
}

@article{akimov2018simple,
  title = {A Simple Phase Correction Makes a Big Difference in Nonadiabatic Molecular Dynamics},
  volume = {9},
  ISSN = {1948-7185},
  url = {http://dx.doi.org/10.1021/acs.jpclett.8b02826},
  DOI = {10.1021/acs.jpclett.8b02826},
  number = {20},
  journal = {The Journal of Physical Chemistry Letters},
  publisher = {American Chemical Society (ACS)},
  author = {Akimov,  Alexey V.},
  year = {2018},
  month = Oct,
  pages = {6096–6102}
}

@article{shakiba2022libra,
  title = {Libra: A modular software library for quantum nonadiabatic dynamics},
  volume = {14},
  ISSN = {2665-9638},
  url = {http://dx.doi.org/10.1016/j.simpa.2022.100445},
  DOI = {10.1016/j.simpa.2022.100445},
  journal = {Software Impacts},
  publisher = {Elsevier BV},
  author = {Shakiba,  Mohammad and Smith,  Brendan and Li,  Wei and Dutra,  Matthew and Jain,  Amber and Sun,  Xiang and Garashchuk,  Sophya and Akimov,  Alexey},
  year = {2022},
  month = Dec,
  pages = {100445}
}

@article{zhang2021dynamics,
  title = {Dynamics of Photoexcited Small Polarons in Transition-Metal Oxides},
  volume = {12},
  ISSN = {1948-7185},
  url = {http://dx.doi.org/10.1021/acs.jpclett.1c00003},
  DOI = {10.1021/acs.jpclett.1c00003},
  number = {9},
  journal = {The Journal of Physical Chemistry Letters},
  publisher = {American Chemical Society (ACS)},
  author = {Zhang,  Lili and Chu,  Weibin and Zhao,  Chuanyu and Zheng,  Qijing and Prezhdo,  Oleg V. and Zhao,  Jin},
  year = {2021},
  month = Feb,
  pages = {2191–2198}
}

@article{setvin2014direct,
  title = {Direct view at excess electrons in TiO 2 rutile and anatase},
  volume = {113},
  ISSN = {1079-7114},
  url = {http://dx.doi.org/10.1103/PhysRevLett.113.086402},
  DOI = {10.1103/physrevlett.113.086402},
  number = {8},
  journal = {Physical Review Letters},
  publisher = {American Physical Society (APS)},
  author = {Setvin,  Martin and Franchini,  Cesare and Hao,  Xianfeng and Schmid,  Michael and Janotti,  Anderson and Kaltak,  Merzuk and Van de Walle,  Chris G. and Kresse,  Georg and Diebold,  Ulrike},
  year = {2014},
  month = Aug,
  pages={086402}
}

@article{reticcioli2018formation,
  title={Formation and dynamics of small polarons on the rutile TiO 2 (110) surface},
  author={Reticcioli, Michele and Setvin, Martin and Schmid, Michael and Diebold, Ulrike and Franchini, Cesare},
  ISSN = {2469-9969},
  url = {http://dx.doi.org/10.1103/PhysRevB.98.045306},
  DOI = {10.1103/physrevb.98.045306},
  journal={Physical Review B},
  volume={98},
  number={4},
  pages={045306},
  year={2018},
  publisher={APS}
}

@article{tully1990molecular,
  title = {Molecular dynamics with electronic transitions},
  volume = {93},
  ISSN = {1089-7690},
  url = {http://dx.doi.org/10.1063/1.459170},
  DOI = {10.1063/1.459170},
  number = {2},
  journal = {The Journal of Chemical Physics},
  publisher = {AIP Publishing},
  author = {Tully,  John C.},
  year = {1990},
  month = Jul,
  pages = {1061–1071}
}

@article{li2024formation,
  title = {Formation and Recombination Dynamics of Polarons in Goethite: A Time-Domain Ab Initio Study},
  volume = {15},
  ISSN = {1948-7185},
  url = {http://dx.doi.org/10.1021/acs.jpclett.4c02493},
  DOI = {10.1021/acs.jpclett.4c02493},
  number = {39},
  journal = {The Journal of Physical Chemistry Letters},
  publisher = {American Chemical Society (ACS)},
  author = {Li,  Hongliang and Zhou,  Zhaohui and Vasenko,  Andrey S. and Chulkov,  Evgueni V. and Fang,  Qiu and Long,  Run},
  year = {2024},
  month = Sep,
  pages = {10018–10025}
}

@article{cheng2022photoinduced,
  title = {Photoinduced small electron polarons generation and recombination in hematite},
  volume = {8},
  ISSN = {2057-3960},
  url = {http://dx.doi.org/10.1038/s41524-022-00814-7},
  DOI = {10.1038/s41524-022-00814-7},
  number = {1},
  journal = {npj Computational Materials},
  publisher = {Springer Science and Business Media LLC},
  author = {Cheng,  Cheng and Zhu,  Yonghao and Zhou,  Zhaohui and Long,  Run and Fang,  Wei-Hai},
  year = {2022},
  month = Jul,
  pages={148}
}

@article{li2020anti,
  title = {Anti-correlation between Band gap and Carrier Lifetime in Lead Halide Perovskites under Compression Rationalized by Ab Initio Quantum Dynamics},
  volume = {32},
  ISSN = {1520-5002},
  url = {http://dx.doi.org/10.1021/acs.chemmater.0c01287},
  DOI = {10.1021/acs.chemmater.0c01287},
  number = {11},
  journal = {Chemistry of Materials},
  publisher = {American Chemical Society (ACS)},
  author = {Li,  Wei and Chen,  Zhi and Tang,  Jianfeng and Prezhdo,  Oleg V.},
  year = {2020},
  month = May,
  pages = {4707–4715}
}

@article{li2019anharmonicity,
  title = {Anharmonicity Extends Carrier Lifetimes in Lead Halide Perovskites at Elevated Temperatures},
  volume = {10},
  ISSN = {1948-7185},
  url = {http://dx.doi.org/10.1021/acs.jpclett.9b02553},
  DOI = {10.1021/acs.jpclett.9b02553},
  number = {20},
  journal = {The Journal of Physical Chemistry Letters},
  publisher = {American Chemical Society (ACS)},
  author = {Li,  Wei and Vasenko,  Andrey S. and Tang,  Jianfeng and Prezhdo,  Oleg V.},
  year = {2019},
  month = Sep,
  pages = {6219–6226}
}

@article{prezhdo2021modeling,
  title = {Modeling Non-adiabatic Dynamics in Nanoscale and Condensed Matter Systems},
  volume = {54},
  ISSN = {1520-4898},
  url = {http://dx.doi.org/10.1021/acs.accounts.1c00525},
  DOI = {10.1021/acs.accounts.1c00525},
  number = {23},
  journal = {Accounts of Chemical Research},
  publisher = {American Chemical Society (ACS)},
  author = {Prezhdo,  Oleg V.},
  year = {2021},
  month = Nov,
  pages = {4239–4249}
}

@article{li2017sulfur,
  title = {Sulfur Adatom and Vacancy Accelerate Charge Recombination in MoS2 but by Different Mechanisms: Time-Domain Ab Initio Analysis},
  volume = {17},
  ISSN = {1530-6992},
  url = {http://dx.doi.org/10.1021/acs.nanolett.7b04374},
  DOI = {10.1021/acs.nanolett.7b04374},
  number = {12},
  journal = {Nano Letters},
  publisher = {American Chemical Society (ACS)},
  author = {Li,  Linqiu and Long,  Run and Bertolini,  Thomas and Prezhdo,  Oleg V.},
  year = {2017},
  month = Dec,
  pages = {7962–7967}
}

@article{cheng2021co,
  title = {CO Adsorbate Promotes Polaron Photoactivity on the Reduced Rutile TiO2 (110) Surface},
  volume = {2},
  ISSN = {2691-3704},
  url = {http://dx.doi.org/10.1021/jacsau.1c00508},
  DOI = {10.1021/jacsau.1c00508},
  number = {1},
  journal = {JACS Au},
  publisher = {American Chemical Society (ACS)},
  author = {Cheng,  Cheng and Zhu,  Yonghao and Fang,  Wei-Hai and Long,  Run and Prezhdo,  Oleg V.},
  year = {2021},
  month = Dec,
  pages = {234–245}
}

@article{kruger2008defect,
  title = {Defect states at the TiO 2 (110) surface probed by resonant photoelectron diffraction},
  volume = {100},
  ISSN = {1079-7114},
  url = {http://dx.doi.org/10.1103/PhysRevLett.100.055501},
  DOI = {10.1103/physrevlett.100.055501},
  number = {5},
  journal = {Physical Review Letters},
  publisher = {American Physical Society (APS)},
  author = {Kr\"{u}ger,  P. and Bourgeois,  S. and Domenichini,  B. and Magnan,  H. and Chandesris,  D. and Le Fèvre,  P. and Flank,  A. M. and Jupille,  J. and Floreano,  L. and Cossaro,  A. and Verdini,  A. and Morgante,  A.},
  year = {2008},
  month = Feb,
  pages={055501}
}

@article{kruger2012intrinsic,
  title = {Intrinsic nature of the excess electron distribution at the TiO 2 (110) surface},
  volume = {108},
  ISSN = {1079-7114},
  url = {http://dx.doi.org/10.1103/PhysRevLett.108.126803},
  DOI = {10.1103/physrevlett.108.126803},
  number = {12},
  journal = {Physical Review Letters},
  publisher = {American Physical Society (APS)},
  author = {Kr\"{u}ger,  P. and Jupille,  J. and Bourgeois,  S. and Domenichini,  B. and Verdini,  A. and Floreano,  L. and Morgante,  A.},
  year = {2012},
  month = Mar,
  pages={126803}
}

@article{itano1990quantum,
  title = {Quantum Zeno effect},
  volume = {41},
  ISSN = {1094-1622},
  url = {http://dx.doi.org/10.1103/PhysRevA.41.2295},
  DOI = {10.1103/physreva.41.2295},
  number = {5},
  journal = {Physical Review A},
  publisher = {American Physical Society (APS)},
  author = {Itano,  Wayne M. and Heinzen,  D. J. and Bollinger,  J. J. and Wineland,  D. J.},
  year = {1990},
  month = Mar,
  pages = {2295–2300}
}

@article{kilina2013quantum,
  title = {Quantum Zeno Effect Rationalizes the Phonon Bottleneck in Semiconductor Quantum Dots},
  volume = {110},
  ISSN = {1079-7114},
  url = {http://dx.doi.org/10.1103/PhysRevLett.110.180404},
  DOI = {10.1103/physrevlett.110.180404},
  number = {18},
  journal = {Physical Review Letters},
  publisher = {American Physical Society (APS)},
  author = {Kilina,  Svetlana V. and Neukirch,  Amanda J. and Habenicht,  Bradley F. and Kilin,  Dmitri S. and Prezhdo,  Oleg V.},
  year = {2013},
  month = May,
  pages={180404}
}

@article{akimov2013theoretical,
  title = {Theoretical Insights into Photoinduced Charge Transfer and Catalysis at Oxide Interfaces},
  volume = {113},
  ISSN = {1520-6890},
  url = {http://dx.doi.org/10.1021/cr3004899},
  DOI = {10.1021/cr3004899},
  number = {6},
  journal = {Chemical Reviews},
  publisher = {American Chemical Society (ACS)},
  author = {Akimov,  Alexey V. and Neukirch,  Amanda J. and Prezhdo,  Oleg V.},
  year = {2013},
  month = Apr,
  pages = {4496–4565}
}

@inbook{doltsinis2006molecular,
title = "Molecular dynamics beyond the Born-Oppenheimer approximation: mixed quantum-classical approaches",
author = "Doltsinis, \{N L\}",
year = "2006",
language = "English",
pages = "389 -- 409",
editor = "S Blugel and J Grotendorst and D Marx",
booktitle = "Computational nanoscience: do it yourself!",
publisher = "NIC",
}

@article{barbatti2011nonadiabatic,
  title = {Nonadiabatic dynamics with trajectory surface hopping method},
  volume = {1},
  ISSN = {1759-0884},
  url = {http://dx.doi.org/10.1002/wcms.64},
  DOI = {10.1002/wcms.64},
  number = {4},
  journal = {WIREs Computational Molecular Science},
  publisher = {Wiley},
  author = {Barbatti,  Mario},
  year = {2011},
  month = May,
  pages = {620–633}
}

@article{shluger1993small,
  title = {Small polarons in real crystals: concepts and problems},
  volume = {5},
  ISSN = {1361-648X},
  url = {http://dx.doi.org/10.1088/0953-8984/5/19/007},
  DOI = {10.1088/0953-8984/5/19/007},
  number = {19},
  journal = {Journal of Physics: Condensed Matter},
  publisher = {IOP Publishing},
  author = {Shluger,  A L and Stoneham,  A M},
  year = {1993},
  month = May,
  pages = {3049–3086}
}

@article{tully1971trajectory,
  title = {Trajectory Surface Hopping Approach to Nonadiabatic Molecular Collisions: The Reaction of H+ with D2},
  volume = {55},
  ISSN = {1089-7690},
  url = {http://dx.doi.org/10.1063/1.1675788},
  DOI = {10.1063/1.1675788},
  number = {2},
  journal = {The Journal of Chemical Physics},
  publisher = {AIP Publishing},
  author = {Tully,  John C. and Preston,  Richard K.},
  year = {1971},
  month = Jul,
  pages = {562–572}
}

@article{chng2026ab,
  title = {Ab Initio Polariton Transport Dynamics with the Classical Path Approximation},
  volume = {26},
  ISSN = {1530-6992},
  url = {http://dx.doi.org/10.1021/acs.nanolett.6c00383},
  DOI = {10.1021/acs.nanolett.6c00383},
  number = {18},
  journal = {Nano Letters},
  publisher = {American Chemical Society (ACS)},
  author = {Chng,  Benjamin X. K. and Weight,  Braden M. and Mondal,  M. Elious and Huo,  Pengfei},
  year = {2026},
  month = Apr,
  pages = {6055–6062}
}

@article{augustin1978classical,
  title = {The classical path approximation in time-dependent quantum collision theory},
  volume = {69},
  ISSN = {1089-7690},
  url = {http://dx.doi.org/10.1063/1.437100},
  DOI = {10.1063/1.437100},
  number = {9},
  journal = {The Journal of Chemical Physics},
  publisher = {AIP Publishing},
  author = {Augustin,  Stuart D. and Rabitz,  Herschel},
  year = {1978},
  month = Nov,
  pages = {4195–4200}
}

@misc{dai2025polarons,
  doi = {10.48550/ARXIV.2512.06176},
  url = {https://arxiv.org/abs/2512.06176},
  author = {Dai,  Zhenbang and Lafuente-Bartolome,  Jon and Giustino,  Feliciano},
  title = {Polarons from first principles},
  publisher = {arXiv},
  year = {2025},
  copyright = {arXiv.org perpetual,  non-exclusive license}
}

@article{zhang2019state,
  title = {State-selective dynamics of TiO2 charge-carrier trapping and recombination},
  volume = {10},
  ISSN = {1948-7185},
  url = {http://dx.doi.org/10.1021/acs.jpclett.9b02153},
  DOI = {10.1021/acs.jpclett.9b02153},
  number = {17},
  journal = {The Journal of Physical Chemistry Letters},
  publisher = {American Chemical Society (ACS)},
  author = {Zhang,  Yu and Payne,  Daniel T. and Pang,  Chi L. and Cacho,  Cephise and Chapman,  Richard T. and Springate,  Emma and Fielding,  Helen H. and Thornton,  Geoff},
  year = {2019},
  month = Aug,
  pages = {5265–5270}
}

@article{tanner2022tio2,
  title = {TiO2 polarons in the time domain: implications for photocatalysis},
  volume = {13},
  ISSN = {1948-7185},
  url = {http://dx.doi.org/10.1021/acs.jpclett.1c03677},
  DOI = {10.1021/acs.jpclett.1c03677},
  number = {2},
  journal = {The Journal of Physical Chemistry Letters},
  publisher = {American Chemical Society (ACS)},
  author = {Tanner,  Alex J. and Thornton,  Geoff},
  year = {2022},
  month = Jan,
  pages = {559–566}
}

@article{kowalski2010charge,
  title={Charge localization dynamics induced by oxygen vacancies on the TiO 2 (110) surface},
  author={Kowalski, Piotr M and Camellone, Matteo Farnesi and Nair, Nisanth N and Meyer, Bernd and Marx, Dominik},
  journal={Physical review letters},
  volume={105},
  ISSN = {1079-7114},
  url = {http://dx.doi.org/10.1103/PhysRevLett.105.146405},
  DOI = {10.1103/physrevlett.105.146405},
  number={14},
  pages={146405},
  year={2010},
  publisher={APS}
}

@inproceedings{batatia2022mace,
  series = {NeurIPS 2022},
  title = {MACE: Higher Order Equivariant Message Passing Neural Networks for Fast and Accurate Force Fields},
  url = {http://dx.doi.org/10.52202/068431-0830},
  DOI = {10.52202/068431-0830},
  booktitle = {Advances in Neural Information Processing Systems 35},
  publisher = {Neural Information Processing Systems Foundation,  Inc. (NeurIPS)},
  author = {Batatia,  Ilyes and Kovacs,  David P and Simm,  Gregor and Ortner,  Christoph and Csanyi,  Gabor},
  year = {2022},
  pages = {11423–11436},
  collection = {NeurIPS 2022}
}

@article{batzner20223,
  title = {E(3)-equivariant graph neural networks for data-efficient and accurate interatomic potentials},
  volume = {13},
  ISSN = {2041-1723},
  url = {http://dx.doi.org/10.1038/s41467-022-29939-5},
  DOI = {10.1038/s41467-022-29939-5},
  number = {1},
  journal = {Nature Communications},
  publisher = {Springer Science and Business Media LLC},
  author = {Batzner,  Simon and Musaelian,  Albert and Sun,  Lixin and Geiger,  Mario and Mailoa,  Jonathan P. and Kornbluth,  Mordechai and Molinari,  Nicola and Smidt,  Tess E. and Kozinsky,  Boris},
  year = {2022},
  month = May,
  pages={2453}
}

@article{falletta2022hubbard,
  title = {Hubbard U through polaronic defect states},
  volume = {8},
  ISSN = {2057-3960},
  url = {http://dx.doi.org/10.1038/s41524-022-00958-6},
  DOI = {10.1038/s41524-022-00958-6},
  number = {1},
  journal = {npj Computational Materials},
  publisher = {Springer Science and Business Media LLC},
  author = {Falletta,  Stefano and Pasquarello,  Alfredo},
  year = {2022},
  month = Dec,
  pages={263}
}

@article{ambrosio2025charge,
  title = {Charge localization in optoelectronic and photocatalytic applications: Computational perspective},
  volume = {126},
  ISSN = {1077-3118},
  url = {http://dx.doi.org/10.1063/5.0257250},
  DOI = {10.1063/5.0257250},
  number = {13},
  journal = {Applied Physics Letters},
  publisher = {AIP Publishing},
  author = {Ambrosio,  Francesco and Wiktor,  Julia},
  year = {2025},
  month = Mar,
  pages = {130501}
}

@article{groenhof2020observe,
  title = {Observe while it happens: catching photoactive proteins in the act with non-adiabatic molecular dynamics simulations},
  volume = {61},
  ISSN = {0959-440X},
  url = {http://dx.doi.org/10.1016/j.sbi.2019.12.013},
  DOI = {10.1016/j.sbi.2019.12.013},
  journal = {Current Opinion in Structural Biology},
  publisher = {Elsevier BV},
  author = {Groenhof,  Gerrit and Modi,  Vaibhav and Morozov,  Dmitry},
  year = {2020},
  month = Apr,
  pages = {106–112}
}

@article{landau1933electron,
  title={Electron motion in crystal lattices},
  url = "https://cds.cern.ch/record/444288",
  author={Landau, Lev Davidovich},
  journal={Phys. Z. Sowjet.},
  volume={3},
  pages={664},
  year={1933}
}

@inproceedings{jonsson1998nudged,
  title = {Nudged elastic band method for finding minimum energy paths of transitions},
  url = {http://dx.doi.org/10.1142/9789812839664_0016},
  DOI = {10.1142/9789812839664_0016},
  booktitle = {Classical and Quantum Dynamics in Condensed Phase Simulations},
  publisher = {WORLD SCIENTIFIC},
  author = {JÓNSSON,  HANNES and MILLS,  GREG and JACOBSEN,  KARSTEN W.},
  year = {1998},
  month = Jun,
  pages = {385–404}
}

@article{henkelman2000climbing,
  title = {A climbing image nudged elastic band method for finding saddle points and minimum energy paths},
  volume = {113},
  ISSN = {1089-7690},
  url = {http://dx.doi.org/10.1063/1.1329672},
  DOI = {10.1063/1.1329672},
  number = {22},
  journal = {The Journal of Chemical Physics},
  publisher = {AIP Publishing},
  author = {Henkelman,  Graeme and Uberuaga,  Blas P. and Jónsson,  Hannes},
  year = {2000},
  month = Dec,
  pages = {9901–9904}
}

@article{mai2018nonadiabatic,
  title = {Nonadiabatic dynamics: The SHARC approach},
  volume = {8},
  ISSN = {1759-0884},
  url = {http://dx.doi.org/10.1002/wcms.1370},
  DOI = {10.1002/wcms.1370},
  number = {6},
  journal = {WIREs Computational Molecular Science},
  publisher = {Wiley},
  author = {Mai,  Sebastian and Marquetand,  Philipp and González,  Leticia},
  year = {2018},
  month = May,
  pages={e1370}
}

@article{polli2010conical,
  title = {Conical intersection dynamics of the primary photoisomerization event in vision},
  volume = {467},
  ISSN = {1476-4687},
  url = {http://dx.doi.org/10.1038/nature09346},
  DOI = {10.1038/nature09346},
  number = {7314},
  journal = {Nature},
  publisher = {Springer Science and Business Media LLC},
  author = {Polli,  Dario and Altoè,  Piero and Weingart,  Oliver and Spillane,  Katelyn M. and Manzoni,  Cristian and Brida,  Daniele and Tomasello,  Gaia and Orlandi,  Giorgio and Kukura,  Philipp and Mathies,  Richard A. and Garavelli,  Marco and Cerullo,  Giulio},
  year = {2010},
  month = Sep,
  pages = {440–443}
}

@article{dou2020nonadiabatic,
  title = {Nonadiabatic Molecular Dynamics at Metal Surfaces},
  volume = {124},
  ISSN = {1520-5215},
  url = {http://dx.doi.org/10.1021/acs.jpca.9b10698},
  DOI = {10.1021/acs.jpca.9b10698},
  number = {5},
  journal = {The Journal of Physical Chemistry A},
  publisher = {American Chemical Society (ACS)},
  author = {Dou,  Wenjie and Subotnik,  Joseph E.},
  year = {2020},
  month = Jan,
  pages = {757–771}
}

@article{akimov2014nonadiabatic,
  title = {Nonadiabatic dynamics of charge transfer and singlet fission at the pentacene/C60 interface},
  volume = {136},
  ISSN = {1520-5126},
  url = {http://dx.doi.org/10.1021/ja411800n},
  DOI = {10.1021/ja411800n},
  number = {4},
  journal = {Journal of the American Chemical Society},
  publisher = {American Chemical Society (ACS)},
  author = {Akimov,  Alexey V. and Prezhdo,  Oleg V.},
  year = {2014},
  month = Jan,
  pages = {1599–1608}
}

@article{pastor2022electronic,
  title = {Electronic defects in metal oxide photocatalysts},
  volume = {7},
  ISSN = {2058-8437},
  url = {http://dx.doi.org/10.1038/s41578-022-00433-0},
  DOI = {10.1038/s41578-022-00433-0},
  number = {7},
  journal = {Nature Reviews Materials},
  publisher = {Springer Science and Business Media LLC},
  author = {Pastor,  Ernest and Sachs,  Michael and Selim,  Shababa and Durrant,  James R. and Bakulin,  Artem A. and Walsh,  Aron},
  year = {2022},
  month = May,
  pages = {503–521}
}

@article{meggiolaro2020polarons,
  title={Polarons in metal halide perovskites},
  author={Meggiolaro, Daniele and Ambrosio, Francesco and Mosconi, Edoardo and Mahata, Arup and De Angelis, Filippo},
  url = {http://dx.doi.org/10.1002/aenm.201902748},
  DOI = {10.1002/aenm.201902748},
  journal={Advanced Energy Materials},
  volume={10},
  number={13},
  pages={1902748},
  year={2020},
  publisher={Wiley Online Library}
}

@article{morita2023models,
  title = {Models of Polaron Transport in Inorganic and Hybrid Organic–Inorganic Titanium Oxides},
  volume = {35},
  ISSN = {1520-5002},
  url = {http://dx.doi.org/10.1021/acs.chemmater.3c00322},
  DOI = {10.1021/acs.chemmater.3c00322},
  number = {9},
  journal = {Chemistry of Materials},
  publisher = {American Chemical Society (ACS)},
  author = {Morita,  Kazuki and Golomb,  Matthias J. and Rivera,  Miguel and Walsh,  Aron},
  year = {2023},
  month = Apr,
  pages = {3652–3659}
}

@article{cheng2023time,
  title = {Time-Domain View of Polaron Dynamics in Metal Oxide Photocatalysts},
  volume = {14},
  ISSN = {1948-7185},
  url = {http://dx.doi.org/10.1021/acs.jpclett.3c02869},
  DOI = {10.1021/acs.jpclett.3c02869},
  number = {49},
  journal = {The Journal of Physical Chemistry Letters},
  publisher = {American Chemical Society (ACS)},
  author = {Cheng,  Cheng and Zhou,  Zhaohui and Long,  Run},
  year = {2023},
  month = Dec,
  pages = {10988–10998}
}

@article{brunin2019high,
  title = {High-performance transparent conducting oxides through small-polaron transport},
  volume = {3},
  ISSN = {2475-9953},
  url = {http://dx.doi.org/10.1103/PhysRevMaterials.3.064602},
  DOI = {10.1103/physrevmaterials.3.064602},
  number = {6},
  journal = {Physical Review Materials},
  publisher = {American Physical Society (APS)},
  author = {Brunin,  Guillaume and Rignanese,  Gian-Marco and Hautier,  Geoffroy},
  year = {2019},
  month = Jun,
  pages={064602}
}

@article{sombut2022role,
  title = {Role of Polarons in Single-Atom Catalysts: Case Study of Me1 [Au1,  Pt1,  and Rh1] on TiO2(110)},
  volume = {65},
  ISSN = {1572-9028},
  url = {http://dx.doi.org/10.1007/s11244-022-01651-0},
  DOI = {10.1007/s11244-022-01651-0},
  number = {17-18},
  journal = {Topics in Catalysis},
  publisher = {Springer Science and Business Media LLC},
  author = {Sombut,  Panukorn and Puntscher,  Lena and Atzmueller,  Marlene and Jakub,  Zdenek and Reticcioli,  Michele and Meier,  Matthias and Parkinson,  Gareth S. and Franchini,  Cesare},
  year = {2022},
  month = Jun,
  pages = {1620–1630}
}

@article{rousseau2020theoretical,
  title = {Theoretical insights into the surface physics and chemistry of redox-active oxides},
  volume = {5},
  ISSN = {2058-8437},
  url = {http://dx.doi.org/10.1038/s41578-020-0198-9},
  DOI = {10.1038/s41578-020-0198-9},
  number = {6},
  journal = {Nature Reviews Materials},
  publisher = {Springer Science and Business Media LLC},
  author = {Rousseau,  Roger and Glezakou,  Vassiliki-Alexandra and Selloni,  Annabella},
  year = {2020},
  month = May,
  pages = {460–475}
}

@article{chen2020small,
  title = {Small polarons and the Janus nature of TiO 2 (110)},
  volume = {101},
  ISSN = {2469-9969},
  url = {http://dx.doi.org/10.1103/PhysRevB.101.115402},
  DOI = {10.1103/physrevb.101.115402},
  number = {11},
  journal = {Physical Review B},
  publisher = {American Physical Society (APS)},
  author = {Chen,  Ji and Penschke,  Christopher and Alavi,  Ali and Michaelides,  Angelos},
  year = {2020},
  month = Mar,
  pages={115402}
}

@article{wu2026potential,
  title = {Potential-dependent polaron formation activates TiO2 for the hydrogen evolution reaction},
  volume = {17},
  ISSN = {2041-1723},
  url = {http://dx.doi.org/10.1038/s41467-026-68892-5},
  DOI = {10.1038/s41467-026-68892-5},
  number = {1},
  journal = {Nature Communications},
  publisher = {Springer Science and Business Media LLC},
  author = {Wu,  Tongwei and Guo,  Xiaoxi and Zhang,  Guangjie and Zhang,  Yanning and Song,  Li and Liu,  Zheng and Zhang,  Hui and Shi,  Shucheng and Liu,  Limin and Melander,  Marko M. and Honkala,  Karoliina},
  year = {2026},
  month = Jan 
}

@article{c1998mixed,
  title = {Mixed quantum–classical dynamics},
  volume = {110},
  ISSN = {1364-5498},
  url = {http://dx.doi.org/10.1039/a801824c},
  DOI = {10.1039/a801824c},
  journal = {Faraday Discussions},
  publisher = {Royal Society of Chemistry (RSC)},
  author = {Tully,  John C.},
  year = {1998},
  pages = {407–419}
}

@article{xu2023polaron,
  title = {Polaron-assisted nonadiabatic dynamics in protonated TiO2 with surface water molecule},
  volume = {2},
  ISSN = {2772-5715},
  url = {http://dx.doi.org/10.1016/j.chphma.2023.02.005},
  DOI = {10.1016/j.chphma.2023.02.005},
  number = {4},
  journal = {ChemPhysMater},
  publisher = {Elsevier BV},
  author = {Xu,  Zhongfei and Tong,  Chuanjia and Si,  Rutong and Teobaldi,  Gilberto and Liu,  Limin},
  year = {2023},
  month = Oct,
  pages = {331–336}
}

@article{marcus1956theory,
  title = {On the Theory of Oxidation-Reduction Reactions Involving Electron Transfer. I},
  volume = {24},
  ISSN = {1089-7690},
  url = {http://dx.doi.org/10.1063/1.1742723},
  DOI = {10.1063/1.1742723},
  number = {5},
  journal = {The Journal of Chemical Physics},
  publisher = {AIP Publishing},
  author = {Marcus,  R. A.},
  year = {1956},
  month = May,
  pages = {966–978}
}

@article{marcus2020electron,
  title = {Electron transfer reactions in chemistry. Theory and experiment},
  volume = {65},
  ISSN = {1539-0756},
  url = {http://dx.doi.org/10.1103/RevModPhys.65.599},
  DOI = {10.1103/revmodphys.65.599},
  number = {3},
  journal = {Reviews of Modern Physics},
  publisher = {American Physical Society (APS)},
  author = {Marcus,  Rudolph A.},
  year = {1993},
  month = Jul,
  pages = {599–610}
}

@article{nabi2025impact,
  title = {Impact of hole polaron formation on excitonic transitions in MgO from first principles},
  volume = {9},
  ISSN = {2475-9953},
  url = {http://dx.doi.org/10.1103/mpb2-3wrb},
  DOI = {10.1103/mpb2-3wrb},
  number = {12},
  journal = {Physical Review Materials},
  publisher = {American Physical Society (APS)},
  author = {Nabi,  A. Sameer and Sharifzadeh,  Sahar},
  year = {2025},
  month = Dec,
  pages={124605}
}

@article{smith2017intrinsic,
  title = {Intrinsic Conductivity in Magnesium–Oxygen Battery Discharge Products: MgO and MgO2},
  volume = {29},
  ISSN = {1520-5002},
  url = {http://dx.doi.org/10.1021/acs.chemmater.7b00217},
  DOI = {10.1021/acs.chemmater.7b00217},
  number = {7},
  journal = {Chemistry of Materials},
  publisher = {American Chemical Society (ACS)},
  author = {Smith,  Jeffrey G. and Naruse,  Junichi and Hiramatsu,  Hidehiko and Siegel,  Donald J.},
  year = {2017},
  month = Mar,
  pages = {3152–3163}
}

@article{Friederich2021,
  title = {Machine-learned potentials for next-generation matter simulations},
  volume = {20},
  ISSN = {1476-4660},
  url = {http://dx.doi.org/10.1038/s41563-020-0777-6},
  DOI = {10.1038/s41563-020-0777-6},
  number = {6},
  journal = {Nature Materials},
  publisher = {Springer Science and Business Media LLC},
  author = {Friederich,  Pascal and H\"{a}se,  Florian and Proppe,  Jonny and Aspuru-Guzik,  Alán},
  year = {2021},
  month = May,
  pages = {750–761}
}

@article{Fan2026,
  title = {qNEP: A Highly Efficient Neuroevolution Potential with Dynamic Charges for Large-Scale Atomistic Simulations},
  volume = {22},
  ISSN = {1549-9626},
  url = {http://dx.doi.org/10.1021/acs.jctc.6c00146},
  DOI = {10.1021/acs.jctc.6c00146},
  number = {9},
  journal = {Journal of Chemical Theory and Computation},
  publisher = {American Chemical Society (ACS)},
  author = {Fan,  Zheyong and Tang,  Benrui and Berger,  Esmée and Berger,  Ethan and Fransson,  Erik and Xu,  Ke and Yan,  Zihan and Liu,  Zhoulin and Song,  Zichen and Dong,  Haikuan and Chen,  Shunda and Li,  Lei and Wang,  Ziliang and Zhu,  Yizhou and Wiktor,  Julia and Erhart,  Paul},
  year = {2026},
  month = Apr,
  pages = {4787–4801}
}

@article{Ko2021,
  title = {A fourth-generation high-dimensional neural network potential with accurate electrostatics including non-local charge transfer},
  volume = {12},
  ISSN = {2041-1723},
  url = {http://dx.doi.org/10.1038/s41467-020-20427-2},
  DOI = {10.1038/s41467-020-20427-2},
  number = {1},
  journal = {Nature Communications},
  publisher = {Springer Science and Business Media LLC},
  author = {Ko,  Tsz Wai and Finkler,  Jonas A. and Goedecker,  Stefan and Behler,  J\"{o}rg},
  year = {2021},
  month = Jan,
  pages={398}
}

@article{Leoni2026,
  title = {Machine-Learned Dynamics of Surface Polarons at Reduced Oxide Surfaces},
  ISSN = {1948-7185},
  url = {http://dx.doi.org/10.1021/acs.jpclett.6c02034},
  DOI = {10.1021/acs.jpclett.6c02034},
  journal = {The Journal of Physical Chemistry Letters},
  publisher = {American Chemical Society (ACS)},
  author = {Leoni,  Luca and Franchini,  Cesare},
  year = {2026},
  month = 7,
  pages = {9237–9245}
}

@article{zhang2025advancing,
  title = {Advancing nonadiabatic molecular dynamics simulations in solids with E(3) equivariant deep neural hamiltonians},
  volume = {16},
  ISSN = {2041-1723},
  url = {http://dx.doi.org/10.1038/s41467-025-57328-1},
  DOI = {10.1038/s41467-025-57328-1},
  number = {1},
  journal = {Nature Communications},
  publisher = {Springer Science and Business Media LLC},
  author = {Zhang,  Changwei and Zhong,  Yang and Tao,  Zhi-Guo and Qin,  Xinming and Shang,  Honghui and Lan,  Zhenggang and Prezhdo,  Oleg V. and Gong,  Xin-Gao and Chu,  Weibin and Xiang,  Hongjun},
  year = {2025},
  month = Feb,
  pages={2033}
}

@article{wu2026efficient,
    author = {Wu, Yifan and Wang, Bipeng and Chaudhary, Mohit and Casanova, David and Prezhdo, Oleg V.},
    title = {Efficient Nonadiabatic
Molecular Dynamics with Machine
Learning Hamiltonian Interpolation},
    journal = {The Journal of Physical Chemistry Letters},
    volume = {17},
    number = {26},
    pages = {7328-7335},
    year = {2026},
    month = {06},
    issn = {1948-7185},
    doi = {10.1021/acs.jpclett.6c01666},
    url = {https://doi.org/10.1021/acs.jpclett.6c01666}
}

@article{Tao2023,
  title = {Electron and Hole Polaron Formation and Transport in Monoclinic WO3 Studied by Hybrid Functional Approach},
  volume = {127},
  ISSN = {1932-7455},
  url = {http://dx.doi.org/10.1021/acs.jpcc.3c03991},
  DOI = {10.1021/acs.jpcc.3c03991},
  number = {32},
  journal = {The Journal of Physical Chemistry C},
  publisher = {American Chemical Society (ACS)},
  author = {Tao,  Junyan and Liu,  Taifeng},
  year = {2023},
  month = Aug,
  pages = {16204–16210}
}

@article{Jiang2025,
  title = {Understanding polaron dynamics in CeO2 for advanced catalytic material design},
  volume = {27},
  ISSN = {1463-9084},
  url = {http://dx.doi.org/10.1039/d5cp02631h},
  DOI = {10.1039/d5cp02631h},
  number = {40},
  journal = {Physical Chemistry Chemical Physics},
  publisher = {Royal Society of Chemistry (RSC)},
  author = {Jiang,  Xiao and Cheng,  Xu and Liu,  Zhanqi and Ding,  Liyun and Han,  Weihua},
  year = {2025},
  pages = {21673–21682}
}

@article{Redondo2024,
  title = {Real-space investigation of polarons in hematite Fe2O3},
  volume = {10},
  ISSN = {2375-2548},
  url = {http://dx.doi.org/10.1126/sciadv.adp7833},
  DOI = {10.1126/sciadv.adp7833},
  number = {44},
  journal = {Science Advances},
  publisher = {American Association for the Advancement of Science (AAAS)},
  author = {Redondo,  Jesus and Reticcioli,  Michele and Gabriel,  Vit and Wrana,  Dominik and Ellinger,  Florian and Riva,  Michele and Franceschi,  Giada and Rheinfrank,  Erik and Sokolović,  Igor and Jakub,  Zdenek and Kraushofer,  Florian and Alexander,  Aji and Belas,  Eduard and Patera,  Laerte L. and Repp,  Jascha and Schmid,  Michael and Diebold,  Ulrike and Parkinson,  Gareth S. and Franchini,  Cesare and Kocan,  Pavel and Setvin,  Martin},
  year = {2024},
  month = Nov,
  pages = {eadp7833}
}

@article{Mahiba2026,
  title = {A review on multifunctional applications of MgO nanostructures: from material science to environmental and agricultural innovations},
  volume = {16},
  ISSN = {2046-2069},
  url = {http://dx.doi.org/10.1039/d5ra07016c},
  DOI = {10.1039/d5ra07016c},
  number = {15},
  journal = {RSC Advances},
  publisher = {Royal Society of Chemistry (RSC)},
  author = {Mahiba,  G. Georgelin Jeba and Prabakaran,  A. and Balraj,  Babu},
  year = {2026},
  pages = {13040–13061}
}

@article{Mastronardo2017,
  title = {Strategies for the enhancement of heat storage materials performances for MgO/H 2 O/Mg(OH) 2 thermochemical storage system},
  volume = {120},
  ISSN = {1359-4311},
  url = {http://dx.doi.org/10.1016/j.applthermaleng.2017.04.004},
  DOI = {10.1016/j.applthermaleng.2017.04.004},
  journal = {Applied Thermal Engineering},
  publisher = {Elsevier BV},
  author = {Mastronardo,  Emanuela and Bonaccorsi,  Lucio and Kato,  Yukitaka and Piperopoulos,  Elpida and Lanza,  Maurizio and Milone,  Candida},
  year = {2017},
  month = Jun,
  pages = {626–634}
}

@article{Goyal2025,
  title = {Enhanced electrochemical activity of MgO nanoparticles for high-performance supercapacitors},
  volume = {15},
  ISSN = {2046-2069},
  url = {http://dx.doi.org/10.1039/d5ra04329h},
  DOI = {10.1039/d5ra04329h},
  number = {31},
  journal = {RSC Advances},
  publisher = {Royal Society of Chemistry (RSC)},
  author = {Goyal,  Megha and Pandey,  Sarvesh Kumar and Bhatnagar,  Nitu},
  year = {2025},
  pages = {25209–25220}
}

@article{Oi2016,
  title = {Recent advances of titanium dioxide (TiO2
                    ) for green organic synthesis},
  volume = {6},
  ISSN = {2046-2069},
  url = {http://dx.doi.org/10.1039/c6ra22894a},
  DOI = {10.1039/c6ra22894a},
  number = {110},
  journal = {RSC Advances},
  publisher = {Royal Society of Chemistry (RSC)},
  author = {Oi,  Lee Eng and Choo,  Min-Yee and Lee,  Hwei Voon and Ong,  Hwai Chyuan and Hamid,  Sharifah Bee Abd and Juan,  Joon Ching},
  year = {2016},
  pages = {108741–108754}
}

@article{Jiang2017,
  title = {Photoelectrochemical devices for solar water splitting – materials and challenges},
  volume = {46},
  ISSN = {1460-4744},
  url = {http://dx.doi.org/10.1039/c6cs00306k},
  DOI = {10.1039/c6cs00306k},
  number = {15},
  journal = {Chemical Society Reviews},
  publisher = {Royal Society of Chemistry (RSC)},
  author = {Jiang,  Chaoran and Moniz,  Savio J. A. and Wang,  Aiqin and Zhang,  Tao and Tang,  Junwang},
  year = {2017},
  pages = {4645–4660}
}

@article{marcelis2024titanium,
  title={Titanium dioxide: a versatile earth-abundant optical material for photovoltaics},
  author={Marcelis, Eloy J and ten Elshof, Johan E and Morales-Masis, Monica},
  DOI={10.1002/adom.202401423},
  journal={Advanced Optical Materials},
  volume={12},
  number={36},
  pages={2401423},
  year={2024},
  publisher={Wiley Online Library}
}

@article{Glasbrenner2023,
  title = {The Landau–Zener formula made simple},
  volume = {56},
  ISSN = {1361-6455},
  url = {http://dx.doi.org/10.1088/1361-6455/acc774},
  DOI = {10.1088/1361-6455/acc774},
  number = {10},
  journal = {Journal of Physics B: Atomic,  Molecular and Optical Physics},
  publisher = {IOP Publishing},
  author = {Glasbrenner,  Eric P and Schleich,  Wolfgang P},
  year = {2023},
  month = Apr,
  pages = {104001}
}

@article{Landau:1932vnv,
    author = "Landau, Lev Davidovich",
    editor = "ter Haar, D.",
    title = "{A theory of energy transfer. 2.}",
    doi = "10.1016/B978-0-08-010586-4.50014-6",
    journal = "Phys. Z. Sowjetunion",
    volume = "2",
    year = "1932"
}

@article{Zener1932,
  title = {Non-adiabatic crossing of energy levels},
  volume = {137},
  ISSN = {2053-9150},
  url = {http://dx.doi.org/10.1098/rspa.1932.0165},
  DOI = {10.1098/rspa.1932.0165},
  number = {833},
  journal = {Proceedings of the Royal Society of London. Series A,  Containing Papers of a Mathematical and Physical Character},
  publisher = {The Royal Society},
  author = {Zener,  Clarence},
  year = {1932},
  month = Sep,
  pages = {696–702}
}

@article{Jain2022,
  title = {Pedagogical Overview of the Fewest Switches Surface Hopping Method},
  volume = {7},
  ISSN = {2470-1343},
  url = {http://dx.doi.org/10.1021/acsomega.2c04843},
  DOI = {10.1021/acsomega.2c04843},
  number = {50},
  journal = {ACS Omega},
  publisher = {American Chemical Society (ACS)},
  author = {Jain,  Amber and Sindhu,  Aarti},
  year = {2022},
  month = Dec,
  pages = {45810–45824}
}

\end{document}

%\title{Supplemental Material for:\\ [Titolo del tuo Paper Principale]}

\title{Supplemental Material for:\\ Nonadiabatic polaron dynamics from machine-learned hopping trajectories}% Force line breaks with \\

\author{Daniela Mangano}
\affiliation{University of Vienna, Faculty of Physics and Center for Computational Materials Science, Vienna, Austria}
\affiliation{Department of Physics and Astronomy “Augusto Righi”, Alma Mater Studiorum - Università di Bologna, Bologna, 40127 Italy}
\author{Luca Leoni}
\affiliation{Department of Physics and Astronomy “Augusto Righi”, Alma Mater Studiorum - Università di Bologna, Bologna, 40127 Italy}

\author{Zhongfei Xu}
\affiliation{College of Environmental Science and
Engineering, North China Electric Power University, Beijing
102206, China}

\author{Li-Min Liu}
\affiliation{School of Physics, Beihang University,
Beijing 100191, China}

\author{Cesare Franchini}
\affiliation{University of Vienna, Faculty of Physics and Center for Computational Materials Science, Vienna, Austria}
\affiliation{Department of Physics and Astronomy “Augusto Righi”, Alma Mater Studiorum - Università di Bologna, Bologna, 40127 Italy}

\maketitle

\section{Analysis of the eigenvalues and the nonadiabatic couplings}
In the main text are reported, as an example, data for $\text{TiO}_{2}$ bulk at 500\,K, MgO bulk at 500\,K and $\text{TiO}_{2}$ surface at 700\,K; we show here the results for the systems at the other temperatures that we investigated: rutile bulk and MgO at 400\,K and 600\,K,  rutile surface at 600\,K.

\begin{figure*}[htbp]
    \centering
     \begin{subfigure}[t]{.9\textwidth}
         \centering       
         \includegraphics[width=\linewidth]{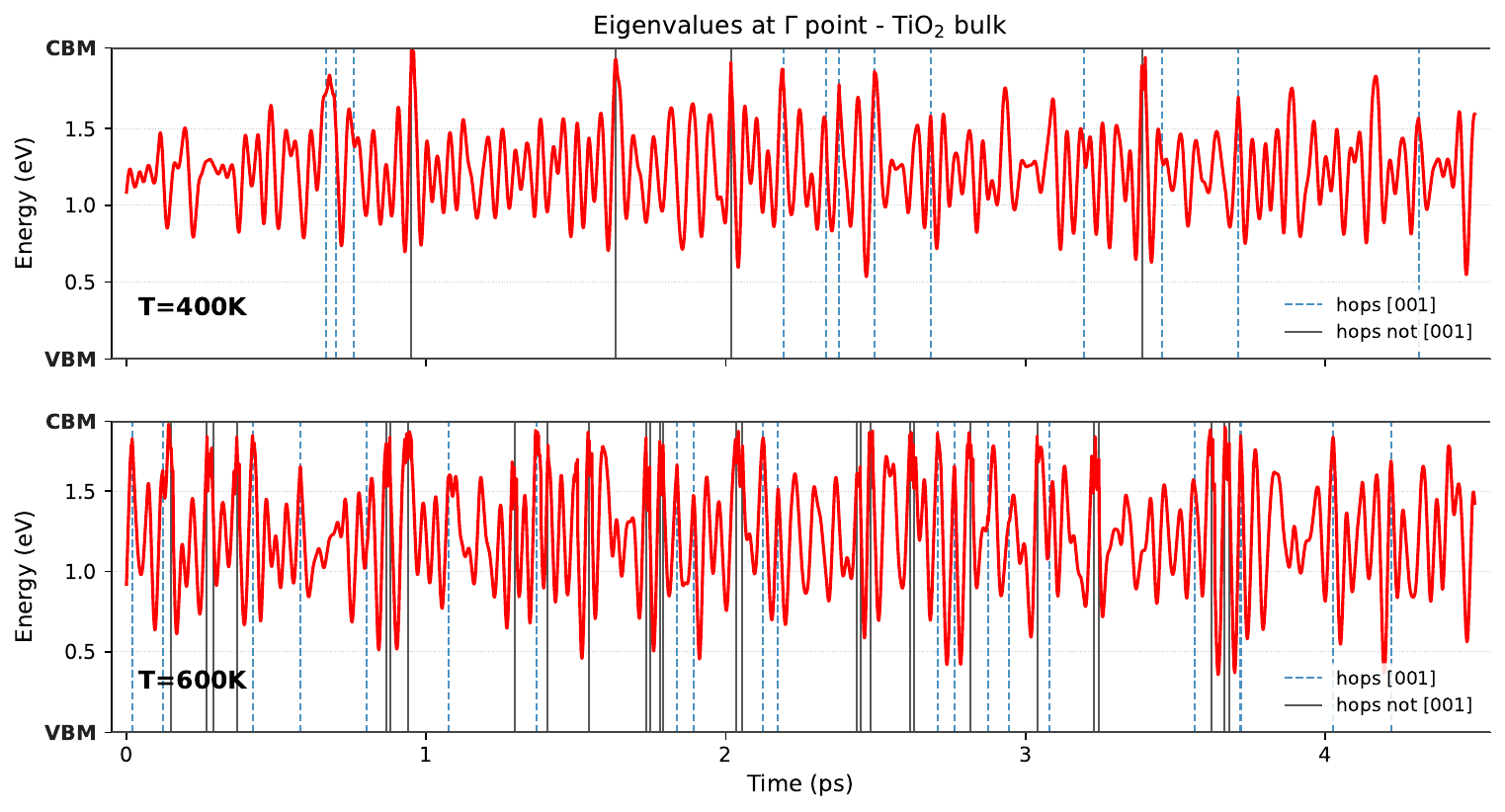}
            \put(-400,240){\large (a)}
       % \end{overpic}
         \phantomsubcaption
         \label{fig:FigS1a_Eigenvals_TiO2}
    \end{subfigure}
     %\vspace{-10mm}
    \begin{subfigure}[t]{.9\textwidth}
         \centering
       
         \includegraphics[width=\linewidth]{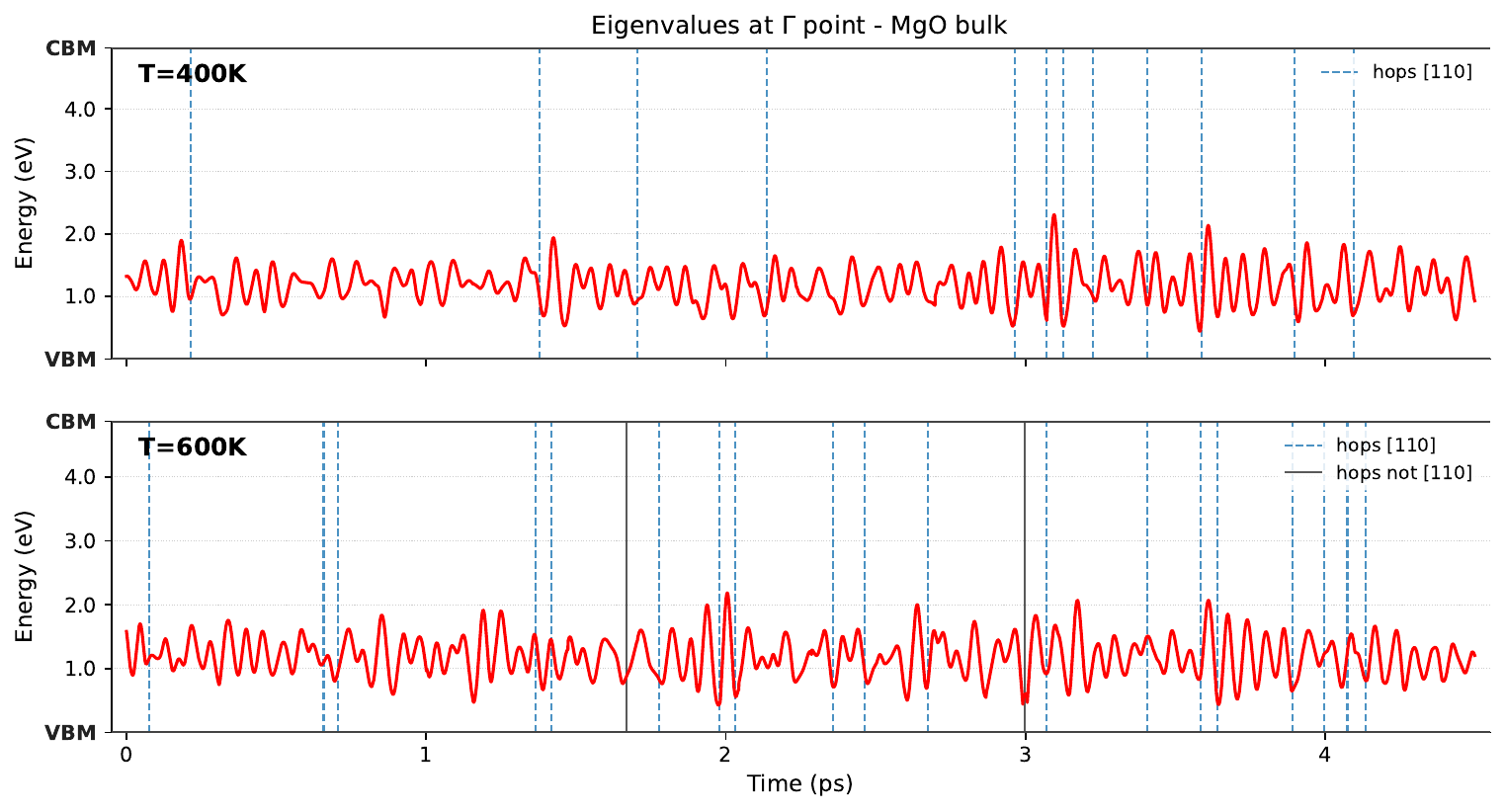}
            \put(-400,240){\large (b)}
        %\end{overpic}
         \phantomsubcaption
         \label{fig:FigS1b_Eigenvals_MgO}
    \end{subfigure}
    %\vspace{-4mm}
    \begin{subfigure}{.9\textwidth}
         \centering
     \includegraphics[width=\linewidth]{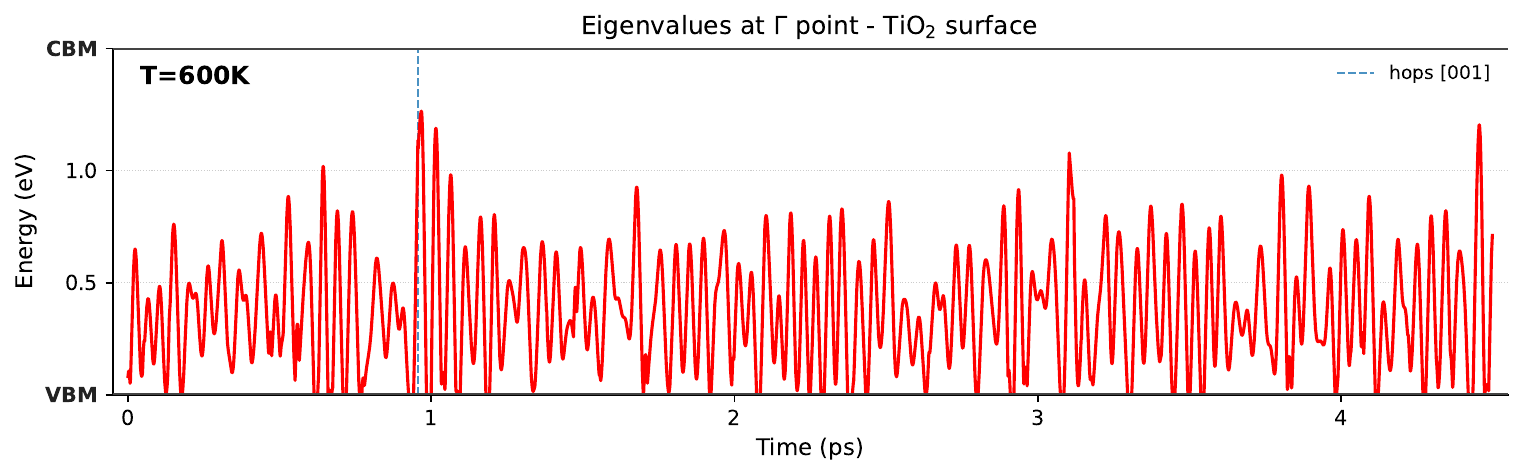}
            \put(-400,135){\large (c)}
        %\end{overpic}
         \phantomsubcaption
         \label{fig:FigS1c_Eigenvals_TiO2-surf}
    \end{subfigure}
        \caption{\justifying Band energies of the CBM, polaron level and VBM for the considered systems. In rutile polaron hoppings occur when the defect level and the CBM approach; similarly, the hole polaron hoppings in MgO are generally accompanied by the approaching of the defect level and VBM. }
        \label{fig:SM_Eigenvals}
\end{figure*}

\begin{figure*}[htbp]
    \centering
     \begin{subfigure}[t]{.9\textwidth}
         \centering
         
         \includegraphics[width=\linewidth]{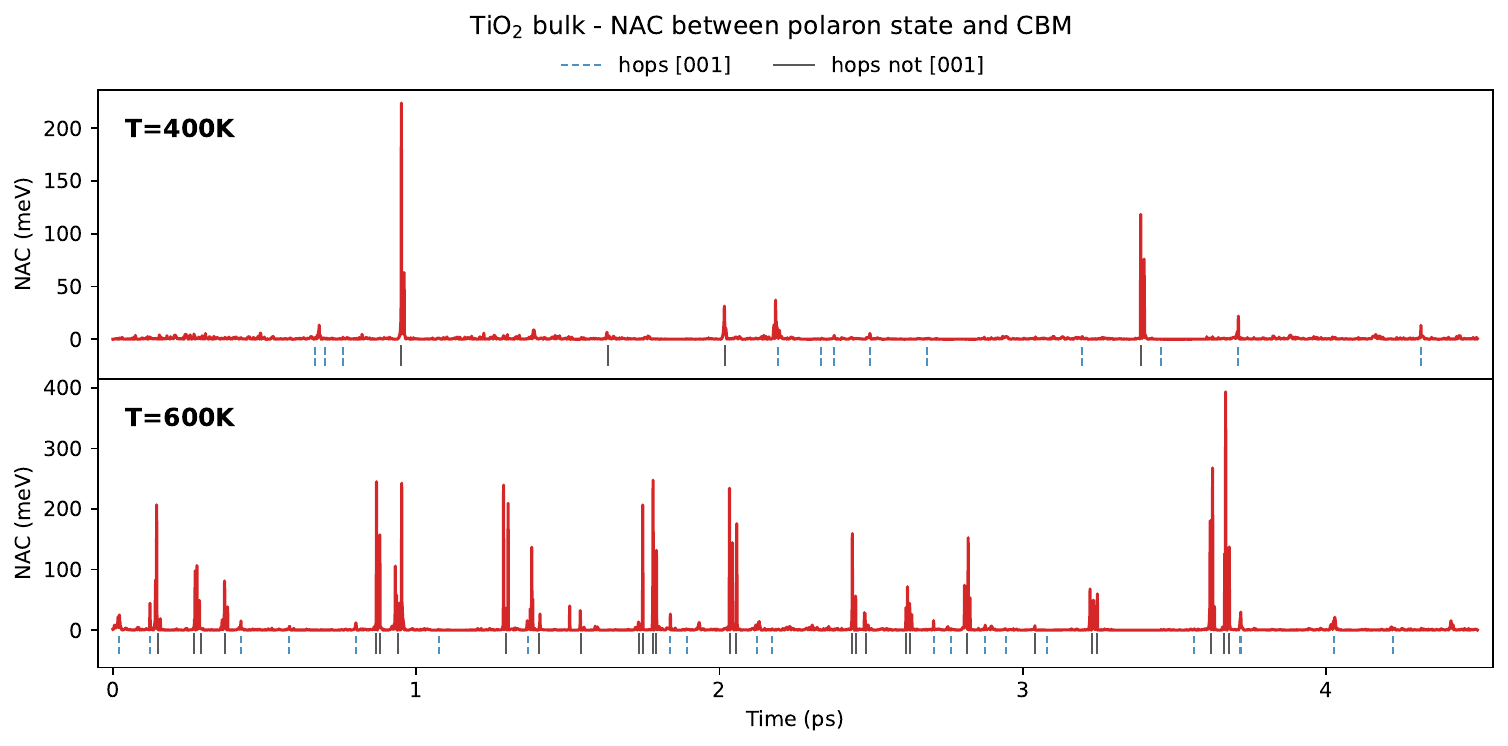}
            \put(-400,215){\large (a)}
       % \end{overpic}
         \phantomsubcaption
         \label{fig:FigS2a_NAC_TiO2}
    \end{subfigure}
     %\vspace{-10mm}
    \begin{subfigure}[t]{.9\textwidth}
         \centering
     \includegraphics[width=\linewidth]{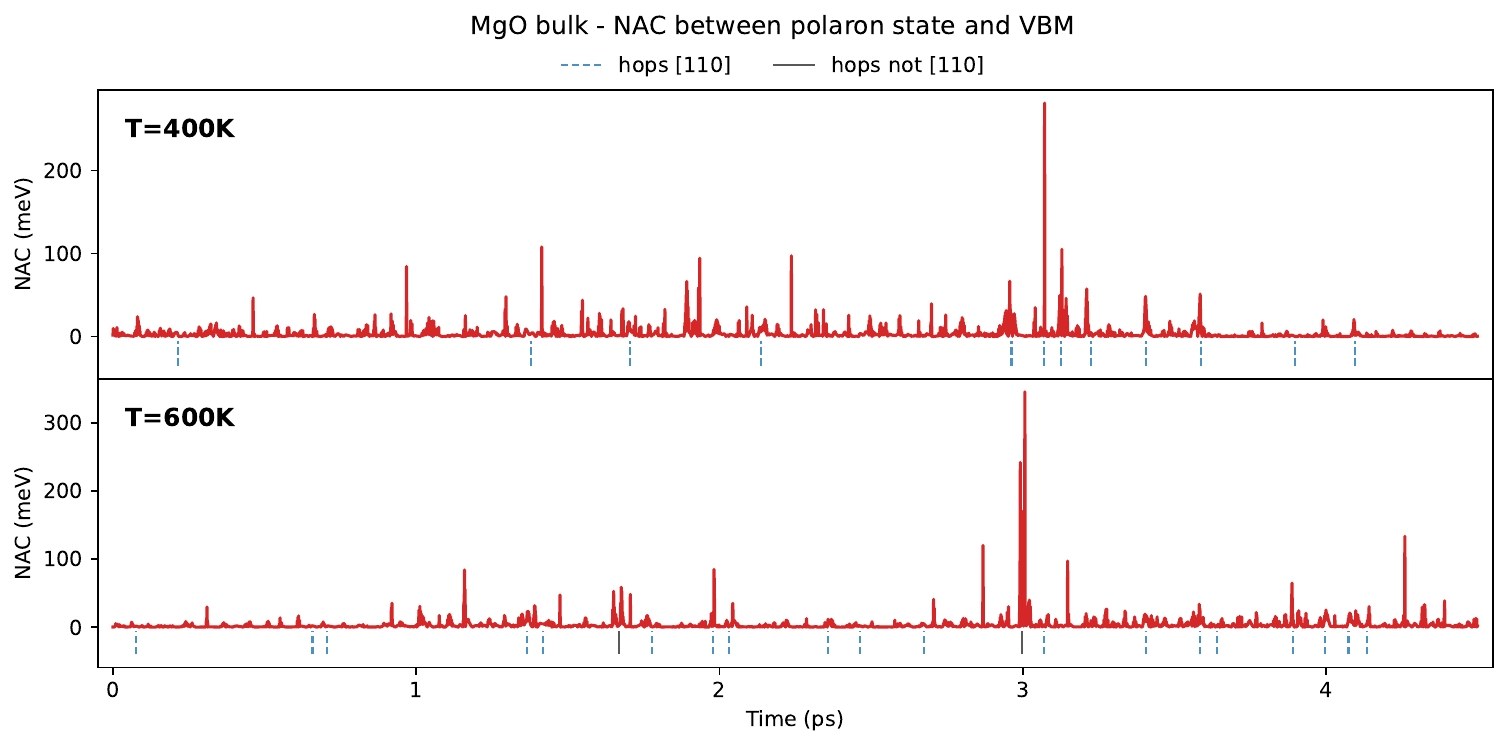}
            \put(-400,215){\large (b)}
        %\end{overpic}
         \phantomsubcaption
         \label{fig:FigS2b_NAC_MgO.pdf}
    \end{subfigure}
    %\vspace{-4mm}
    \begin{subfigure}{.9\textwidth}
         \centering
   \includegraphics[width=\linewidth]{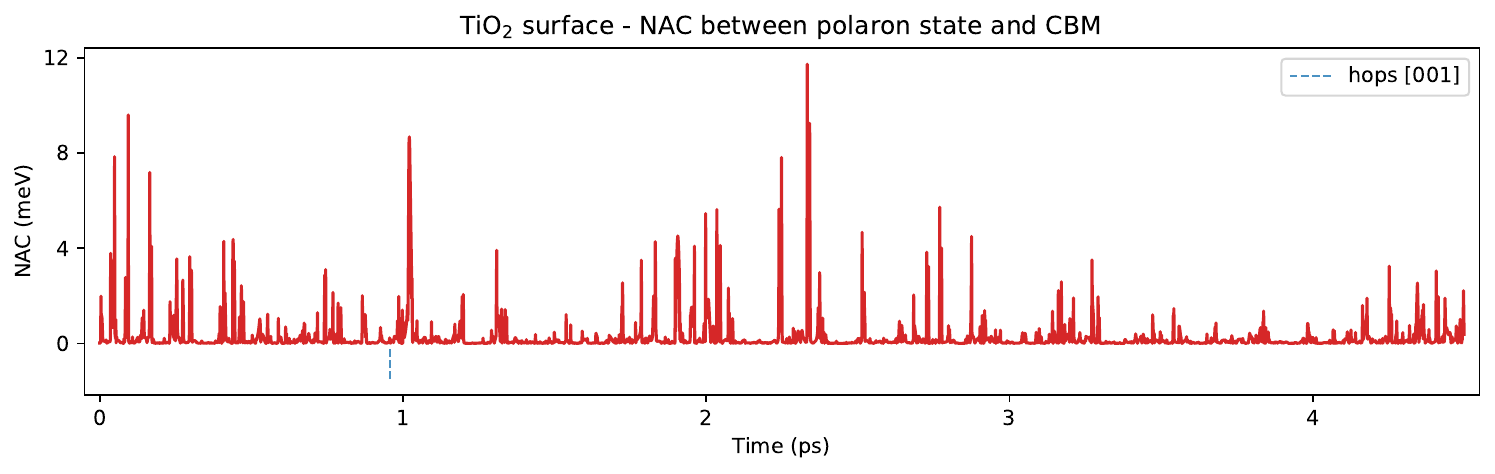}
            \put(-400,137){\large (c)}
        %\end{overpic}
         \phantomsubcaption
         \label{fig:FigS2c_NAC_TiO2-surf}
    \end{subfigure}
        \caption{\justifying Absolute value of NAC for the considered systems. (a) In bulk rutile evident peaks can be associated with polaron hoppings, especially for not [001]. Regarding (b) MgO and (c) rutile surface the couplings show less strong fluctuations.}
        \label{fig:SM_NAC}
\end{figure*}

In Fig.~\ref{fig:SM_Eigenvals} we study the evolution of the energies for the VBM, polaron level and CBM as a function of time for the different systems. Fig.~\ref{fig:SM_NAC} shows, in absolute value, the NAC between the polaron state and the CBM for the rutile material, the NAC between the polaron state and the VBM for MgO. 

\begin{figure*}[htbp]
    \centering
     \begin{subfigure}{.45\textwidth}
         \centering

         \begin{overpic}[width=\linewidth]{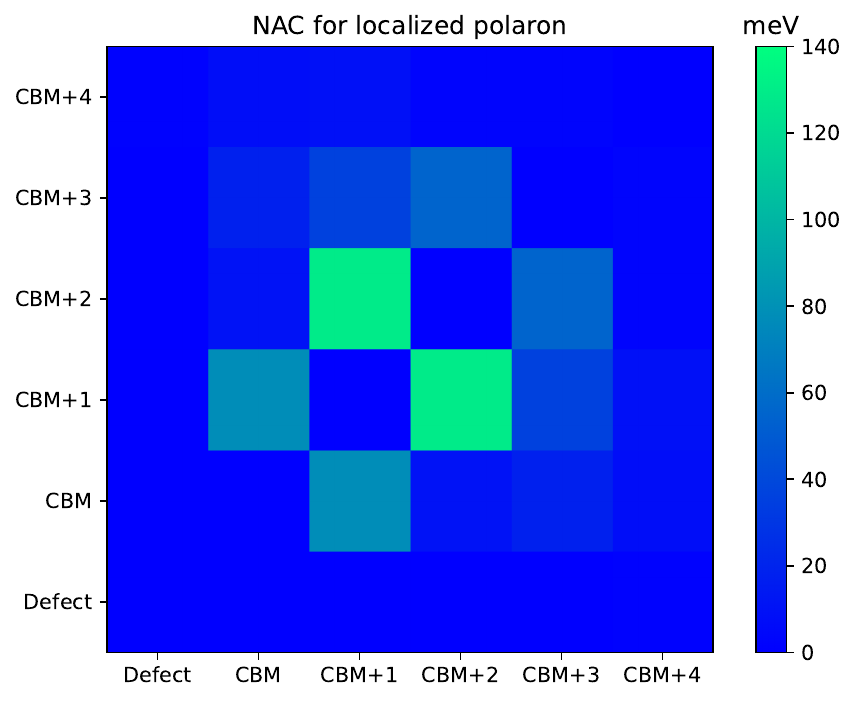}
            \put(0,80){\large (a)}
        \end{overpic}
         \phantomsubcaption
\label{fig:FigS3_Colormap_localized_TiO2}
    \end{subfigure}
     %\hspace{0.cm}
\begin{subfigure}{.45\textwidth}
         \centering
         \begin{overpic}[width=\linewidth]{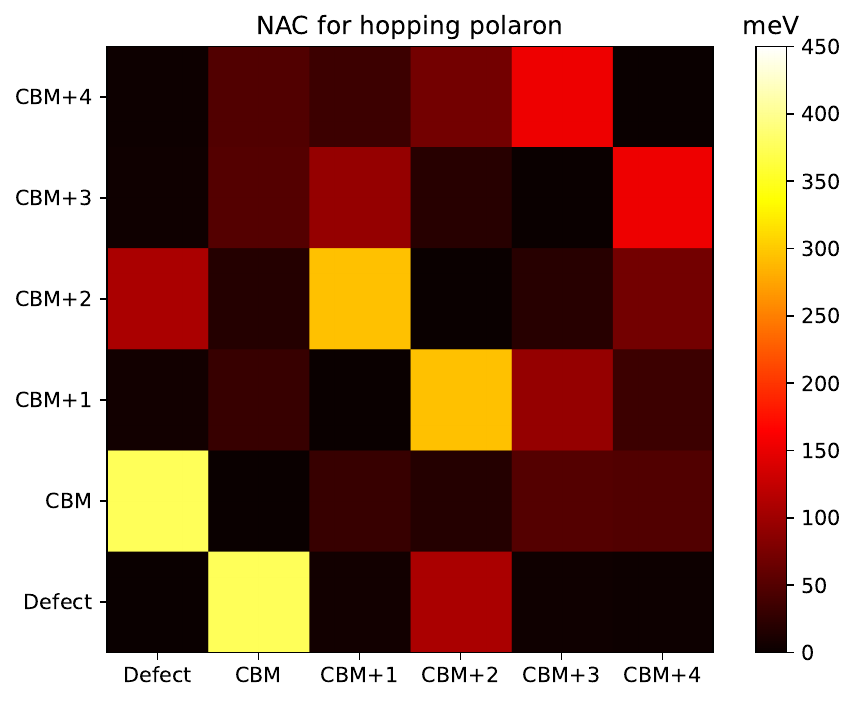}
            \put(0,80){\large (b)}
        \end{overpic}
        \phantomsubcaption
    \label{fig:FigS3_Colormap_hopping_TiO2}
    \end{subfigure}
%\vspace{0.5cm}
        \caption{\justifying NAC colormap of TiO$_2$ bulk at 500~K when the polaron is (a) localized at a site or (b) undergoing a nonadiabatic hopping event. The color scale represents the absolute value of the NAC strength. Different color scales are used for the localized and hopping configurations, to strike the difference in the NAC strengths, especially the value of the coupling between the polaron in-gap state and the CBM.}
        \label{fig:NAC_colormap_TiO2_500}
\end{figure*}

Fig.~\ref{fig:NAC_colormap_TiO2_500} shows the NAC matrix for bulk TiO$_2$ system, in a configuration in which the polaron is localized at a site and when the polaron is undergoing a nonadiabatic hopping.

%------------------------------------------------------

%---------------------------------------------------

\section{Excited states dynamics and decoherence}
\begin{figure*}[htbp]
    \centering
     \begin{subfigure}[t]{.99\textwidth}
         \centering
         \begin{overpic}[width=\linewidth]{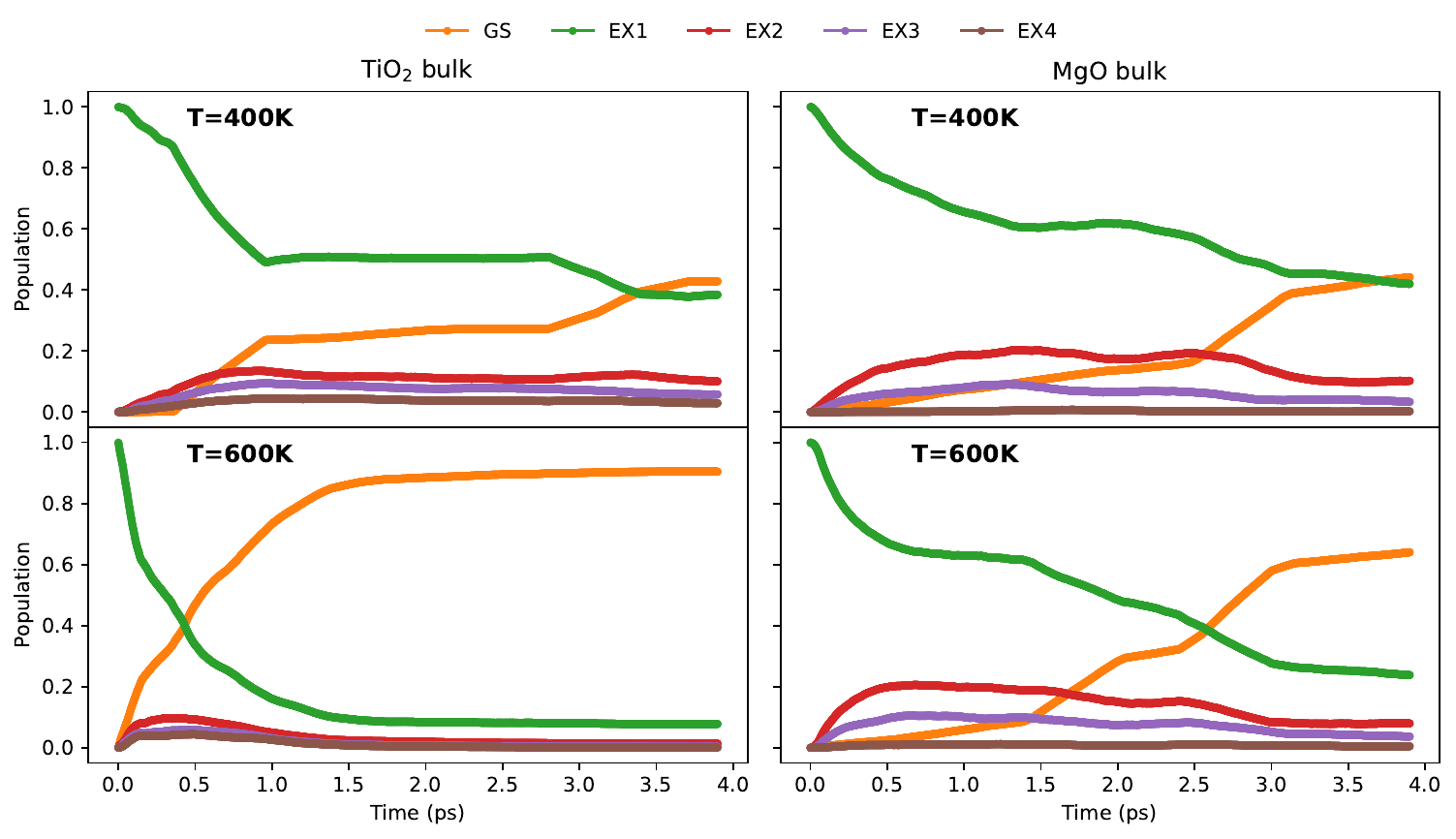}
            %\put(6,55){\large (a)}
        \end{overpic}
         \phantomsubcaption
         \label{fig:FigS4_NAMD_TiO2_MgO}
    \end{subfigure}
     %\hspace{0.5cm}
    \begin{subfigure}[t]{.6\textwidth}
         \centering
         
         \begin{overpic}[width=\linewidth]{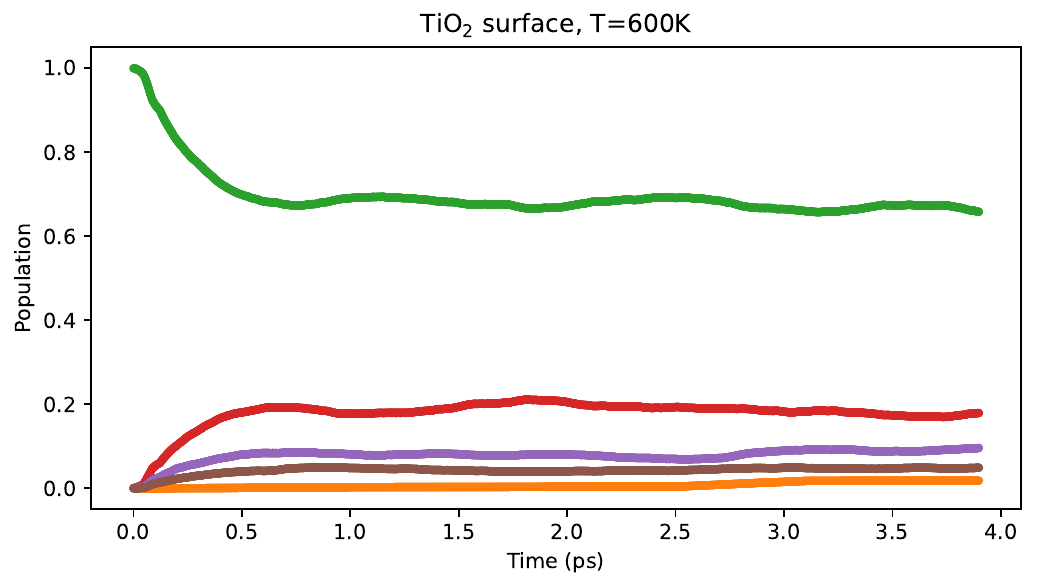} 
           % \put(-6,50){\large (b)}
        \end{overpic}
         \phantomsubcaption
         \label{fig:FigS4_NAMD_TiO2-surf}
    \end{subfigure}

        \caption{\justifying Evolution of the populations for the considered systems initially in the first excited state. The dominant transition is towards
the ground state.   }
        \label{fig:SM_Excited_states}
\end{figure*}

\begin{figure*}[htbp]
    \centering
     \begin{subfigure}[t]{.49\textwidth}
         \centering
     \includegraphics[width=\linewidth]{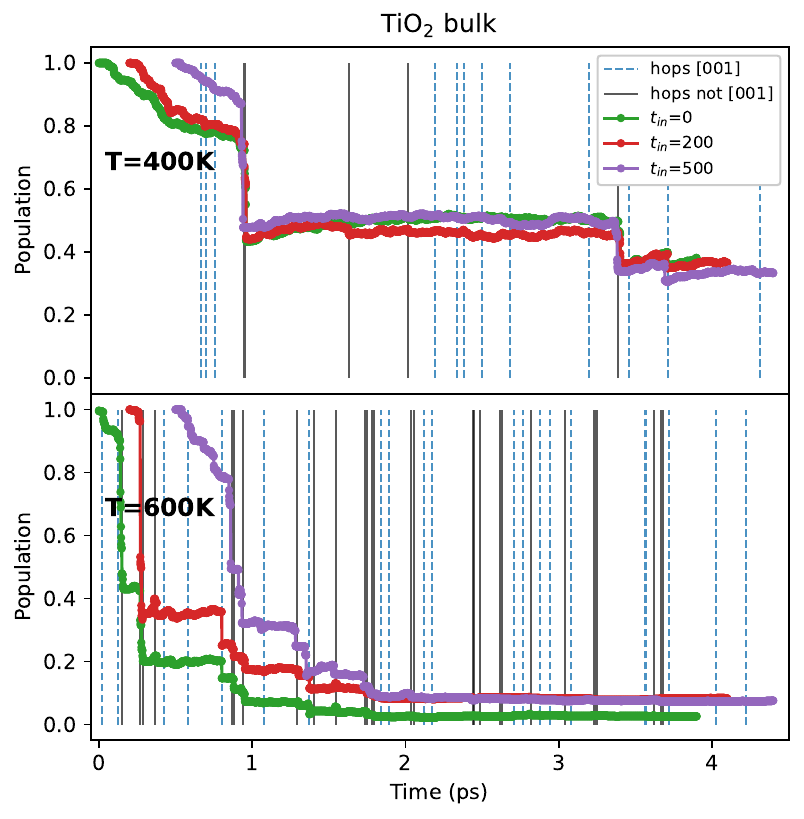}
           % \put(-200,230){\large (a)}
       % \end{overpic}
         \phantomsubcaption
         \label{fig:FigS5_3iconds_TiO2_EX1}
    \end{subfigure}
     %\vspace{-10mm}
    \begin{subfigure}[t]{.49\textwidth}
         \centering
    
         \includegraphics[width=\linewidth]{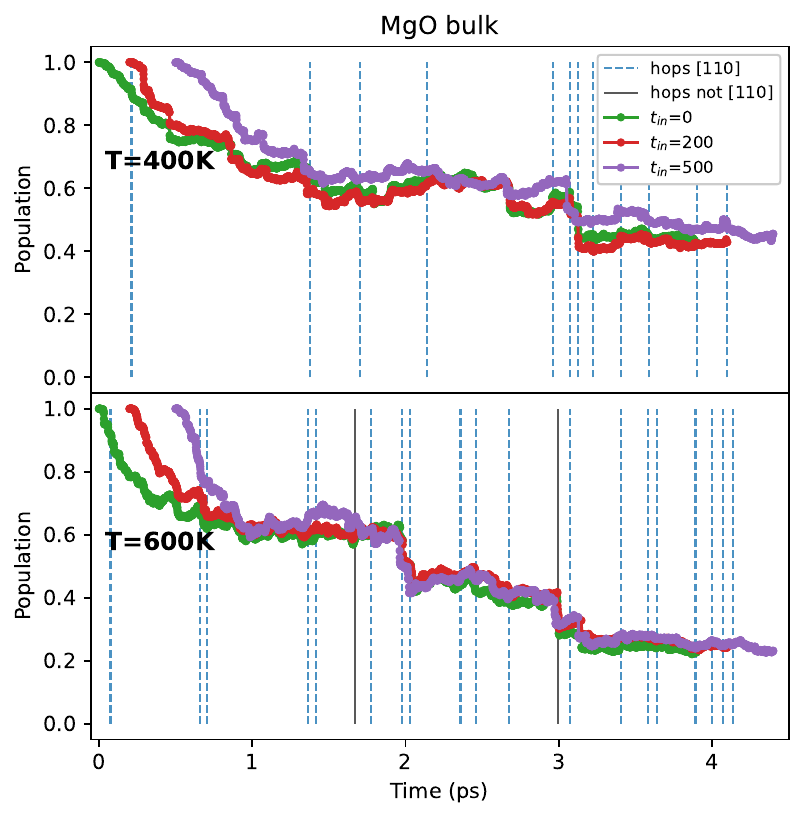}
       
         \phantomsubcaption
         \label{fig:FigS5_3iconds_MgO_EX1}
    \end{subfigure}
    %\vspace{-4mm}
    \begin{subfigure}[t]{.9\textwidth}
         \centering
         
         \begin{overpic}[width=\linewidth]{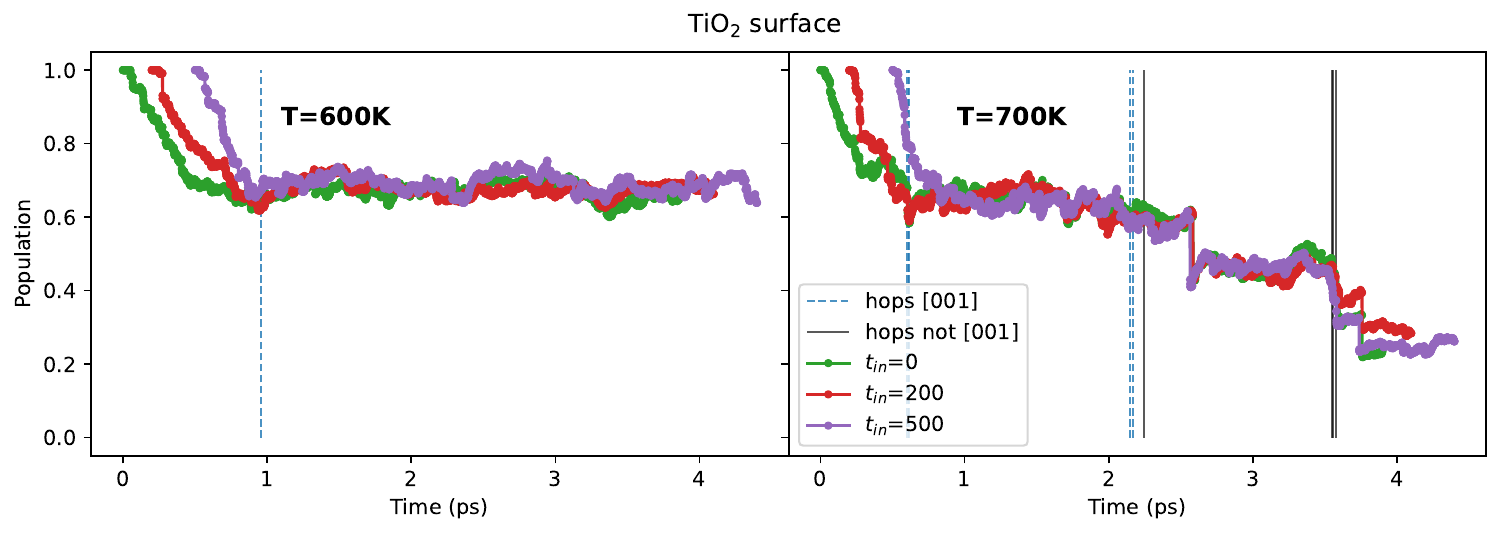}%
        \end{overpic}
         \phantomsubcaption
         
         \label{fig:FigS5_3iconds_TiO2-surf_EX1}
    \end{subfigure}
    
        \caption{\justifying For every system are plotted three representative trajectories (namely starting from initial time t$_{\text{in}}$=0, 200, and 500\,fs). In bulk rutile the visible sudden drops in the population can be associated with the not [001] polaron hoppings. Regarding MgO, some population
drops can be noticed, but the decay is in general smoother than in rutile. }
        \label{fig:SM_Single_iconds_EX1}
\end{figure*}

\begin{figure*}[htbp]
    \centering
     \begin{subfigure}{.49\textwidth}
         \centering

         \begin{overpic}[width=\linewidth]{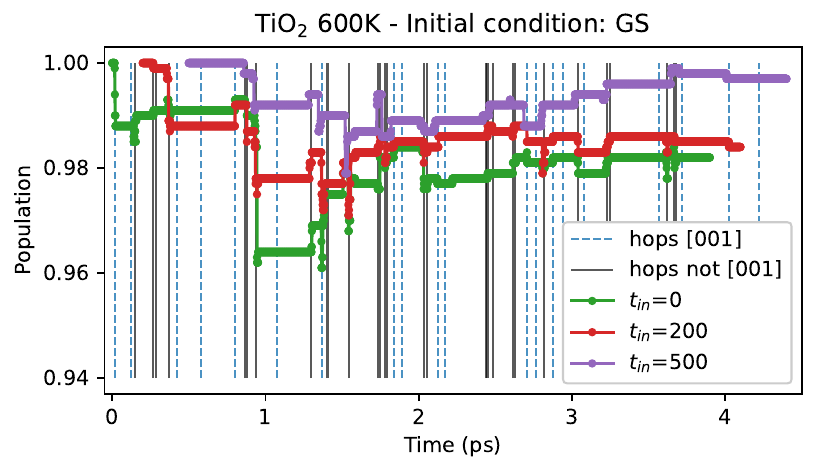}
            \put(0,53){\large (a)}
        \end{overpic}
         \phantomsubcaption
\label{fig:FigS6a_3iconds_TiO2_GS}
    \end{subfigure}
     %\hspace{0.cm}
\begin{subfigure}{.49\textwidth}
         \centering
         \begin{overpic}[width=\linewidth]{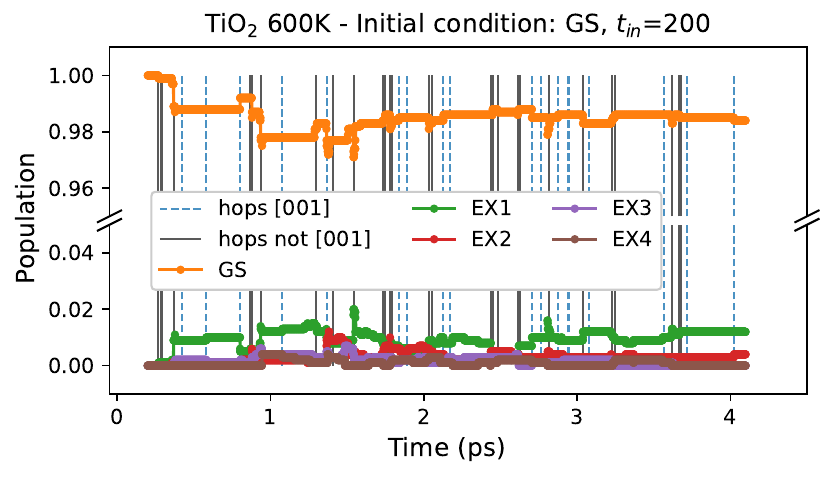}
            \put(0,53){\large (b)}
        \end{overpic}
        \phantomsubcaption
    \label{fig:FigS6b_1iconds_TiO2_GS}
    \end{subfigure}
%\vspace{0.5cm}

\begin{subfigure}{.49\textwidth}
         \centering
         
                  \begin{overpic}[width=\linewidth]{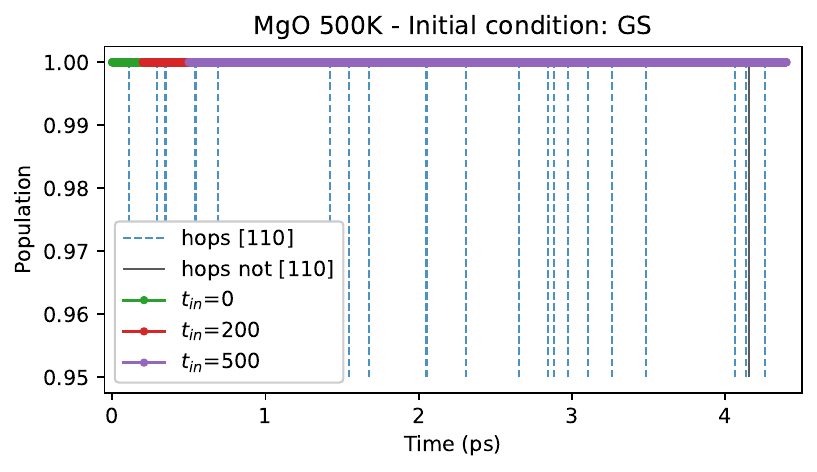}
            \put(0,53){\large (c)}
        \end{overpic}
        \phantomsubcaption
        \label{fig:FigS6c_3iconds_MgO_GS}
    \end{subfigure}
%\hspace{0.5cm}     
    \begin{subfigure}{.49\textwidth}
         \centering
         \begin{overpic}[width=\linewidth]{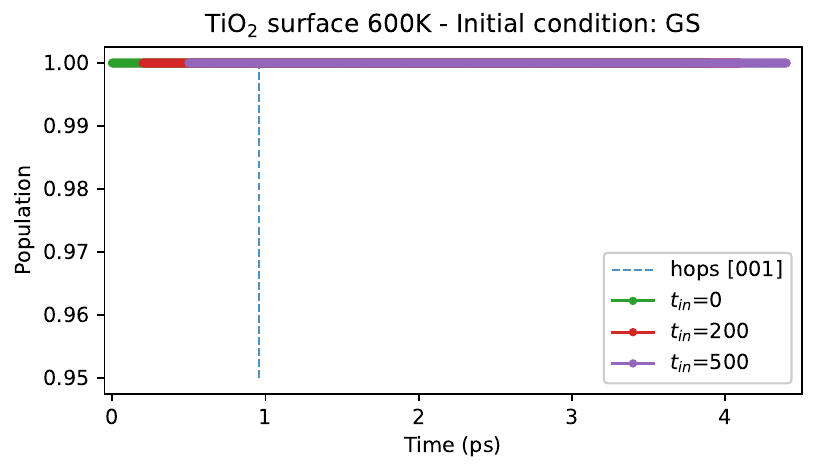}
            \put(0,53){\large (d)}
        \end{overpic}
        \phantomsubcaption
        \label{fig:FigS6d_3iconds_TiO2-surf_GS}
    \end{subfigure}

        \caption{\justifying  (a) Bulk rutile $\text{TiO}_{2}$ is prepared in the GS and evolved: its population changes slightly (1-3\%): it means that some electronic transitions occur, namely the electron is promoted towards the excited states. These discontinuities in the population plot can be associated to not [001] polaron hoppings. (b) Considering a representative trajectory, with initial time $\text{t}_{\text{in}}$=200\,fs, we can say that 2\% of GS population is promoted to excited states, specifically roughly 1.3\% to EX1 and 0.7\% to EX2. 
        On the contrary, (c) in MgO (at every temperature) and (d) in rutile surface at 600\,K, a GS$\rightarrow$EX1 is never observed.  }
        \label{fig:SM_single_iconds_GS}
\end{figure*}

Fig.~\ref{fig:SM_Excited_states} shows the de-excitation process of rutile bulk and MgO at 400 K and 600 K, rutile surface
at 600\,K. The systems have been initialized in the EX1 state, to study the evolution towards the GS. Fig.~\ref{fig:SM_Single_iconds_EX1} shows the populations for the systems prepared in EX1, along a few representative individual trajectories, each starting from a different initial geometry. Fig.~\ref{fig:SM_single_iconds_GS} shows the evolution in time of rutile bulk at 600\,K, MgO at 500\,K and rutile surface at 600\,K, initialized in the GS.

\begin{figure*}[htbp]
    \centering
     \begin{subfigure}{.45\textwidth}
         \centering

         \begin{overpic}[width=\linewidth]{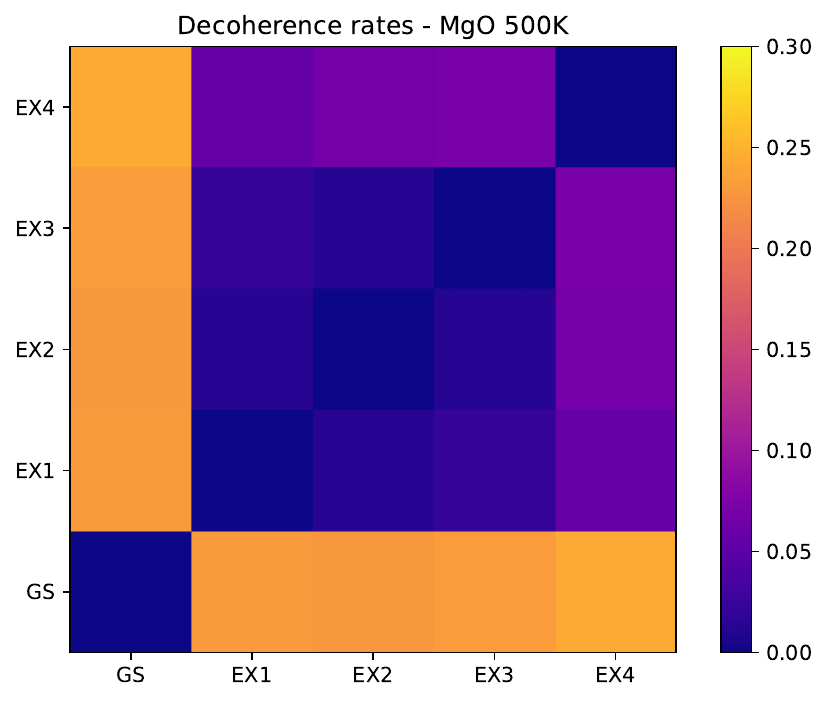}
            \put(0,80){\large (a)}
        \end{overpic}
         \phantomsubcaption
\label{fig:FigS7_Decoherence_rates_MgO}
    \end{subfigure}
     %\hspace{0.cm}
\begin{subfigure}{.45\textwidth}
         \centering
         \begin{overpic}[width=\linewidth]{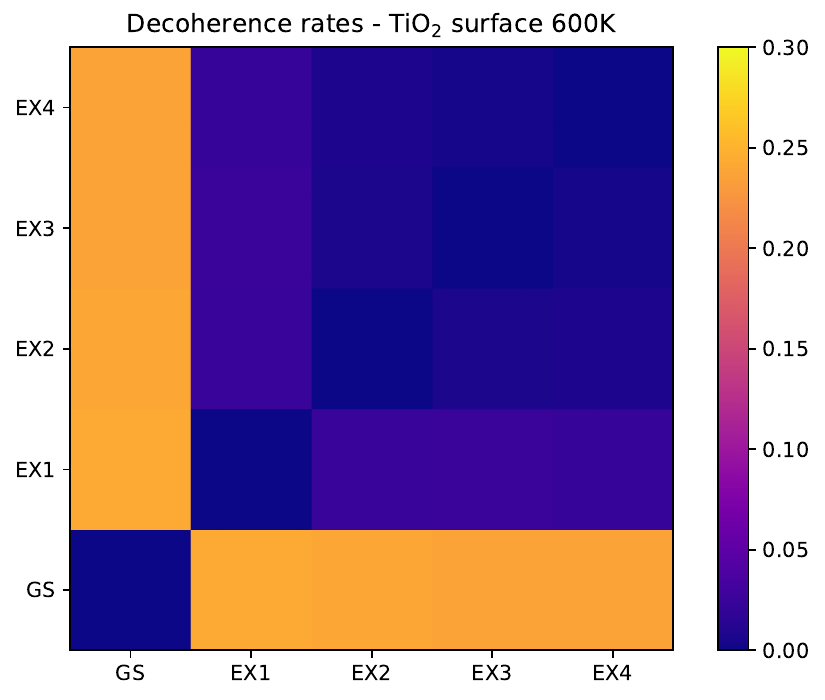}
            \put(0,80){\large (b)}
        \end{overpic}
        \phantomsubcaption
    \label{fig:FigS7_Decoherence_rates_TiO2-surf}
    \end{subfigure}
%\vspace{0.5cm}
        \caption{\justifying  Decoherence rates color map, for the simulation of (a) MgO at 500~K and (b) TiO$_2$ surface at 600~K. The
values reported in the plot have been obtained by averaging the decoherence matrices over the different initial geometries. In complete analogy with bulk rutile, the values are higher for pair of states involving the defect level.}
        \label{fig:decoherence_rates_colormap}
\end{figure*}

Fig.~\ref{fig:decoherence_rates_colormap} shows the decoherence rates matrix for MgO at 500~K and TiO$_2$ surface at 600~K.

%-----------------------------------------------

%-----------------------------------------------

